\documentclass{elsart}

\usepackage{graphicx}
\usepackage[latin1]{inputenc}
\usepackage{amssymb}
\usepackage{bm}
\usepackage[sort&compress]{natbib}
\journal{Progress in Particle and Nuclear Physics}

\newcommand{\rd}{\mathrm{d}}
\newcommand{\Zmoy}{\langle Z \rangle}
\bibpunct{[}{]}{,}{n}{}{}
\begin{document}

\begin{frontmatter}
\title
{Liquid-Gas Phase Transition in Nuclei}
\author[ipno]{B.~Borderie\corauthref{cor}},
\corauth[cor]{Corresponding author - borderie@ipno.in2p3.fr}
\author[ganil]{J.~D.~Frankland\ead{john.frankland@ganil.fr}},

\address[ipno]{Institut de Physique Nucl\'eaire, CNRS/IN2P3, Univ. Paris-Sud,
Université Paris-Saclay, F-91406 Orsay Cedex, France}
\address[ganil]{GANIL, (CEA/DRF-CNRS/IN2P3), BP 55027 F-14076 Caen, France}

\begin{abstract}
This review article takes stock of the progress made in understanding
the phase transition in hot nuclei and highlights the coherence of
observed signatures. 
\end{abstract}


\end{frontmatter}

\tableofcontents



\normalsize

\section{Introduction}

  Phase transitions are universal properties of interacting matter
  and traditionally they have been studied in the thermodynamic limit
  of macroscopic systems. A phase transition occurs when a phase becomes
  unstable in given thermodynamical conditions described with
  intensive variables like temperature, pressure {\ldots} 
  The interaction between nucleons in nuclei is similar to the interaction
  between molecules in a van der Waals fluid: a short-distance repulsive core and a
  long-distance attractive tail. It is the reason why Bertsch and
  Siemens~\cite{Ber83} suggested that the nuclear interaction should
  lead to a liquid-gas (LG) phase transition in nuclei. This original work
  also suggested that if the equation of state of nuclear matter is of van 
  der Waals type, nucleus-nucleus collision experiments may bring
  excited nuclei into the spinodal region of the phase diagram in
  which spinodal instabilities may develop exponentially and
  lead to the spectacular break-up of nuclei commonly called
  multifragmentation. Starting from this work, considerable theoretical 
  and experimental efforts were made to yield a better understanding
  of possible scenarios~\cite{WCI06}. In particular, one part of the theoretical
  effort was devoted to the consequences of finite size effects as far
  as the phase transition signatures are concerned~\cite{Gro01,LNP02,Cho04}. 
  With isolated
  finite systems like nuclei, the concept of thermodynamic limit
  cannot apply and extensive variables like energy and entropy are no
  longer additive due to the important role played by surfaces.
  On the experimental side, studies are performed using heavy-ion collisions
  at intermediate and relativistic energies and hadron-nucleus
  collisions at relativistic
  energies. Detailed studies of reaction products are obtained
  with powerful multidetectors~\cite{Sou06}
  allowing the detection of a large amount of the many fragments and light
  particles produced. It also appears that further progress is linked to 
  the knowledge of many observables which gives the possibility to study
  correlations inside the multifragment events and to realize very
  constrained simulations.
  
  This review is exclusively focused on manifestations of the
  nuclear LG phase transition in hot nuclei.
  A variety of reviews of nuclear multifragmentation and of related
  dynamical and statistical models are available for a thorough
  description and analysis of the
  field~\cite{MoWo93,Bon95,Gro97,Mor97,Das05,Vio06,Bor08,Giu14}.
  The present review is organized as follows. In sections~\ref{why},
  \ref{thermo} and~\ref{how}
  we explain why and how to study a phase transition in hot nuclei. 
  Section~\ref{phases} illustrates the liquidlike behaviour of nuclei
  in their ground states or at low excitation energies and the
  experimental evidence that at very high excitation energies
  they behave like a gas. Signatures of a first-order
  phase transition in hot nuclei are discussed in
  section~\ref{transition}; we present the wide range of predicted 
  behaviours and their experimental observations, 
  before concluding with coherency of the different signals.

\section{Why study a phase transition in hot nuclei?}\label{why}

Before presenting the phases of nuclear matter including the effects of
 different proton and neutron concentrations and the influence of
surface and Coulomb effects when going from infinite matter to nuclei,
we want to address some  general comments related to thermodynamics of
nuclei.
We all learned that phase transitions exist only in large systems,
strictly in the thermodynamic limit. However multifragmentation has 
long been known to be the dominant decay mode of a nucleus with $A$ nucleons
at excitation energy between around $3A$ and $10A$ MeV. From this
observation it became evident that concepts like entropy and phase
transitions apply to such very small many-body systems typically
composed of a few hundred of nucleons. Therefore an extension of conventional
macroscopic and homogeneous thermodynamics to such finite systems was needed.
A few words now about the concept of temperature, which was largely and successfully
used at low excitation energies. At high excitation energies, to use it,
one has to admit that nuclei have enough time to thermalize during
collisions. From the theoretical side it was shown that energy
relaxation can be totaly fulfilled depending on bombarding
energies~\cite{Ber78,Toe82,Cug84,Cas87} and experimental results have
confirmed these expectations~\cite{Jou91,I15-Bor99}.
Another more conceptual point which is of relevance to the nuclear
decay problem is concerned with the ergodic hypothesis which is used
in connection with single systems. The essential idea behind the
ergodic hypothesis is that a system in equilibrium evolves through a
representative set of all accessible microstates over a time interval
associated with a measurement. For ergodic systems, a
theoretical treatment of equilibrium can be constructed either in
terms of the properties of a single system measured over an infinite
time or, more conveniently, in terms of properties of a pseudo/fictive
ensemble of constrained systems which provides a representative sample
of all attainable configurations. This last possibility is relevant
for studying the phase transition of nuclei even though the ergodic
hypothesis does not apply in this case. Indeed the chaotic character
of collisions involved to produce hot nuclei favors a large covering of statistical
partitions when an homogeneous event sample is studied. This
discussion will be developed and deepened in the next section.

\subsection{Nuclear matter: the liquid-gas phase
transition}\label{nucmat}
\begin{figure}[htb]
\begin{center}
 \includegraphics[scale=0.75]{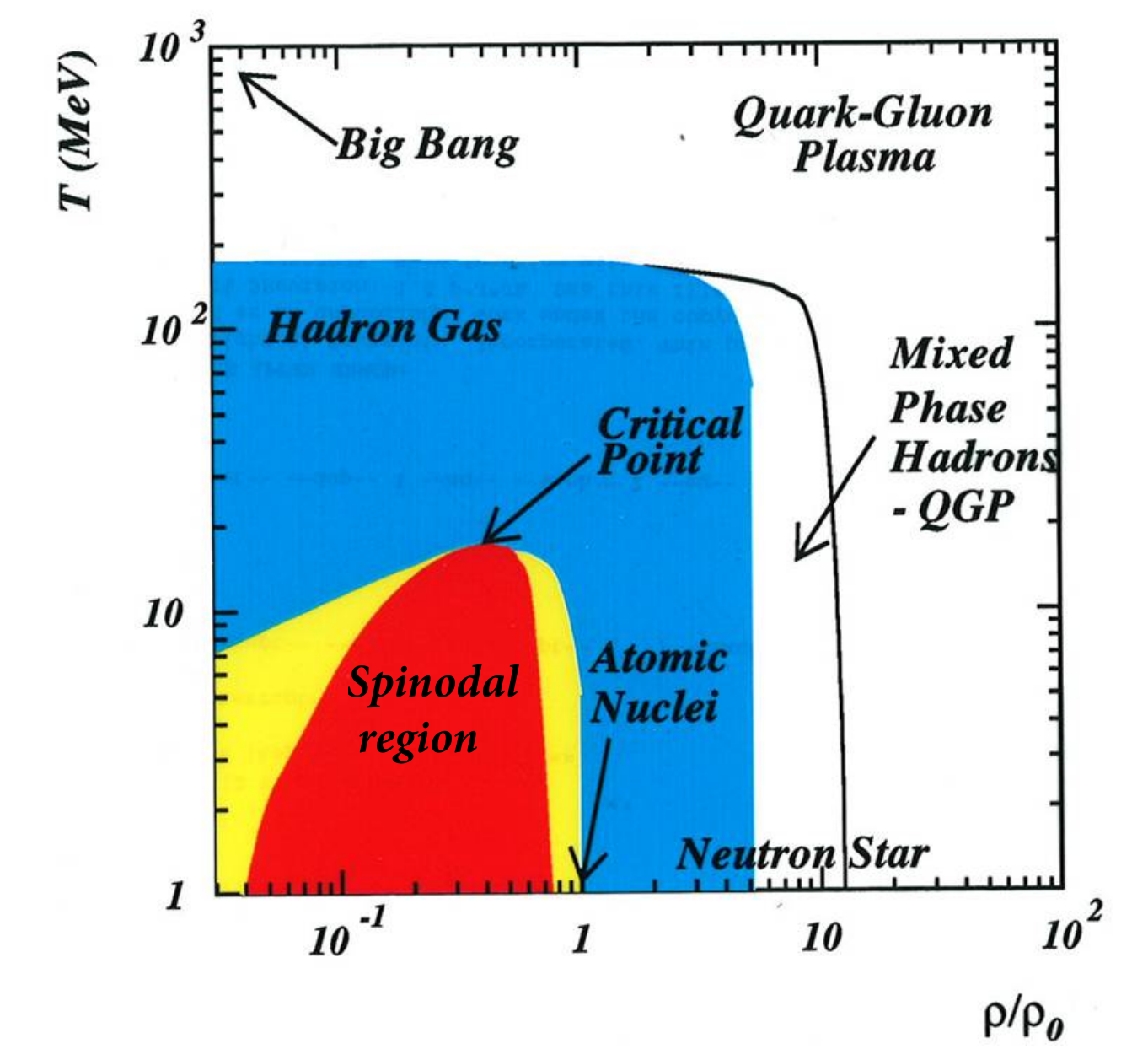}
\end{center}
\caption{Phase diagram of nuclear matter: the horizontal axis shows
the matter density, and the vertical axis shows the temperature. The
density $\rho$ is given normalized to the saturation density $\rho_{0}$.
The liquid-gas mixed phase region (yellow area) which ends up at the critical
point contains the spinodal region (red area).
}
\label{fig:diagphas}
\end{figure}

Nuclear physics is a field that interconnects very much to adjacent
fields such as elementary particle physics (at the higher energy end) or
astrophysics. We will concentrate on the nuclear region below
30 MeV per nucleon excitation energy (equivalently $\thicksim$ 25 MeV temperature)
and below density $\rho$ equal to 2-3 times the normal nuclear density
$\rho_{0}$, which is deduced from the maximum of saturation density of
finite nuclei and estimated as 0.155~$\pm$0.005 
nucleons fm$^{-3}$~\cite{Mar18}.
This represents only a rather small
portion of the nuclear matter phase diagram, as predicted theoretically
and displayed in Fig.~\ref{fig:diagphas}, if we note that on the figure
both axes are shown in logarithmic scale.

\subsubsection{Symmetric matter}\label{symmat}
Symmetric nuclear matter is an idealized macroscopic system with an equal number
of neutrons and protons. It interacts via nuclear forces, and Coulomb
forces are ignored due to its size. Its density $\rho$ is spatially uniform.
The nucleon-nucleon interaction is comprised of two components according
to their radial interdistance : a very short-range repulsive part
which takes into account the incompressibility of the medium and a long-range
attractive part. Changed by five orders of magnitude the nuclear interaction is
very similar to van der Waals forces acting in molecular media and
consequently
the phase transition in nuclear matter resembles  the LG
phase transition in classical fluids. 
However, as compared to classical fluids the main difference comes from the
gas composition.
For nuclear matter the gas phase is predicted to be composed not only of
single nucleons, neutrons and protons, but also of complex particles
like alpha-particles and light
fragments depending on temperature conditions~\cite{Bug00,Bug01}.
In some sense, strictly speaking, one should speak of a liquid-vapour
phase transition.

\begin{figure}[htb]
\begin{center}
 \includegraphics[scale=0.75]{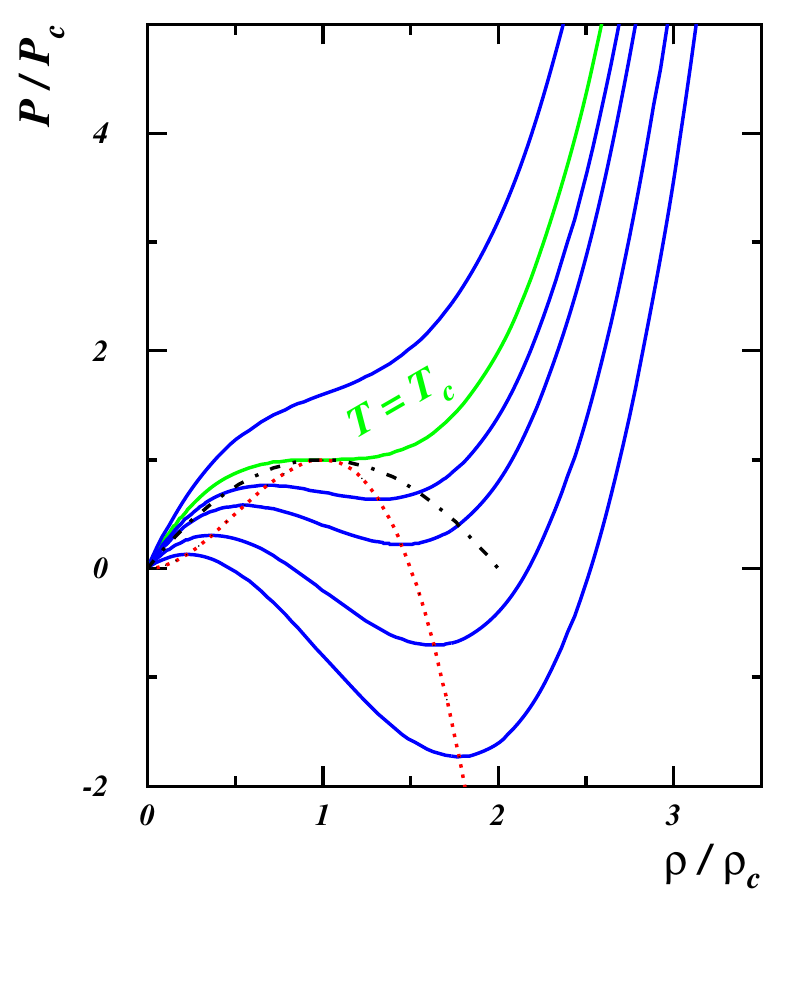}
 \includegraphics[scale=0.75]{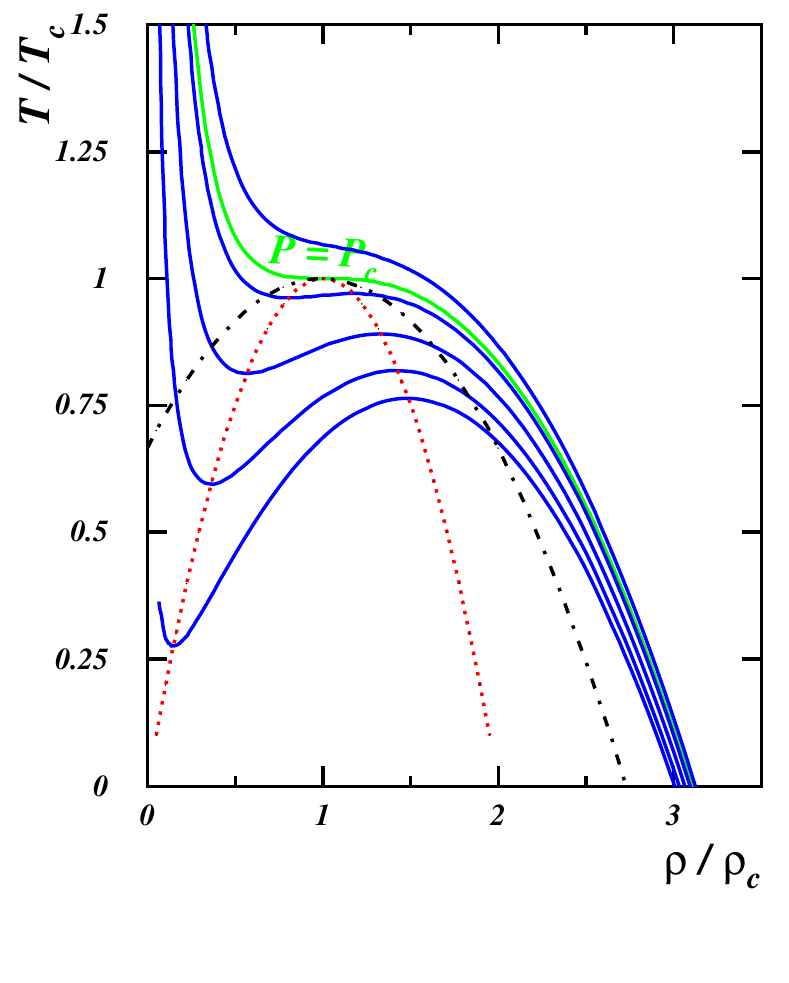}
\end{center}
\caption{Equation of state relating the pressure (left) or the temperature
(right) and the density (normalised to critical values) in nuclear matter.
The curves represent isotherms (left) and isobars (right). The dashed-dotted
lines are the boundaries of the coexistence region and the red dotted lines the 
boundaries of the spinodal region.
From~\protect\cite{I46-Bor02}.}
\label{fig:eos}
\end{figure}

A set of isotherms for an
equation of state (pressure versus density) corresponding to
 nuclear forces (Skyrme effective interaction and finite temperature
Hartree-Fock theory~\cite{Jaq83}) is shown
in Fig.~\ref{fig:eos}.
It exhibits the maximum-minimum structure typical of van der Waals
equation of state.
Depending on the effective interaction chosen and on the
model~\cite{Jaq83,Jaq84,Cse86,Mul95}, the nuclear
equation of state (EOS)
exhibits a critical point at $\rho_{c}\approx$ 0.3-0.4$\rho_{0}$ and
$T_{c}\approx$ 16-18~MeV. 
The region below the dotted line in Fig.~\ref{fig:eos}
corresponds to a domain of negative compressibility: at constant temperature
an increase of density is associated to a decrease of pressure.
Therefore in this region density fluctuations will be catastrophically amplified
until matter becomes inhomogeneous, separated into domains of high
(normal) liquid density and low density gas, which finally form two
coexisting phases in equilibrium. It is the so-called
spinodal region and spinodal fragmentation (decomposition) is the dynamics
of the phase transition. Instability growth times are equal to around
30-50~fm/c (30~fm/c~=~10$^{-22}s$) depending on density ($\rho_{0}$/2 - $\rho_{0}$/8) and
temperature (0 - 9MeV)~\cite{Colo97}. 
Spinodal instabilities have long been proposed as the
mechanism responsible for multifragmentation~\cite{Ber83,Hei88,Lope89}. 
The spinodal region constitutes the major part of the coexistence region
(dashed-dotted line in Fig.~\ref{fig:eos}) which also contains two
metastable regions: one at density below
$\rho_{c}$ for the nucleation of drops and one above $\rho_{c}$
for the nucleation of bubbles (cavitation). 

\subsubsection{Asymmetric matter}\label{asymmat}
Asymmetric nuclear matter, i.e. when the ratio of neutrons to protons is no more
equal to one, is evidently a richer subject of research because its
equation of state is relevant for both nuclear physics and astrophysics.
In recent years, given the stimulating perspectives
offered by new radioactive ion beam facilities and nuclear
astrophysics, an important theoretical activity has been
developed and reviews are available~\cite{Bar05,Bao08,Hor14,Bal16}.
Thermodynamic properties have been studied starting from
non-relativistic and relativistic effective interactions and, in general, 
the physics is not dependent on the theoretical framework.
In asymmetric matter, the energy per nucleon, i.e. the EOS, 
is a functional of the total ($\rho = \rho_{n} + \rho_{p}$) and
isospin ($\rho_{3} = \rho_{n} - \rho_{p}$) densities. In the usual
parabolic form in terms of the asymmetry parameter
$I \equiv \rho_{3}/\rho = (N-Z)/A$ we can define a symmetry energy
$\frac{E_{sym}}{A}(\rho)$:
 \begin{center} 
 $\frac{E}{A}(\rho,I) =\frac{E}{A}(\rho,I=0)
   +\frac{ E_{sym}}{A}(\rho) I^2$   
 \end{center}
 The first term is the isoscalar term, invariant under proton and neutron
 exchange,  while the second (isovector) one gives the correction brought by
 neutron/proton asymmetry. For $I$=1 this term gives the equation of state
 of neutron matter. Note that because $I$ is, for most nuclei, smaller than 
 0.3, the isovector term is much smaller than the symmetric part, which
 implies that isospin effects should be rather small and all the more
 difficult to evidence.
 The symmetry term (Eq. (\ref{eq:esym})) gets a kinetic contribution directly
 from Pauli  correlations and a potential contribution from the properties of 
 the isovector part 
 of the effective in-medium nuclear interactions used in models.
\begin{equation}
\frac{E_{sym}}{A}(\rho)=\frac{\varepsilon_F (\rho)}{3}+ 
\frac{C}{2} F(\rho /\rho _0)
\label{eq:esym}
 \end{equation}
 $\epsilon_F$ is the Fermi energy, $F(1)$=1 and $C\approx$32 MeV.
 For convenience in comparing different
 implementations, symmetry energy is commonly approximated as :
\begin{equation}
\frac{E_{sym}}{A}(\rho)=\frac{C_{s,k}}{2} (\frac{\rho}{\rho_0})^{2/3}
+ \frac{C_{s,p}}{2} (\frac{\rho}{\rho _0})^{\gamma} \label{gamma}
\label{eq:asy}
 \end{equation}
$\gamma$ defines the ``asy-stiffness'' of the EOS around normal density. 
The symmetry energy is said to be
``asy-soft'' if  $E_{sym}^{pot}$ presents a maximum (between 
$\rho_0$ and $2\rho_0$), followed by a decrease and vanishing 
($\gamma<$1)
and ``asy-stiff'' if it monotonically increases with $\rho$ ($\gamma \geq$1).
Constraining the density dependence of the symmetry energy from nuclear
structure measurements, heavy ion collisions and astronomical
observations is in progress~\cite{Hor14,Bal16}.
 
\begin{figure}[htb]
\begin{minipage}[b]{0.49\textwidth}
 \includegraphics[width=\textwidth]{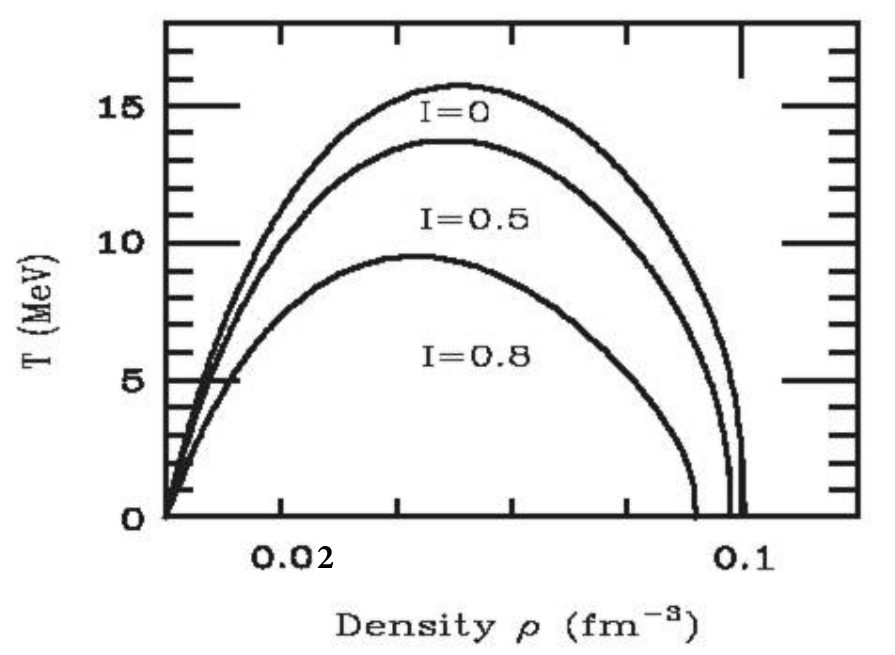}
\caption{Spinodal boundaries in the density-temperature plane for
different asymmetries $I$. Instability regions are under the curves.
From~\cite{Bar98}.}
\label{fig:spinoas}
\end{minipage}%
\hspace*{0.02\textwidth}
\begin{minipage}[b]{0.49\textwidth} 
\includegraphics[width=1.1\textwidth]{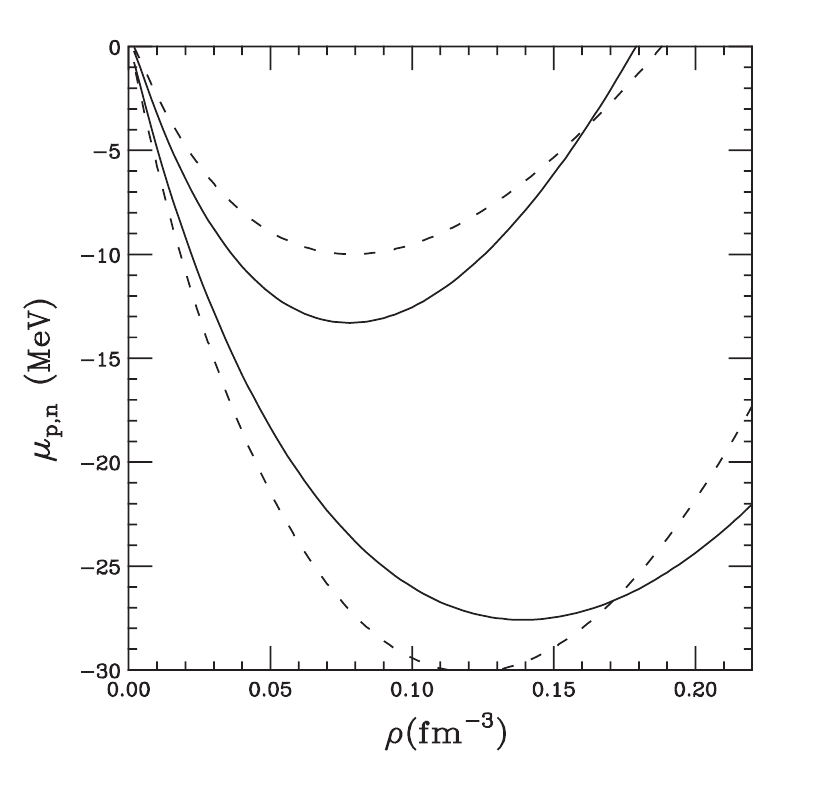}
\caption{Density dependence, for $I$=0.2, of neutron (upper curves) 
and proton (lower curves) chemical potentials for asy-superstiff ($\gamma
\sim$1.6 - solid lines)  and asy-soft ($\gamma$=0.5 - dashed lines) EOS.
From~\cite{Bar05}.}
\label{fig:chempot}
\end{minipage}
\end{figure}

There are two qualitative new features of the LG phase transition
in asymmetric matter. Firstly the asymmetry leads to shrinking of the region of
instabilities, the spinodal region, with a reduction of
critical temperature and density~\cite{Mul95,Bar98} (see
Fig.~\ref{fig:spinoas}). Note a peculiarity of asymmetric nuclear matter,
the direct correspondence between the nature of fluctuations and the
occurrence of mechanical or chemical instabilities is lost and we face
a more complicated scenario with the uniqueness of the unstable modes
in the spinodal region; the instability is always dominated by total
density fluctuations even for large asymmetries. This has been clearly
shown in the framework of linear response theory and in full transport
simulations~\cite{Mar03,Bao02}.
Such a result is due to gross properties of the $n/p$ interaction.
The second new feature is what is called an isospin distillation,
 strictly speaking neutron distillation, which produces a liquid phase
composed of more symmetric matter (minimization of symmetry energy in
the dense phase) and a neutron rich gas.
The origin of this phenomenon
is easily understood when looking at the evolution of the neutron 
and proton chemical potentials with density, as displayed in
Fig.~\ref{fig:chempot}. We recall that the chemical
potential is the derivative of the energy with respect to the number of
particles of the system. The differences of the local chemical potentials,
for neutrons and protons, which can be expressed as
$\mu_n - \mu_p = 4 E_{sym}(\rho) I/A$, governs the mass flow in non
equilibrium systems. 
In the density region corresponding to the LG coexistence 
($\rho \lesssim 0.6\rho_0$, i.e. $\rho \lesssim$ 0.10 in the figure ) 
one can observe that neutrons and protons move in phase, both towards 
higher $\rho$. The slope of $\mu_p$ is however steeper than that of 
$\mu_n$. This means that the liquid clusters (high density) produced by
bulk instability will be more symmetric while the gas phase 
(low density) will get enriched in neutrons. As the difference between 
the chemical potential slopes is more marked for an asy-soft EOS 
(dashed lines), the distillation effect will be stronger in that case.

\subsection{From nuclear matter to hot nuclei}\label{hotnuc}
Evidently the hot piece of nuclear matter produced in any nuclear
collision  has at most a few hundred nucleons and so is not adequately
described by the properties of infinite nuclear matter; surface
and Coulomb effects cannot be ignored. These effects have been
evaluated and lead to a sizeable reduction of
the critical temperature~\cite{Jaq83,Jaq84,De99}.
Finite size effects have been found to reduce the critical temperature by
2-6 MeV depending on the size of nuclei while the Coulomb force is responsible
for a further reduction of 1-3 MeV. However large reductions due to small
sizes are associated with small reductions from Coulomb. Consequently,
in the range $A$ = 50-400 a total reduction of about 7 MeV is calculated leading
to a ``critical'' temperature of about 10 MeV for nuclei or nuclear
systems produced in collisions between very heavy nuclei. 
The authors of reference~\cite{Jaq84}  indicate
that, due to some approximations, the derived values
can be regarded as upper limits. Finally we can recall that,
in infinite nuclear matter, the binding energy per particle is 16 MeV 
whereas it
is about 8 MeV in a finite nucleus. Clearly these values well compare
with the  $T_{c}$ values for infinite nuclear matter and finite systems just
discussed. 

\begin{figure}[htb]
\begin{minipage}[b]{0.49\textwidth}
 \includegraphics[width=\textwidth]{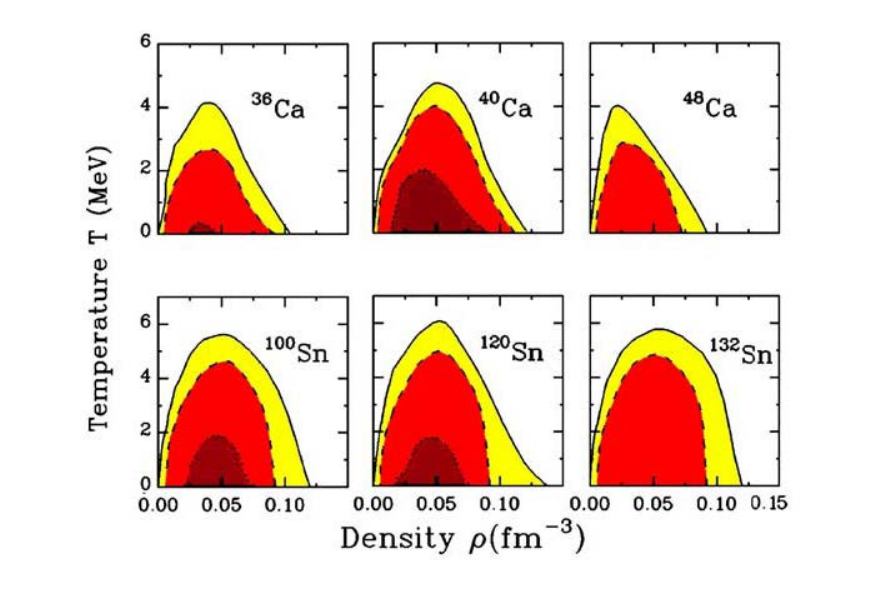}
\caption{Spinodal border (full line) in the density-temperature plane
associated to $L$ = 3, for Ca and Sn isotopes. Points having the same
growth time equal to either 100fm/c (dashed) or 50fm/c (dots) are also
delineated. From~\cite{Col02}.}
\label{fig:spinonuc}
\end{minipage}%
\hspace*{0.02\textwidth}
\begin{minipage}[b]{0.49\textwidth} 
\includegraphics[width=1.1\textwidth]{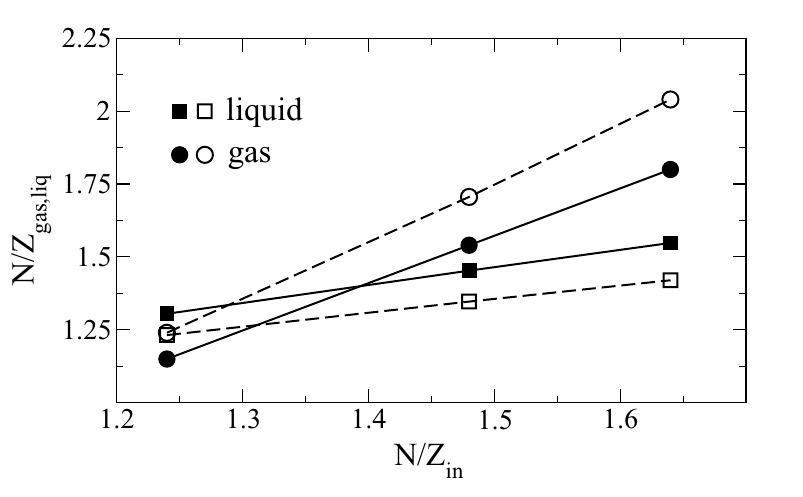}
\caption{The asymmetry N/Z of the gas (circles) and of the liquid 
(squares) phase for central Sn+Sn collisions with different initial 
N/Z. Solid lines and solid symbols refer to the asystiff parametrization,
and dashed lines and open symbols refer to the asy-soft parametrization.
From~\cite{Colo08}.}
\label{fig:LGNZ}
\end{minipage}
\end{figure}

For finite systems composed of asymmetric matter a quantal approach
has been used to determine the spinodal region~\cite{Col02}. A quite
complex structure of the unstable modes is observed in which volume
and surface instabilities are generally coupled and cannot be easily
disentangled. For each multipolarity, $L$, several unstable modes appear.
Fig.~\ref{fig:spinonuc} shows, for octupole instabilities, spinodal regions
in the density-temperature plane for Ca and Sn isotopes. Heavier
systems have a larger instability region than the lighter ones. Moreover,
more asymmetric systems are less unstable.

For isospin distillation, dynamical simulations were performed for
central ($b=$2~fm) symmetric Sn+Sn collisions, with masses 112, 124 
and 132, at 50~MeV per nucleon incident energy~\cite{Colo08} by using two forms of 
the symmetry energy in the interaction, a stiff one
corresponding to $\gamma \sim$ 1.6 
and a very soft one ($\gamma \sim$ 0.2-0.3). The isospin content of the
liquid and gas phases (here assimilated to fragments 
with 3~$\leq Z\leq$~10 and light particles, respectively)
is depicted as a function of the initial $N/Z$
in Fig.~\ref{fig:LGNZ}. The fragments here are the primary hot ones.
It appears that the $N/Z$ of the gas phase is 
larger than that of the liquid; the difference increases with the 
initial $N/Z$, and is larger in the asy-soft case because the symmetry 
energy at low density is larger. For the less neutron-rich system, 
the liquid phase is more neutron-rich than the gas in the asy-stiff 
case; this inversion is caused by Coulomb effects which become dominant 
over symmetry effects, leading to a strong proton emission. Finally 
one can notice that $I_{frag} < I_{syst}$ for n-rich systems and 
conversely $I_{frag} > I_{syst}$ for ``n-poor'' systems. 

\section{Applications of thermodynamic concepts to
heavy-ion collisions and hot nuclei}\label{thermo}

In this section we will attempt to provide the reader with the necessary
theoretical background to understand, as will be presented in the
following sections, how it is possible to study a phase transition
in atomic nuclei. The two major obstacles to this endeavour concern
the problem of phase transitions in finite systems, and the application
of statistical mechanics to processes occurring in the dynamics of
finite, open systems.

Any experiment we can
perform is obviously far from the thermodynamic limit: the largest
possible hot nucleus/nuclear system that could conceivably be produced experimentally
would have less than five hundred nucleons ($^{238}$U + $^{238}$U
collisions); and in actual fact is more realistically limited to
$\sim$~200 - 300
nucleons due to reaction dynamics and the Coulomb repulsion between
protons. Finite (small) systems require a specific statistical mechanical
treatment, for which there now exists a vast literature: apart from
advances specifically concerning small systems, this includes also
the wider field of statistical mechanics of systems with long-range
interactions. Long-range systems interact with a potential
which decays at large
distances like $r^{-\alpha}$, where $\alpha\leq d$; $d$ is the
dimension of the space where the system is embedded. 
For such systems the total energy per particle diverges in the thermodynamic
limit. Small systems can be seen as a special case of the latter where
the interaction range, although short, is of the order of the system
size. We will try to present in this section a review of the essential
aspects of this field, many of which may still be relatively new to
non-practitioners.

The formation and decay of nuclear systems undergoing multifragmentation
or vaporization occurs, according to various dynamical simulations
(see for example~\cite{Riz07,Bon14,Xu16}),
on timescales of between a few tens and a few hundreds of fm/c
($10^{-22}-10^{-21}$
seconds). Although transport models predict that nucleon-nucleon collisions
can rapidly thermalize nucleon momentum distributions at Fermi energies
and above, the application of statistical equilibrium concepts seems
counter-intuitive when dealing with highly-excited systems which disintegrate
almost as soon as they are formed.
Given that reaction
products are produced on a timescale which is comparable with the
time for the projectile to \textquoteleft cross\textquoteright{} the
target, the success of equilibrium models could imply that the dynamical
evolution of the system prior to multifragmentation is important only
insofar as it determines the constraints which are required to characterize
effective statistical ensembles in order to understand the 
data \citep{Col97,Cho06}.
To end this section, we will further develop these points and explain
the paradigm shift required in order to progress with the identification
of a phase transition in hot nuclei.

\subsection{Statistical mechanics for finite systems}\label{statfs}
In the beginning was thermodynamics. Thermodynamics is an empirical
science created to understand the functioning of steam engines (Carnot
cycles, thermal equilibrium, entropy, etc.) - macroscopic systems
with short-range interactions. Then came statistical mechanics, whose
fathers (Boltz\-mann and Gibbs) sought to give a microscopic grounding
for thermodynamics by relating the microscopic properties of $N$-body
interacting systems to their macroscopic behaviour, thus introducing
the concept of statistical ensembles. Most of the applications of
statistical mechanics during the first century of its existence were
used to explain or predict the macroscopic behaviour of matter starting
from well-established microscopic interactions. These were invariably
short-ranged interactions, and always in the thermodynamic limit.
In this case the time-averaged properties of a single system can be
calculated using a statistical ensemble of equivalent fictitious systems
(the property known as \emph{ergodicity}), and the physicist is free
to choose whichever statistical ensemble is the easiest to work with
in order to find the result (this is called \emph{ensemble equivalence}). 

As this situation lasted for nearly a century, it is quite natural
that the \emph{assumptions} that always worked in these cases lost
their original significance and became seen as \emph{prerequisites}
for statistical mechanics to be valid. In terms of education, as these
were the only cases which were widely known, they were also the only
ones to be widely taught, thus perpetuating the deeply-held conviction
that they were \emph{sine qua non} conditions for the validity of
statistical mechanics. It was mostly forgotten that statistical mechanics,
as it was used and taught, was nothing but an approximation, whose
validity depended on certain assumptions, such as additivity and the
existence of the thermodynamic limit, which is an application of the
law of large numbers.

Obviously, for macroscopic systems of particles interacting with short-range
interactions, statistical mechanics (and thermodynamics) is a very
\emph{good} approximation; indeed the approximation is so good that
it is to all intents and purposes an \emph{exact} description of the
macroscopic properties of such systems. In this case, using the true
exact method to calculate the properties of such systems, i.e. $N$-body
molecular dynamics where $N$ is of the order of the Avogadro number,
would both be intractable and, frankly, overkill: there is no need
to calculate the exact dynamics of such systems when 3 thermodynamic
variables are sufficient to describe their behaviour with a level
of precision which is far superior to the resolution of any experimental
measurement.

Cracks appeared in the foundations of thermodynamics when people tried
to apply it to something that was not a steam engine: for example self-gravitating
systems \emph{i.e.} stars \citep{Lyn68,Thi70},
or phase transitions in small systems such as atomic 
clusters \citep{Lab90,Wal94,Kun94,Che92},
and, of course, hot nuclei \citep{Gro97,Gulm02}.
The suggestion that such systems could exhibit a negative heat capacity
when described microcanonically, whereas the canonical heat capacity
is always positive by construction, thus violating ensemble equivalence,
provoked a crisis of statistical mechanics which was almost of the
same order as the crisis of physics itself at the turn of the twentieth
century. Reactions varied from violent rejection to the conviction
that the theory must quite simply be wrong, or that the apparent ensemble
\emph{inequivalence} must be a simple artefact due to some inappropriate
approximation or hypothesis.

Nowadays, when such phenomena have been explored using many different
approaches (and, in some cases, even measured) for many different
systems of different types, both with long-range interactions, or,
as in the case which particularly interests us in this review article,
finite systems, and with a solid general theoretical grounding to
explain their existence \citep{Tou09,Cam09}
(even though this has only reached fruition over the last ten years),
the particular properties of their statistical mechanics should no
longer be an affront to the sensibilities of even the most hardened
thermodynamicist.

\subsubsection{Non-additivity, ensemble inequivalence and non-concave 
entropies\label{subsec:Non-extensivity-and-non-additivi}}
One of the most important differences between short-range and/or macroscopic
systems and long-range or finite systems is non-additivity. In a macroscopic
system with short-range interactions, if $X$ is some extensive variable
characterizing the system (extensive quantities are proportional to
the system size), then splitting the system into two (macroscopic)
subsystems, $A$ and $B$, they will be characterized by the quantities
$X_{A}$ and $X_{B}$, with $X_{A\cup B}=X_{A}+X_{B}$. To be more
rigorous, we can write
\[
\lim_{N\rightarrow\infty}X_{A}+X_{B}+X_{AB}\rightarrow X_{A\cup B}\;\mathrm{(additive)}
\]
where $X_{AB}$ is the contribution from the interaction or surface
between the two subsystems. In the thermodynamic limit, for short-range
systems,
\[
\lim_{N\rightarrow\infty}X_{AB}\rightarrow0
\]
because in this case the interaction only occurs at the surface between
the two subsystems, which becomes negligible compared to the bulk
for a macroscopic system.

On the other hand, for systems with long-range interactions, the interaction
contribution $X_{AB}$ concerns the whole system and never disappears,
even in the thermodynamic limit. For small systems, on the other hand,
even if interactions are short-range, the contribution from the surface
between the two subsystems can be of the same order as that of the
``bulk'', and so cannot be neglected. In this case, 
\[
X_{A\cup B}\neq X_{A}+X_{B}\;\mathrm{(non-additive)}
\]

It is important to understand the subtle difference between additivity
and extensivity. Some early works on the statistical mechanics of
small systems~\cite{Gro01,She06}
mistakenly identified non-extensivity as the key to understanding
their behaviour, but it is in fact non-additivity which is responsible
for the unusual properties of both long-range and finite systems~\cite{Mor13}.
A system may well be extensive (for example, with a total energy proportional
to the number of particles in the system) and yet be non-additive
(total energy of system not equal to the sum of energies of its subsystems):
for example, the Curie-Weiss model of interacting spins on a lattice
(see~\cite{Cam09}). On the other hand,
non-extensive systems can never be additive.

Non-additivity has profound consequences for statistical mechanics.
The most important and far-reaching is the possibility for different
thermodynamic ensembles to give different predictions of the system's
behaviour: this is called 
\emph{ensemble inequivalence}~\cite{Elli00,Barr01,Tou11}.
This is at variance with the still widely-held - and widely taught
- belief that the microcanonical and the canonical ensembles should
always predict the same equilibrium properties of many-body systems
in the thermodynamic limit. This is in fact a special case, albeit
one which holds for most macroscopic systems: those with short-range
interactions.

Most striking are the differences observed between microcanonical
and canonical ensembles. To see how this comes about, let us first
consider the textbook method to derive the canonical probability distribution
by imagining a system divided into a subsystem $A$ of interest and
a (much larger) subsystem $B$ which plays the role of a heat reservoir.
Central to the derivation is the assumption that the energy is additive,
which allows to write $E_{B}=E-E_{A}$. Obviously, when the interaction
energy between subsystems $E_{AB}$ is not negligible because of non-additivity,
this assumption breaks down; this does not mean that the 
canonical ensemble cannot be defined, however, as we will see below.

The van Hove theorem \cite{Hov49} states that for thermodynamic
stability, thermodynamic potentials such as the entropy must be everywhere
concave. If a system's microcanonical entropy were locally non-concave
in some energy interval $[E_{1},E_{2}]$, the argument goes, it would
maximize its entropy (and thus recover concavity) at any intermediate
energy $E=\lambda E_{1}+(1-\lambda)E_{2}$ (with $0\leq\lambda\leq1$)
by dividing into subsystems with energies $E_{1}$ and $E_{2}$ and
a combined entropy $S_{12}=\lambda S(E_{1})+(1-\lambda)S(E_{2})\geq S(E)$
(in other words it would undergo phase separation). 
As discussed in
\citep{Chal88,Cho06}, van Hove's theorem does not
apply to non-additive (finite) systems. For non-additive systems phase
separation at fixed energy is not possible because of the non-negligible
interaction energy $E_{12}$, and therefore in the microcanonical
ensemble the convex region of the entropy corresponds to equilibrium
states; on the other hand, in the canonical ensemble where the energy
\emph{is} free to fluctuate, such states are highly improbable and
practically unobservable: ensemble equivalence is violated.

\subsubsection{The large deviation theoretical picture of statistical mechanics}
In recent years Ellis, Touchette et al~\cite{Elli00,Tou04}
have provided the most comprehensive and sound basis for the understanding
of the relation between ensemble inequivalence, non-concave entropies
and phase transitions. Indeed, they have provided almost a re-foundation
of statistical mechanics which ensures, among other things, the correct
description of systems with long-range interactions, or, equivalently,
finite systems, both in and out of equilibrium. 

Defining statistical mechanics as a tool to find the most probable
(macro)states of a random system of particles in interaction, i.e.
equilibrium states, they turn to the mathematical theory of large
deviations which is concerned with limiting forms of the probability
distributions of fluctuations. If the probability that some random
variable $A_{n}$ takes a value in the set $A$ can be expressed as
\begin{equation}
P_{n}(A_{n}\in A)\approx\exp-nI_A\label{eq:large-deviation}
\end{equation}
where $I_A$ is a positive constant, then it is said to satisfy a \emph{large
deviation principle}~\cite{Tou09}. $n$ is a parameter which is assumed
to be large; it could, for example, be the number of particles or some
other measure of the system size. In this
case, the most probable value(s) of $A_{n}$ will be determined by
the minimum(a) of the \emph{rate function}, $I_A$, defined by the limit 
\[
\lim_{n\rightarrow\infty}-\frac{1}{n}\ln P(A_{n}\in B)=I_A
\]
which must exist if Eq.~(\ref{eq:large-deviation}) holds.

\begin{figure}
\includegraphics[clip,width=1\textwidth]{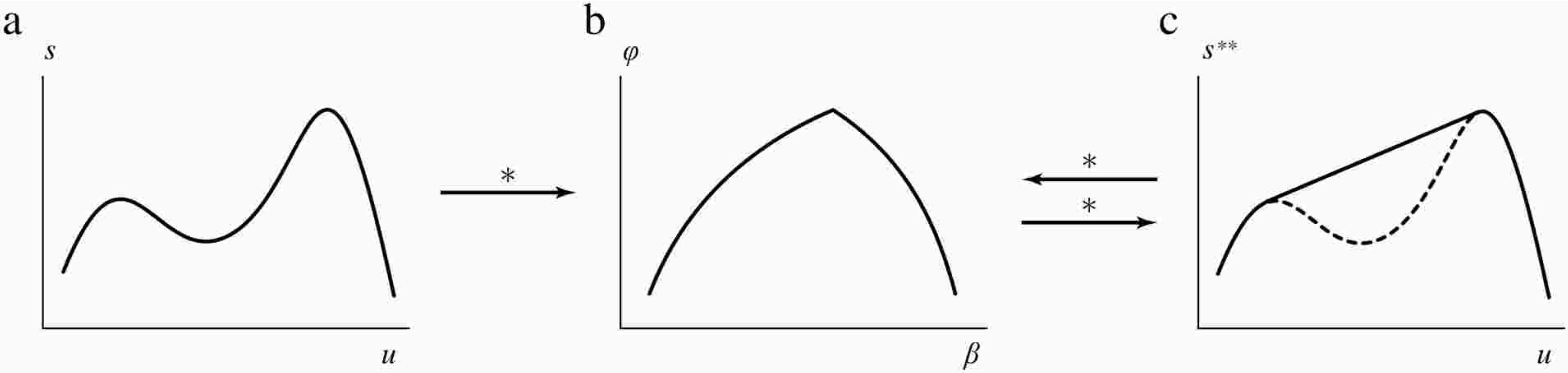}

\caption{Ensemble inequivalence due to non-concavity of the entropy. (a) An
entropy which is not concave everywhere, i.e. with a convex intruder
(note that here it is the entropy per particle $s=S/N$ which is presented
as a function of energy per particle $u=E/N$); (b) The LFT of $s(u)$
(represented by the arrow with an asterisk) is the (Massieu) free
energy per particle $\phi(\beta)$: the non-concave part of $s(u)$
results in a non-differentiable point in $\phi(\beta)$; (c) a further
LFT of $\phi(\beta)$ produces the concave hull of $s$, $s^{**}(u)$
(solid line). From~\cite{Tou09}\label{fig:Ensemble-inequivalence}.}

\end{figure}

Applying this theory to determine the probability of measuring the
mean energy of a thermodynamic system at a given fixed value, it turns
out that the rate function in this case is the (negative) microcanonical
entropy, as defined by Boltzmann: in other words, the fact that the
most probable (equilibrium) state of a system at fixed energy maximizes
the entropy is a natural consequence of the large deviation principle,
Eq.~(\ref{eq:large-deviation}). Similarly, the most probable state
of a system whose energy can fluctuate is determined by the minima
of a rate function which is closely related to the canonical free
energy. 

Furthermore in this framework the canonical free energy (or, more
precisely, the Massieu potential $\Phi(\beta)=\beta F(\beta)$) is
obtained quite naturally from the microcanonical entropy by a \emph{Legendre-Fenchel
transform }(LFT)
\begin{equation}
\Phi(\beta)=\inf_{E}\left[\beta E-S(E)\right]\label{eq:legendre-fenchel-free-energy}
\end{equation}
which is valid even if the entropy $S(E)$ is non-differentiable,
and whether or not it is everywhere concave. If $S(E)$ \emph{is}
everywhere concave, it can be obtained by LFT from the canonical free
energy: 
\begin{equation}
S^{**}(E)=\inf_{\beta}\left[\beta E-\Phi(\beta)\right]\label{eq:legendre-fenchel-entropy}
\end{equation}
\emph{i.e. }if $S(E)$ is everywhere concave, $S^{**}(E)=S(E)$. In
this case, the canonical and microcanonical ensembles are \emph{equivalent
at the thermodynamic level}. 

For entropies which are not everywhere strictly concave, $S^{**}(E)$
is the concave hull of $S(E)$, \emph{i.e.} in this case the full
physics of the microcanonical ensemble cannot be deduced from the
canonical ensemble. This is illustrated in Fig. \ref{fig:Ensemble-inequivalence},
taken from~\cite{Tou09}. Ensemble non-equivalence
therefore arises from the mathematical properties of the Legendre-Fenchel
transform, and the occurrence of non-concave entropies. As a general
consequence, when entropies are everywhere concave, the ensembles
are always equivalent and one ensemble is as good as another when
calculating thermodynamics of a system. On the other hand, ensemble
inequivalence arises every time that entropy is not strictly globally
concave.

\subsubsection{First order phase transitions in finite systems}

According to the Ehrenfest definition, the canonical free energy function
$\phi(\beta)$ of Fig.~\ref{fig:Ensemble-inequivalence}(b) is that
of a first-order phase transition, as it presents a discontinuity
in its first-order derivative. The microcanonical entropy obtained
by LFT from this free energy (full line in Fig.~\ref{fig:Ensemble-inequivalence}(c))
is that of a first-order phase transition for an additive system,
which in the presence of short range interactions means in the thermodynamic
limit. For these systems, in a certain range of energies, the entropy
of the system is greater when it is divided into two different homogeneous
phases than if it contains a single homogeneous phase: a section of
constant slope in the entropy appears in this energy range because
the total entropy is obtained by a linear combination of the entropies
of the two phases. This linear segment corresponds to a constant temperature
as the system is transformed from one phase into the other, For these
systems ensemble equivalence is not violated (strictly speaking, the
ensembles are said to be partially equivalent, as, while never convex,
the entropy is not everywhere \emph{strictly} concave).

\begin{figure}
\begin{minipage}[b]{0.40\textwidth}
 \includegraphics[width=\textwidth]{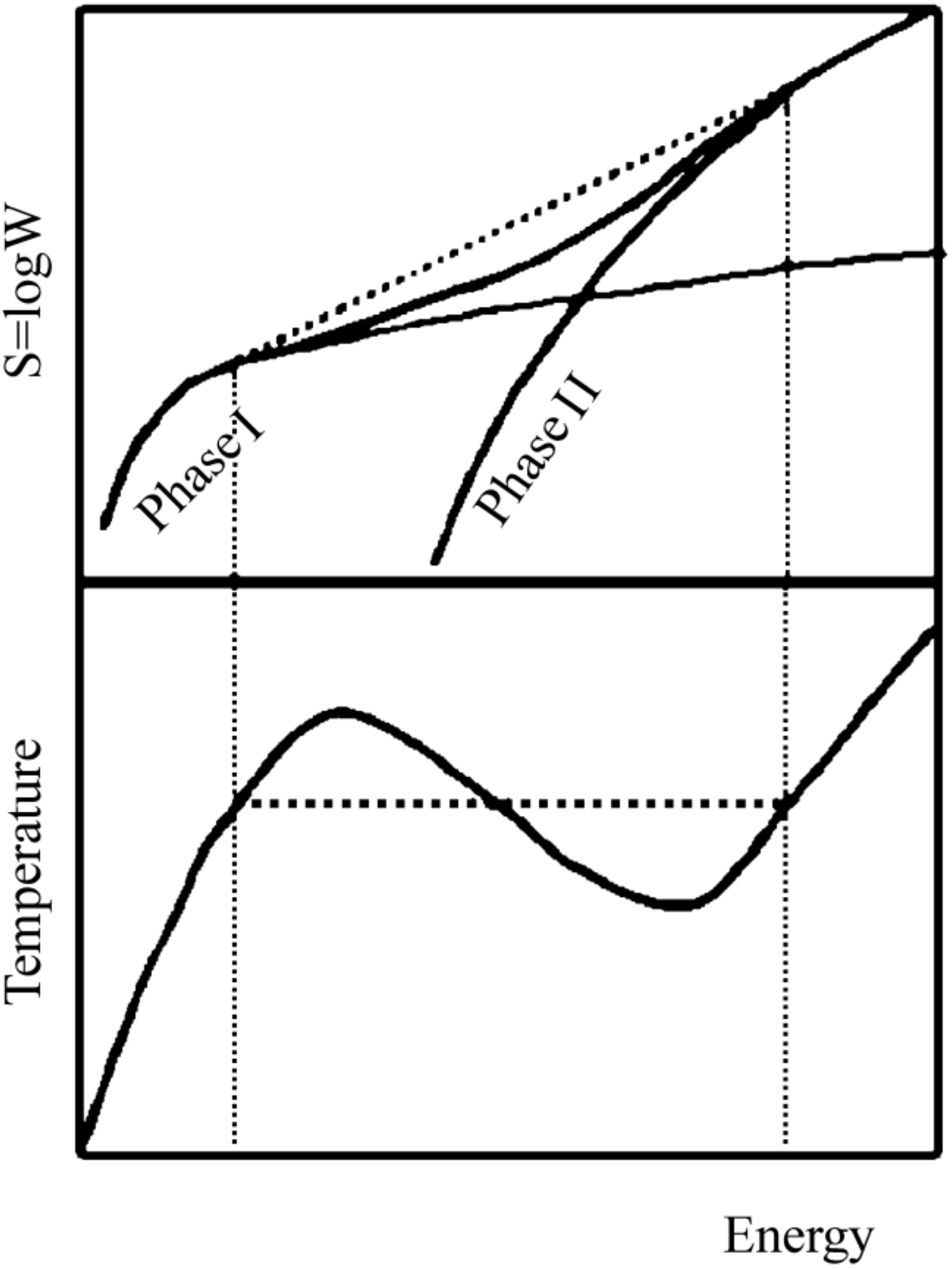}
\caption{Schematic description of a first-order phase transition in a finite
system as the sudden opening of a new disordered phase at a certain
threshold energy. From \citep{Cho02}.\label{fig:disordered-above-threshold}}
\end{minipage}%
\hspace*{0.20\textwidth}
\begin{minipage}[b]{0.40\textwidth} 
\includegraphics[width=\textwidth]{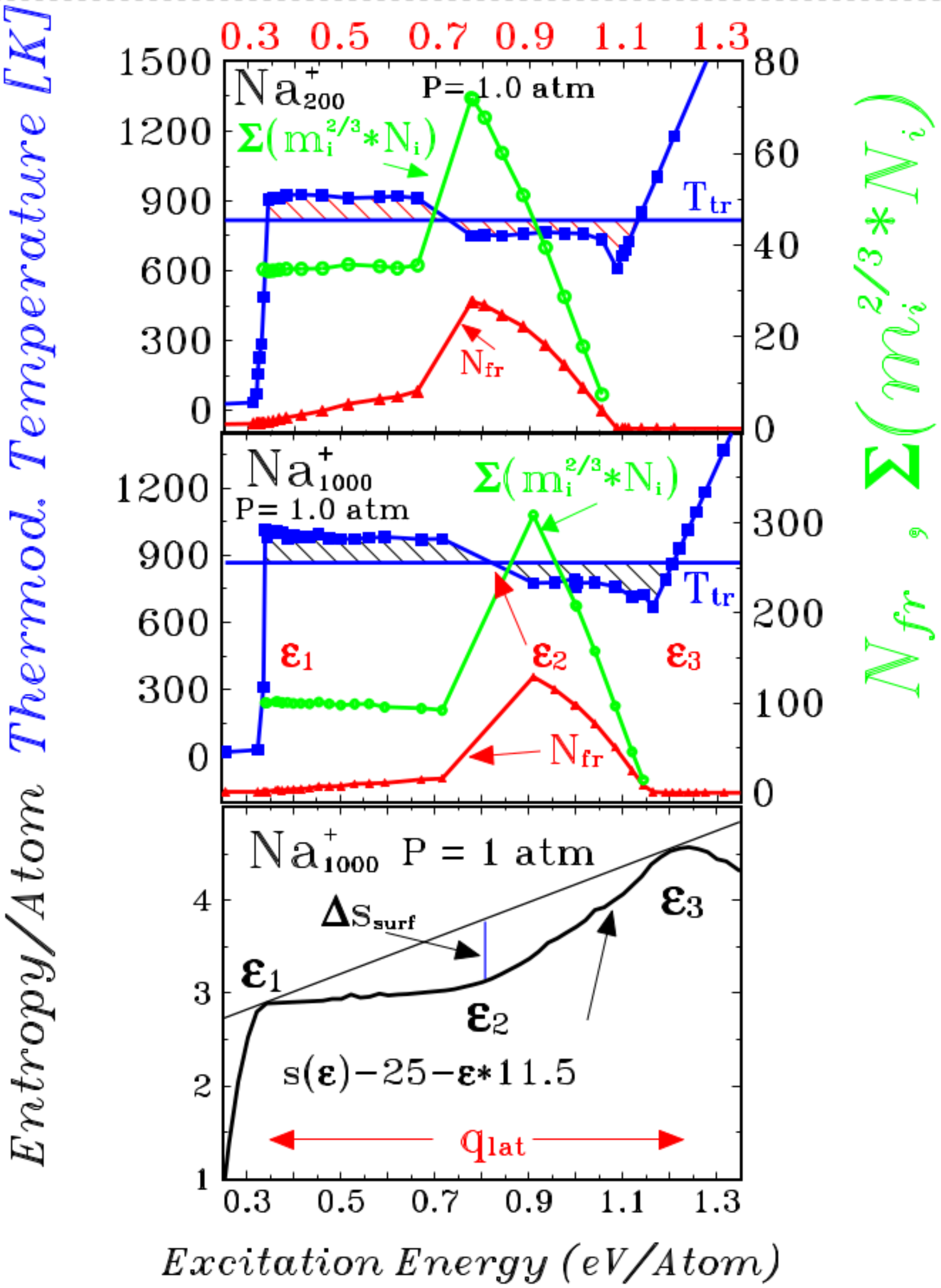}
\caption{Top two panels: (blue squares) constant pressure microcanonical caloric
curves for atomic clusters of 200 or 1000 atoms; (red triangles) number
of fragments $N_{fr}$ with $m_{i}\geq2$ atoms; (green triangles)
effective number of surface atoms $\sum_{m_{i}\geq2}(N_{i}m_{i}^{2/3})$.
Bottom panel: the microcanonical entropy for $Na_{1000}^{+}$ clusters
showing the entropy deficit $\Delta s_{surf}$ due to the increase
in effective surface area of the inhomogeneous fragmented system.
From \citep{Gros97}.\label{fig:fragment-surface-entropy}}
\end{minipage}
\end{figure}
The microcanonical entropy with a convex intruder shown in Figure
\ref{fig:Ensemble-inequivalence}(a) is typical of a first-order phase
transition in non-additive systems, as was first realized in the specific
case of the melting of finite atomic 
clusters \citep{Chal88,Lab90,Kun93,Wal94}
and later developed in the field of phase transitions for hot nuclei
\citep{Gro97,Cho99}. These studies are
of particular interest because in the case of rare gas atoms interacting
through a Lennard-Jones potential the thermodynamic phase transition
is the well-known first order solid-liquid transition. It can be rigorously
shown that the behaviour associated with the convex entropy function
of the finite systems is the embryonic precursor of the infinite system
phase transition \citep{Che92}.

For finite clusters of between 13 and 147 argon atoms, the solid-liquid
transition occurs \emph{without phase separation}: in the ``coexistence''
region corresponding to energies where the entropy is convex, the
clusters are either all ``solid'' or all ``liquid'' \citep{Lab90},
where the two \emph{phaselike forms} can be distinguished energetically
(either at different times when considering the dynamical evolution
of a single cluster, or in different clusters when considering a statistical
ensemble). As mentioned above, these clusters are too small to support
coexistence of multiple phases inside the same system, and therefore
cannot ``heal'' their convex entropy by mixing the two together.
Rather, the different thermodynamic phases of matter first manifest
themselves microscopically as distinct regions of phase space with
their own characteristic temperatures, separated by an energy barrier
\citep{Wal94}. Indeed, the first premises
of a first-order phase transition at a microscopic level can be seen
as the sudden opening of a new disordered phase at a certain threshold
energy (see Fig.~\ref{fig:disordered-above-threshold}), with an
entropy which increases much faster than that of the ordered phase
\citep{Bix89,Cho02},

\[
\left(\frac{\partial S}{\partial E}\right)_{\mathrm{disordered}}>
\left(\frac{\partial S}{\partial E}\right)_{\mathrm{ordered}},
\]
 creating a convex intruder in the total entropy of the system. As
$\partial S/\partial E$ is of course nothing but the inverse temperature,
this implies a lower temperature for the higher energy disordered
phase at the onset, and indeed such transitions are always accompanied
by back-bending caloric curves where the temperature first decreases
before resuming a monotonic increase in the disordered phase \citep{Lab90,Wal94}.

Gross studied the embryonic liquid-gas transition for metallic clusters
of 200 to 3000 atoms \citep{Gros97}. They manifest
another way in which a finite system can undergo a phase transition
of this type without bulk phase coexistence: the clusters undergo
fragmentation into a mixture of smaller clusters (fragments) and monomers.
Fig.~\ref{fig:fragment-surface-entropy} shows two examples of the
evolution of the number of fragments $N_{fr}=\sum_{m_{i}\geq2}N_{i}$
with energy for the clusters $Na_{200}^{+}$ and $Na_{1000}^{+}$.
Within the transition region (\emph{i.e. }between the energies $\epsilon_{1}$
and $\epsilon_{3}$) $N_{fr}$ steadily increases, reaches a maximum
and then decreases as all fragments are transformed into monomers.
The effective increase in the amount of surface in this inhomogeneous
system due to the presence of the fragments is represented by the
total number of surface atoms in the fragments, $\sum_{m_{i}\geq2}(N_{i}m_{i}^{2/3})$
(green curve in Fig.~\ref{fig:fragment-surface-entropy}). Like
$N_{fr}$, it too reaches a maximum inside the transition region,
and leads to an entropy decrease $\Delta s$ with respect to the concave
hull which would be achieved for bulk phase coexistence (bottom panel
in Fig.~\ref{fig:fragment-surface-entropy}). The microcanonical
entropy therefore presents a convex intruder which signals the presence
of a first-order phase transition.

As system size increases, but still far from the thermodynamic limit,
it will be constituted of sufficient bulk material so that different
phase regions can coexist within it. However some non-additivity still
remains as long as $N<\infty$: in this case although the part of
the entropy corresponding to the bulk (which increases like $N$)
is maximized by the phase coexistence, there are other terms which
increase with the size of the interphase surface (which increases
like $N^{(d-1)/d}$, \emph{i.e. }$N^{2/3}$ in 3 dimensions). The
surface contribution must be negative: if not, the surface area would
maximize and the two phases would become one fog-like phase \citep{Chal88}.
The size of the surface depending on the proportions of the two phases,
it will first increase with energy, reach a maximum when each phase
occupies 50\% of the bulk, and then decrease as the energy increases
further. The entropy of a finite two-phase system will therefore fall
below the concave hull shown by the full line in Figure \ref{fig:Ensemble-inequivalence}(c),
and present a convex intruder rather like the dashed line in the same
figure; the convexity disappears (for systems with short-range interactions)
as the thermodynamic limit is approached, like $N^{2/3}/N\sim N^{-1/3}$. 
It is interesting to note here a subtle point made by
Gross~\citep{Gro06}. Although the entropy per particle regains its
concavity in the thermodynamic limit, the curvature of the total entropy
$S$ = $S_{vol} - \Delta S_{surf}$ will remain positive in the
transition region as the bulk entropy  $S_{vol}$ is the
concave hull with zero curvature. Therefore the overall curvature is
given by $-\partial^{2}\Delta S_{surf}/\partial E^{2}$. To quote Gross, ``the
ubiquitous phenomena of phase separation exist only by this
reason''~\citep{Gro06}. However it should be remembered that in the
strict thermodynamic limit the total entropy is diverging and only the
entropy per particle makes sense. In any case, for finite systems,
however large, the convex region is always present.

\subsection{Pseudo-equilibrium}\label{pseudo}
It was Bohr who introduced statistical mechanics to nuclear physics
\citep{Bohr36} and Weisskopf who introduced concepts of
nuclear temperature and entropy with the theory of neutron evaporation
from ``excited'' nuclei \citep{Wei37}.
In the framework of the compound nucleus picture they developed, statistical
equilibrium is justified by the clear separation of timescales between
the formation of a compound nucleus, its equilibration, and subsequent
decay. As pointed out in the introduction to this section, in
collisions at the energies required for multifragmentation or
even vaporization, the separation of timescales for formation and
decay of hot nuclei is not always so clear, and yet models based on
classical equilibrium statistical mechanics are extremely successful
in reproducing or even predicting many observables for these reactions.

A statistical treatment is justified whenever a very large number
of microstates exists for a given set of observables. This is always
the case for the output of a collision, meaning that at
least in principle a statistical approach should always be successful.
An ensemble of events coming from similarly prepared initial systems
and/or selected by sorting always constitutes a statistical ensemble
\citep{Cho06}. To use classical equilibrium statistical mechanics
requires an adequate definition of the relevant microstates i.e. just
that information which ineluctably entails the production of a given
macroscopic event \citep{Col97}. For the multibody
decay of hot nuclei, the microstates relevant to a statistical description
correspond to the microscopic configuration of each reaction at the
\emph{freeze-out instant}: this is defined as the time after which
the characteristics of the fragments and particles produced in the
reaction will no longer significantly change, apart from the effects
of secondary decay (evaporation of light particles due to
residual excitation energy) and Coulombian acceleration due to mutual
repulsion between charged fragments.

Statistical equilibrium means that the probabilities, $\{p_{i}\}$,
of each microstate compatible with the constraints placed upon the
system (conservation laws, etc.) maximize the associated statistical
entropy,
\begin{equation}
S=-\sum_{i}p_{i}\log p_{i}+\sum_{X}\lambda_{X}<X>\label{eq:shannon-entropy}
\end{equation}
where the set of $<X>$ are the constraints and $\lambda_{X}$ are
the associated Lagrange multipliers \citep{Cho02,Gul03}.
In this case we say that the available phase space is uniformly populated.
Any set of microstates for which this population is achieved given
a certain set of constraints corresponds to statistical equilibrium
\emph{at the level of the corresponding statistical ensemble}. In
the case of hot nuclei, we are dealing with an ensemble of freeze-out
configurations produced by many different collisions. For the application
of statistical equilibrium approaches it is unimportant whether each
individual collision had achieved equilibrium at the freeze-out instant,
it is only required that the ensemble of realized configurations be
equivalent to a random sample taken from the available phase space.
This can be achieved by the chaotic nature of the dynamics of the
reactions which in addition are averaged over many different initial
conditions in order to constitute an ensemble of events that covers
the phase space uniformly \citep{Cho06}, all the more so if
the portion of phase space in question is well-defined, i.e.
\emph{when
ensembles are built from homogeneous event selections}. To quote the
fathers of the first statistical model of composite fragment production
in the 20-200 MeV per nucleon bombarding energy range, Randrup and Koonin,
``it is not necessary to argue that equilibrium be reached in any
given collision, since a statistical occupation of the phase space
at the one-fragment inclusive level can occur as a result of averaging
over many separate collision events, each of which can be far from
equilibrium throughout'' \citep{Ran81}. This approach
has been called \emph{pseudo-equilibrium} \citep{Col97}. 

It is important to underline the change of paradigm associated with
this approach. Early on in the development of statistical models for
multifragmentation, Gross suggested that equilibrium might be achieved
\emph{at the level of each reaction}
by ``chaotic mixing'' \citep{Gro90},
a sufficiently intense period of nucleon and energy exchange between
the strongly-interacting nascent fragments as the system expands towards
freeze-out. However, as Cole has pointed out \citep{Col97},
this is a strong hypothesis which can in addition unnecessarily complicate
the interpretation of results. Instead we concern ourselves only with
the equilibrium of statistical ensembles composed of the systems at
freeze-out; more precisely, as exact equilibrium is a theoretical
abstraction which cannot be achieved in the real world, our statistical
ensemble need only be sufficiently close to equilibrium for most observable
properties to be consistent with a uniform population of the phase
space. Residual effects which are directly linked to the collision
dynamics may then reveal themselves in the fine details of the comparison
between model and data.

Before leaving this topic, let us point out an important aspect which
should not be forgotten: the statistical ensembles built from systems
at freeze-out are not ergodic. There is no equivalent single system
which would evolve over time through the ensemble of microstates of
the ensemble. If we were to ``unfreeze'' any of the systems in our
ensemble and let time run on, obviously the particles and fragments
would immediately continue their flight toward the detectors; even
if we were to take one and put it in a (very small) box to try to
keep it in the freeze-out configuration, it would soon cease to resemble
any of the other systems of the statistical ensemble, the Coulomb
repulsion forcing the charged fragments against the walls of the box.
And yet, at the level of the statistical ensemble, it is perfectly
possible to speak of a well-defined characteristic volume, $<V>$,
or mean square radius $<R^{2}>$. The associated Lagrange multiplier
(see Eq. \ref{eq:shannon-entropy}) $\lambda_{V}=\beta p$, thereby
defining the pressure at the level of the ensemble. Therefore defining
thermodynamic properties for dynamically evolving open systems is
not a problem with this approach. The non-ergodicity is not a problem
\emph{per se} for the validity of the approach, but tends to disturb
the unwary as it is at odds with the usual approach where a thermodynamic
system is represented by a fictitious statistical ensemble. When studying
a phase transition in hot nuclei, the statistical ensemble is real
and phase transition is evidenced from the thermodynamics of the
ensemble. 

\subsubsection{Effective statistical ensembles}

The starting point for our studies of nuclear thermodynamics is
therefore a statistical ensemble prepared by the dynamics of collisions.
The question is then: which ensemble is best suited to a study of
the thermodynamics of hot nuclei?

One might be tempted to reply that the microcanonical ensemble is
most apt, as it describes isolated systems of fixed energy and particle
number. However this is not necessarily adapted to the data we have
to analyse. The closest we could come to such a situation would be
in the case of hadron-induced reactions such as
$\pi^{+/-}+X\rightarrow Y*$. Even so, it could be pointed out that
the thermodynamic microcanonical ensemble is defined not only for
fixed $E$, $N$ but also $V$. 
For systems undergoing a LG phase transition the volume $V$ is an
essential degree of freedom. At best, an average size of the
fragmented systems at freeze-out can be inferred from experimental
observables. Indeed the volume is not fixed but
multiplicity and partition-dependent.
From the theoretical point of view one is therefore forced to consider
a statistical ensemble for which the volume can fluctuate from event
to event around an average value~\cite{Bon95}. One comes to a 
microcanonical \emph{isobar} ensemble
in which the average freeze-out volume is used as a
constraint~\cite{Cho00,Gul03}, defined through the partition function
\begin{equation}
Z_{\lambda}(E) = \sum_V W_{V}(E) exp(-\lambda AV),
\label{eq:isobar}
\end{equation}
with the density of states $W_{V}(E)$ having energy $E$ and volume $V$
with $A$ particles.
In this ensemble the $E$ and the Lagrange conjugate of the volume
observable $\lambda$ represent the two state variables of the system.
This is not the microcanonical
ensemble defined by the entropy $S = logW_{V}(E)$ and to avoid
misunderstandings, one should note the temperature, pressure and
average volume by $T_{\lambda}$, $P_{\lambda}$ and $<V>_{\lambda}$ 
with the associated Lagrange multiplier 
$\lambda$ = $P_{\lambda}$ / $T_{\lambda}$. 

The choice of statistical ensemble is best determined by the data.
When one is not interested in an event-by-event analysis and only wants to
calculate mean values at very high excitation energies ($\geq$ 8 - 10~MeV
per nucleon) where the number of particles associated to deexcitation is 
large~\cite{Hah88,Hahn88,Kon94,Bon95,Gul97} (see~\ref{gas}),
then it is clear that a grand-canonical approach is most suited.
The grandcanonical or macrocanonical ensemble corresponds to the rougher
description where the number of particles as well as energy 
of the systems can fluctuate. In this ensemble the temperature and the chemical
potential are fixed variables.  Constraints are only  on the average mass 
and charge of the systems.
On the other hand to perform analyses on an event-by-event basis
or to study, for example, partial energy fluctuations (see~\ref{Cvneg})
the microcanonical ensemble, is the relevant one.
It is used to describe a system which has 
fixed total energy and  particle number~\cite{Rad97,Gro90,Bon95,Koo87}.
In this ensemble the temperature is no longer a natural concept and a
microcanonical temperature can be introduced through the thermodynamic
relation: $T_{micro}^{-1} = \partial S / \partial E$.
Results have to be discussed as a mixing of microcanonical ensembles
in order to be compared to those of canonical ensembles.
Numerical realizations are possible after elaborating specific algorithms
based on the Monte Carlo method. Finally one can conclude about the choice 
of the different ensembles by saying that the excitation energy domain,
the pertinent observable to study  and the event
sorting chosen impose
the dedicated statistical ensemble to be used. For comparison with data
additional  constraints (volume, pressure, average volume\ldots ) 
are added; they correspond with associated Lagrange multipliers 
to isochore and isobar ensembles~\cite{Das01,Cho00}.

\section{How to study a phase transition in hot nuclei}\label{how}
\begin{figure}
\begin{center}
\includegraphics[width=0.6\textwidth]{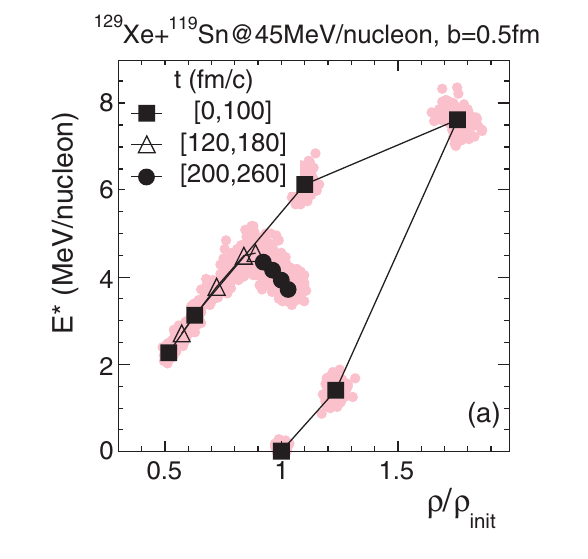}
\end{center}
\caption{Mean trajectory in the phase diagram plane (density -
excitation energy) for central
collisions between Xe and Sn nuclei; each black marker
stands for a time from $t$ = 0 to $t$ = 260 fm/c by a step of 20 fm/c
(30 fm/c = 10$^{-22}$s).
From~\cite{Bon14}. \label{XeSn45}}
\end{figure}

If experiments benefit from a large variety of nuclear collisions
to produce and study hot nuclei, it is essential to underline the
importance of the mutual support between theory and experiments to
progress on the complex subject of phase transition for hot nuclei.
To illustrate this, Fig.~\ref{XeSn45} shows how theory gives precious
information on trajectories in the phase diagram for central collisions
leading to quasifusion. One learns immediately that after a compression
phase due to the initial collisional shock a subsequent expansion
occurs leading to the mixed phase region.
We will see all along this review how this mutual support is present for most of
the aspects and especially to better specify the thermodynamic
variables: excitation/thermal energy, temperature, pressure, density
or average volume at freeze-out. 

Among the existing models some are related to statistical descriptions 
based on multi-body phase space calculations
whereas others describe the dynamic evolution of systems resulting
from collisions between nuclei via molecular dynamics
or stochastic mean field approaches
The first approach uses the techniques of equilibrium statistical 
mechanics with the freeze-out scenario defined in section~\ref{pseudo} and has 
to do with a thermodynamical description of the phase transition for 
finite nuclear systems. The second, in
principle more ambitious, completely describes the time evolution of
collisions and thus helps in learning about nuclear matter (stiffness of 
the effective interaction and in-medium nucleon-nucleon cross-sections), its phase 
diagram, finite size effects and the dynamics of the phase transition
involved.

\subsection{A large choice of collisions to produce hot nuclei}\label{collchoice}
Experimentally to study  a phase transition in hot nuclei we dispose of heavy-ion 
collisions at intermediate and relativistic energies and hadron-nucleus 
collisions at relativistic energies.
Investigations must apply to homogeneous 
samples of events, which requires an appropriate sorting 
mandatory for thermodynamical purposes (section~\ref{pseudo}).
In hadron-nucleus collisions all events have similar 
topological properties independently of the impact parameter, as a 
single hot nucleus is created after a more or less abundant 
preequilibrium emission. Conversely, in heavy-ion collisions, the 
outgoing channel is different depending on the masses and asymmetry 
of the incident partners, the incident energy and the
impact parameter. At intermediate energies 
residual interactions (nucleon-nucleon collisions) strongly compete with mean
field effects; the number of nucleon-nucleon collisions largely fluctuates, leading to
different final reaction channels for the same initial conditions. The
weakening of the mean field hinders, on average, full stopping above 
about 30~MeV per nucleon incident energy;
the large fluctuations mentioned above allow however the observation of
 "quasifusion" at higher energies, although with 
 small cross sections~\cite{I60-Lau06}.
Most of the collisions end up in two remnants coming from the projectile 
and the target, what we call quasi-projectile and 
quasi-target - accompanied by some evaporated particles -, and some 
fragments and particles with velocities intermediate between those of the
remnants: these are called mid-velocity products. They may have several
origins, e.g. direct preequilibrium emission from the overlap region 
between the incident partners, or a neck of matter between them which
may finally separate from quasi-projectile or quasi-target, or from both.
At relativistic energies, mean field effects being negligible,
a geometrical picture - the participant-spectator model -~\cite{Nag88,Sch96}
well describes mid-peripheral and peripheral collisions
which lead to what are called projectile and target spectators instead of
quasi-projectiles and quasi-targets at lower incident energies~\cite{I45-Luk03}.
Whatever the type of reaction, a fraction of the incident translational
energy is transformed into ``excitation energy'', $E^*$, which may be 
shared into thermal energy (heat) and collective energies.
While experimental calorimetry gives a direct access to  
$E^*$, knowing how it is shared between thermal or collective 
energies relies on models.
The sorting of events measured with powerful multidetectors is generally done  
using global variables, which serve to condense the large amount of  
information obtained for each event. Ref.~\cite{I69-Bon08} well
illustrates how to carefully select hot nuclei of similar sizes produced 
in central (quasifusion) and semi-peripheral
(compact quasi-projectiles) collisions.
Two philosophies guide the methods used for event sorting:  
the impact parameter dependence, and the event topology.
Details can be found in~\cite{Bor08}.

To conclude on this part one can say that with central heavy-ion collisions at
intermediate energies leading to quasifusion one can select a well
defined set of events for each incident
energy. For semi-peripheral and peripheral heavy-ion collisions at
both intermediate and relativistic energies and hadron-nucleus collisions at
relativistic energies one can follow, with a
single experiment, the evolution of
deexcitation properties of hot quasi-projectiles, projectile
spectators and selected hot nuclei over a large excitation energy domain through
specific variables like, for example, the size of the heaviest fragment
(quasi-projectiles) or the charge bound in fragments (projectile
spectators).
On the theoretical side statistical and dynamical models are first used to
qualitatively learn about collisions. Then, results of models are
quantitatively compared to experimental data. Models are also used
to bring complementary information when it is missing from experiments.

\subsection{Statistical models}\label{statmod}
As we will see all along the following sections, a large variety of statistical
models are used to predict and support experimental observations related
to a phase transition in hot nuclei and to give complementary
information to data when needed. The present subsection makes a 
brief presentation of those models and gives their spirit.

\subsubsection{Fundamental statistical models and Fisher droplet
model}\label{fundstat}
In this class of models we group those which are not specifically
nuclear in nature: (i) percolation model, (ii) Ising, lattice gas
and Potts models, which are used in the
study of a phase transition in hot nuclei to derive qualitative or
semi-quantitative behaviours, (iii) the Fisher droplet model used
in a more quantitative way to extract critical - pseudo critical
information and free energy. We refer to section~\ref{transition} for
applications of these models.

\paragraph{Percolation, Ising, lattice gas and Potts models}\label{PILP}

Percolation~\cite{Sta94} is the simplest example of a model that
displays critical behaviour. It is purely geometrical and can be
described as a grid of Euclidian dimension $d$ in which the
nodes are randomly populated with probability $p$, which
is called site percolation. If instead of the nodes we activate the
intranode links with probability $p$ one speaks of bond
percolation. Site-bond percolation processes are those in which 
$p$ is  different from 1. In such a model the
phase transition or the critical point is related to the appearance in
the system of a percolating (single) cluster. In such a cluster a set
of nearest-neighbour sites or bonds are active, that goes from 
-~$\infty$ to +~$\infty$. For a finite system with a given geometry
like a box, a possible definition of percolating cluster is that there
exists a set of nearest-neighbour occupied sites (activated bonds)
that extends from one side of the box to the opposite one (other
definitions can also be used). For infinite systems there exists a
sharp critical bond activation probability $p_c$ such that for $p$
above $p_c$ the probability of finding a percolating cluster is 1,
whereas below $p_c$ the probability of finding such a cluster is 0.
For finite lattices the transition from one regime to the other is smooth.
The order parameter for this model is $P_{\infty}$ which is the
fraction of occupied nodes that belong to the percolating cluster
and the distance from criticality $\epsilon$ is ($p$ - $p_c$). 
Since sites/bonds are empty
with probability (1 - $p$), the probability of a node to belong to the
infinite cluster is $p$.$P_{\infty}$ and the probability of belonging
to a finite cluster is $\sum_{s}s.n_s$ where $n_s$ is the yield of the
occupied boxes of size $s$. The critical properties of
percolation are represented by the singular behaviour of moments of
the cluster size distribution which are expressed as a function of
($p$ - $p_c$) (or its absolute value) with exponents that contain
critical exponents $\sigma$, $\beta$, $\gamma$ and $\tau$ related among them.

A remarkably successful model of an interacting system is the Ising model.
A classical spin variable $s_k$, which is allowed to take values
$\pm$1, is placed on each site of a regular lattice, under the influence
of an external magnetic field $h$ and a constant coupling $J$ between
neighbouring sites according to the Hamiltonian
\[
\mathcal{H}_{IS} = -h\sum_{k=1}^{N}s_{k} -
\frac{J}{2}\sum_{k\not=j}^{N}s_{k}s_{j}, 
\]
where the second sum extends to closest neighbours.

The Ising model was originally introduced to give a simple
description of ferromagnetism. In reality the phenomenon of
ferromagnetism is far too complicated to be treated in a satisfactory
way by this oversimplified Hamiltonian. However the fact that the
Ising model is exactly solvable in 1$d$ and 2$d$ and that very accurate
numerical solutions exist for the three dimensional case makes this
model a paradigm of first and second order phase transitions.
The other appeal of the Ising model is its versatility.
It is why it is also well
adapted to describe fluid phase transitions. One can show that a close
link exists between the Ising hamiltonian and the lattice gas
Hamiltonian, which is the simplest modelization of the LG phase
transition
\[
\mathcal{H}_{LG} = \frac{1}{2m}\sum_{k=1}^{N}p_{k}^{2}n_{k} -
\frac{\epsilon}{2}\sum_{k\not=j}^{N}n_{k}n_{j}. 
\]
In the lattice gas model, the same N lattice sites in $d$ dimensions are
characterized by an occupation number, $n_k$ = 0,1, and by a $d$ component vector
$\overrightarrow{p_k}$. Occupied sites (particles) interact with a
constant closest neighbour coupling $\epsilon$. For nuclei the
coupling constant $\epsilon$ = - 5.5 MeV is fixed so as to reproduce
the saturation energy. The relative particle density $\rho$/$\rho_0$ 
is defined as the number of occupied sites divided by the total number 
of sites and is linked to the mean magnetization of the Ising model, $M$, by
$\rho$/$\rho_0$ = 2$M$ - 1. Different choices can be made to measure
the average volume of the system. The most natural measure is obtained by
averaging on the set of events with, 
for each event $e$, the volume observable proportional to the cubic radius
\[
V^{(e)}=\frac{4\pi}{3A}\sum_{i=1}^{N}r_{i}^{3}n_{i},
\]
where $r_i$ is the distance to the centre of the lattice, 
$n_{i}$ is the occupation number and $A$ is the number of particles.
Even for simplified models such as the Ising model no analytical  solution
exists for a number of dimensions larger than 2. This is the reason
why mean field solutions have been developed~\cite{Gul03}.
Moreover the exact solution of three  dimensional Ising-based models
can only be achieved  through numerical Metropolis
simulations~\cite{Gulm99}.

Another classical spin model is the Potts model. To define this model
a $q$-state variable, $\sigma_i$ = 1, 2, 3{\ldots} $q$, is placed on
each lattice site. The interaction between the spins is described by
the Hamiltonian
\[
\mathcal{H} = -J \sum_{<ij>} \delta_{\sigma_{i}\sigma_{j}}.
\]
$\delta$ is a Kronecker delta-function so the energy of two neighbouring
spins is -$J$ if they are in the same state and zero otherwise. Thus,
the Potts model has $q$ equivalent ground states where all the spins
are identical but can take any one of the $q$ values. As the
temperature is increased there is a transition to a paramagnetic phase
which is continuous for $q \leq$ 4 but first-order for $q >$ 4 in two
dimensions.

\paragraph{Fisher's model\label{Fishmod}}
M.E. Fisher~\cite{Fis67} proposed a droplet model to describe the power 
law behaviour of the cluster
mass distribution around the critical point for a LG phase transition.
The vapour coexisting with a liquid in the mixed
phase is schematized as an ideal gas of
clusters, which appears as an approximation to a non-ideal fluid. This
model was applied early on to multifragmentation data~\cite{Hir84,Ell02}
by considering all fragments but the largest in each event as the gas phase, 
the largest fragment being assimilated to the liquid part. 
The yield of a fragment of mass A reads:
\begin{equation}
\mathrm{d} N / \mathrm{d} A = \eta(A) = 
 q_{0}A^{-\tau} \exp((A\Delta\mu(T)-c_{0}(T)\varepsilon A^{\sigma})/T).
\label{fisher}
\end{equation}
In this expression, $\tau$ and $\sigma$ are universal critical exponents, 
$\Delta \mu$
is the difference between the liquid and actual chemical potentials,
$c_0(T)\varepsilon A^{\sigma}$ is the surface free energy of a droplet of
size A, $c_0$ being the zero temperature surface energy coefficient;
$\varepsilon = (T_{c}-T)/T_{c}$ is the control parameter and describes 
the distance of the
actual to the critical temperature. At the critical point 
$\Delta\mu=0$ and surface energy vanishes: $\eta(A)$ follows a
power law. Away from the critical point, 
but along the coexistence line  $\Delta\mu=0$, the
cluster distribution is given by:
$\mathrm{d} N / \mathrm{d} A = \eta(A) = 
 q_{0}A^{-\tau} \exp((-c_{0}(T)\varepsilon A^{\sigma})/T)$.\\
The temperature $T$ is determined by assuming a degenerate Fermi gas. 
The probability of finding a fragment of mass $A$ can be equivalently
and directly calculated from the free energy. For constant
pressure statistical ensembles, the Gibbs free energy is the suitable
quantity to look at, while for a constant volume 
ensemble (as assumed in many models) the Helmhotz free
energy is the relevant one. One or the other prescription gives some
differences especially above the critical point. Assuming a free
energy $F$, the mass yield near the critical point can be 
written~$\eta(A) = y_{0}A^{-\tau} \exp(-F/T)A$. If one introduces the
two constituents, neutrons and protons, a mixing entropy term appears
in the mass-atomic number yield. Details can be found in~\cite{Liu18}.

\subsubsection{Models of nuclear multifragmentation}\label{multifragmod}
In this class of models we group those which take into account specific
nuclear properties such as binding energies, level densities, surface
tension, etc.\emph{ i.e. }models whose physical picture is that of
the production of multiple nuclear fragments, as opposed to
generic clusters. The starting point for such models is
the freeze-out instant previously described in section~\ref{pseudo}.
A highly-excited
nuclear system will arrive, at some point in its evolution, at a moment
commonly known as the freeze-out after which the characteristics
of the fragments produced by its decay will no longer significantly
change.
This
is a more than reasonable assumption: in fact, if we ``play the film
in reverse'' and imagine the final detected products flying back
out of the detectors towards the target, it is clear that such an
instant must exist. 
The freeze-out configuration is commonly assumed to correspond to
a moment at which all fragments produced in the break-up have moved
sufficiently far apart so that they are outside of the range of the
nuclear interaction; otherwise they would experience further dissipative
interactions and possibly nucleon exchange with their neighbours,
as is well known from the study of dissipative nuclear reactions in
the deep-inelastic regime~\cite{Bon74,Gro78,Mor81,Fel87}.
This, too, is a reasonable assumption.

The main hypothesis of these models is that the final products can
be calculated based only on the available phase space at freeze-out,
given a set of constraints such as the total numbers of neutrons and
protons, total energy, angular momentum, etc. (any of which may, depending
on the model, be fixed or allowed to fluctuate). Specific models differ
in their description of the freeze-out configuration, the implementation
of the initial conditions (constraints), and the numerical methods
employed to make predictions based on the corresponding ensembles.
In the following we will try to present the most important distinguishing
aspects of the most successful and well-used models.

The pioneering work of Randrup and Koonin~\cite{Ran81}
is commonly recognized to be the first example of such a model, but
it suffered from limitations such as only treating the production
of light clusters using a grand-canonical approach, and was therefore
limited to excitation energies well above the phase transition domain.
Subsequent models acknowledged and built upon this work in order to
treat more realistically aspects such as the role of the Coulomb repulsion
and the production of heavy fragments, to be able to explore the predicted
coexistence region.

\paragraph{The Copenhagen model (SMM)\label{par:The-Copenhagen-model}}

The Statistical Multifragmentation Model (SMFM~\cite{Bond85,Bondo85,Bon95}),
more commonly known as SMM, is one of the most widely used statistical
models for the interpretation of nuclear multifragmentation data.
It describes the break-up/multifragmentation of an ensemble of 
excited nuclear systems $(A_{0},Z_{0})$
into partitions $\{N_{AZ};1\leq A\leq A_{0},0\leq Z\leq Z_{0}\}$.
The freeze-out stage consists of hot fragments and nucleons or light
clusters ($A<4$) occupying a volume $V$ in thermal equilibrium characterized
by a temperature $T$. After their formation in the freeze-out volume,
the fragments propagate independently in their mutual Coulomb fields
and undergo secondary decays. The deexcitation of the hot primary
fragments proceeds via evaporation, fission, or via Fermi break-up
for primary fragments with $A\leq16$.

The break-up volume $V=(1+\kappa)V_{0}=(1+\kappa)A_{0}/\rho_{0}$
is taken large enough so that no fragments overlap; typical values
are $\kappa\approx2$, \emph{i.e. }$V\approx3V_{0}$. 
$V_{0}$ is the volume of a nucleus of mass $A_0$ at normal density. 
The hot fragments
($A\geq4$) are spherical droplets at normal nuclear density, whose
free energy is described by a charged liquid drop parametrization
containing bulk, symmetry, surface and Coulomb terms. The bulk term
contains a Fermi gas dependence on temperature. The surface term vanishes
at the critical temperature of infinite nuclear matter, usually taken
to be $T_{C}=18$ MeV. The partition temperature $T$ is determined
in order to conserve energy from one partition to another. The free
energy component associated with thermal motion of fragments depends
on a ``free'' volume $V_{f}=\chi V_{0}$ in which they can move
without overlapping. $\chi$ depends on the multiplicity of the partition
and typically varies between 0.2 and 2. The assumption of thermal
equilibrium means that a single temperature $T$ is used to characterize
both the fragments' momenta and their internal excitation energy,
but the degree of equipartition can be modified by treating the inverse
nuclear level density parameter $\epsilon_{0}=A/a$ which appears
in the bulk component of the fragment free energy as a free parameter:
setting $\epsilon_{0}=\infty$ results in a hot gas of cold fragments
with zero excitation energy.
\begin{figure}[htb]
\begin{center}
 \includegraphics[scale=0.85]{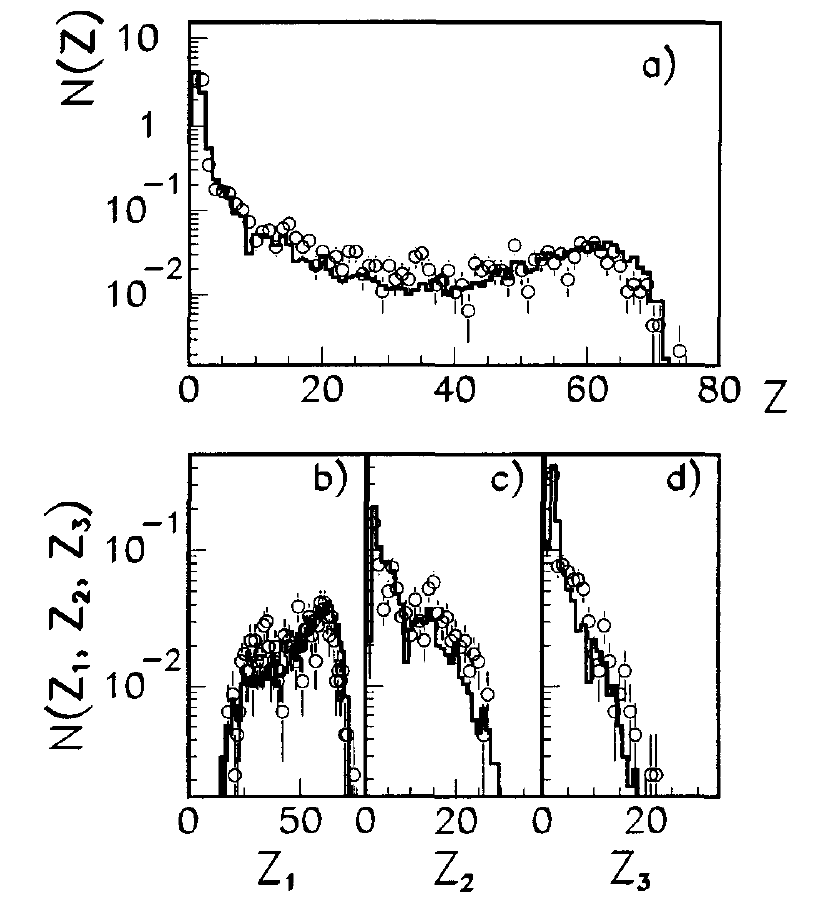}
\end{center}
\caption{Comparison of experimental mean elemental fragment
multiplicity, $N(Z)$ and distribution of the three heaviest fragments
(Z$_1$, Z$_2$ and Z$_3$) in each event (quasi-projectile hot nuclei
produced in peripheral - 0.8$< b/b_{max} \leq$0.9 - $^{197}$Au + $^{197}$Au 
collisions
at 35~MeV per nucleon incident energy with fission events removed) 
with SMM simulations. The
circles show experimental data and the lines the SMM predictions
filtered by the experimental device. From~\protect\cite{MDA99}.}
\label{fig:QPSMM}
\end{figure}
\begin{figure}[htb]
\begin{center}
 \includegraphics[scale=0.40]{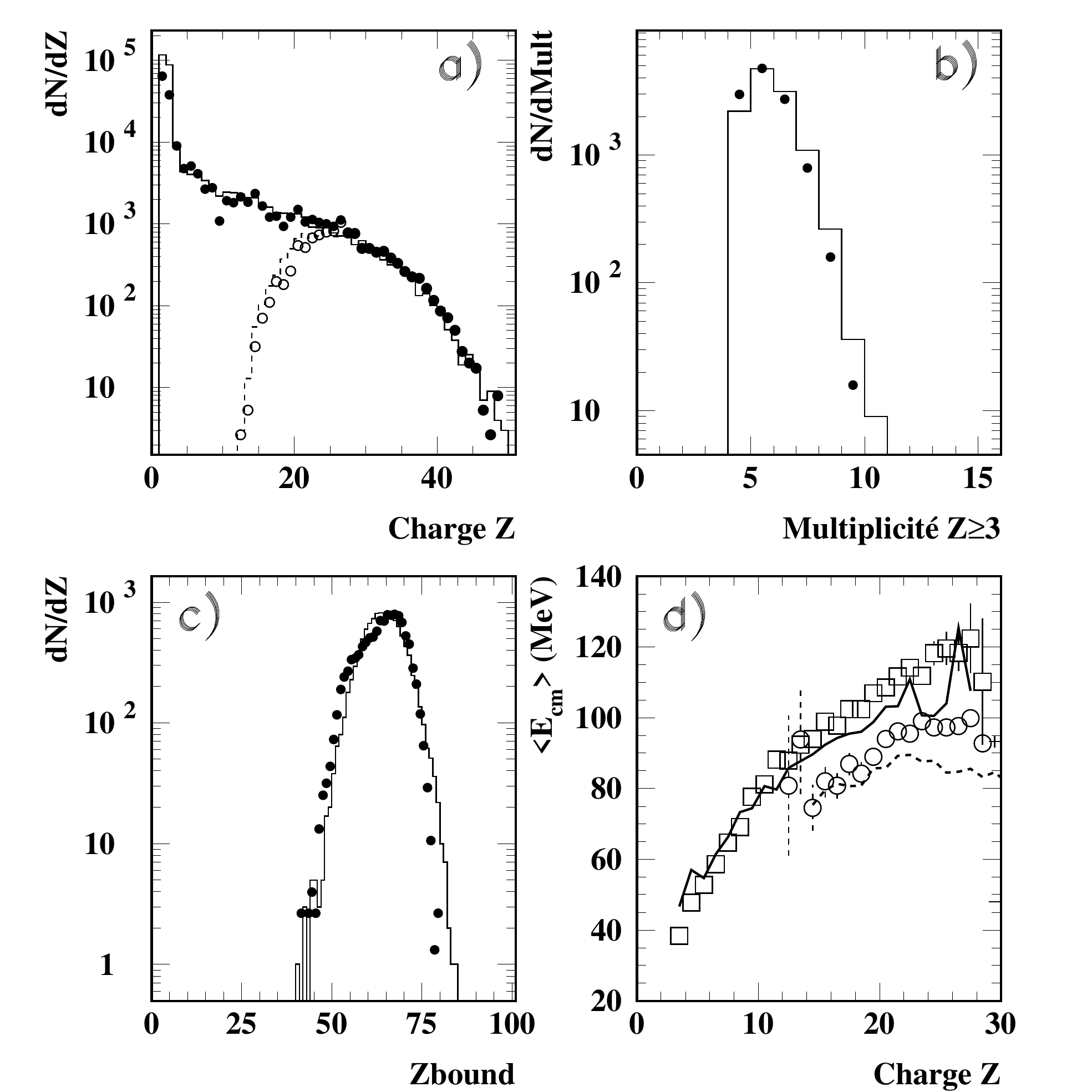}
\end{center}
\caption{Comparison of experimental data (Xe+Sn quasifusion hot nuclei
produced in central collisions at 32~MeV per nucleon incident energy)
with SMM simulations. The lines are for data and symbols for filtered SMM
predictions
(all fragments, except open circles and dashed lines which refer to the
largest fragment of each partition). Z$_{bound}$ represents the sum of the
charges of all fragments. From~\protect\cite{T25NLN99}.}
\label{fig:QFSMM}
\end{figure}

In the original version of the model~\cite{Bon95},
partition generation was performed using a Monte Carlo method. All
possible partitions with low ($M\leq3,4$) multiplicity are directly
generated and the associated mean multiplicity calculated using their
microcanonical statistical weights. If the calculated mean multiplicity
$<M>$ is small enough, one of these partitions is randomly selected
to generate an event. If not, a partition with larger multiplicity
is generated starting from the grand-canonical expression for $<N_{AZ}>$
calculated from the free energy of the partition. It should be noted
that in this version of the model the Coulomb interaction between
fragments was approximated in a Wigner-Seitz approach.

A later improvement to the model was the introduction of Metropolis
sampling, using the so-called ``Markov chain'' approach to efficiently
generate partitions representative of the whole 
phase space~\cite{Botv00}.
Starting from a partition of multiplicity $M$ a new distinct partition
is generated by moving one nucleon of the partition: this corresponds
to either emission or absorption of a free nucleon by one of the fragments,
or to transfer of a nucleon from one fragment to another. 
This procedure was shown to significantly improve the quality of the
statistical sampling compared to the previous method. Moreover, it
allows to calculate directly the Coulomb contribution for each break-up
channel based on actual fragment coordinates in the freeze-out volume,
and to explicitly include conservation of angular momentum in the
model; recently this has allowed to begin systematic theoretical investigations
of the Coulomb and angular momentum effects on multifragmentation
in peripheral heavy-ion collisions at Fermi energies, especially on
the isotope yields, which are crucial for astrophysical 
applications~\cite{Erg15}. 

Radial expansion velocities, fully
decoupled from thermal properties, were also added for a better
comparison with experiments. As for the ``Big Bang'' a self similar
expansion (collective velocity proportional to \textbf{r}) is
observed up to around 80 - 100~fm/c after the beginning of central collisions
in all dynamical models and this is why this prescription was retained.

The quality of agreement with data explains the large success of this
model and this is well illustrated by
Fig.~\ref{fig:QPSMM} and Fig.~\ref{fig:QFSMM} which show
different fragment observable distributions measured for both
quasi-projectiles and quasifusion hot nuclei and compared to SMM
results filtered by the experimental devices.
For quasi-projectiles, from peripheral $^{197}$Au on $^{197}$Au collisions at
35~MeV per nucleon incident energy, SMM predictions are obtained with a source:
$A_0$ = 197, $Z_0$ = 79, a freeze-out volume of 3.3$V_0$, a mean thermal energy of
3.4 MeV per nucleon with a standard deviation of 1.2 MeV per nucleon
and a radial collective energy of 0.3~MeV per nucleon which can be
attributed mainly here to thermal pressure~\cite{Fri90,I69-Bon08}.
For central Xe+Sn collisions at 32~MeV per nucleon incident energy,
to get the observed agreement (Fig.~\ref{fig:QFSMM}), the input parameters of the
source are the following: $A_0$ = 202, $Z_0$ = 85 as compared to A=248 and Z=104 
for the total system, which indicates preequilibrium emission, freeze-out volume 3$V_0$, 
partitions fixed at thermal excitation energy of 5~MeV per nucleon and added
radial expansion energy of 0.6~MeV per nucleon.

\paragraph{The canonical thermodynamical model (CTM)}

Das Gupta, Mekjian and co-workers~\cite{Gup98,Das05}
developed a model for nuclear multifragmentation with a very similar
underlying physical picture to that of SMM. However the numerical
implementation is greatly simplified by the use of the canonical ensemble.
The canonical partition function for $A$ nucleons, $\mathcal{Z}_{A}$,
can be easily obtained starting from $\mathcal{Z}_{0}=1$ thanks to
the recursion relation
\[
\mathcal{Z}_{A}=\frac{1}{A}\sum_{k=1}^{A}k\omega_{k}\mathcal{Z}_{A-k}
\]
where $\omega_{A}$ is the partition function for a fragment with
$A$ nucleons, given by
\[
\omega_{A}=\frac{V_{f}}{h^{3}}\left(2\pi mT\right)^{3/2}A^{3/2}\exp\left(\frac{-F_{AZ}}{T}\right)
\]
Here $V_{f}$ is the free volume as in SMM, but unlike in that model
it is taken simply equal to the break-up volume minus the excluded
volume of the fragments themselves, \emph{i.e. }$V_{f}=V-V_{0}$ which
means that in CTM the two volume parameters of SMM are identical:
$\chi=\kappa$.
\begin{figure}[htb]
\begin{center}
 \includegraphics[scale=0.30]{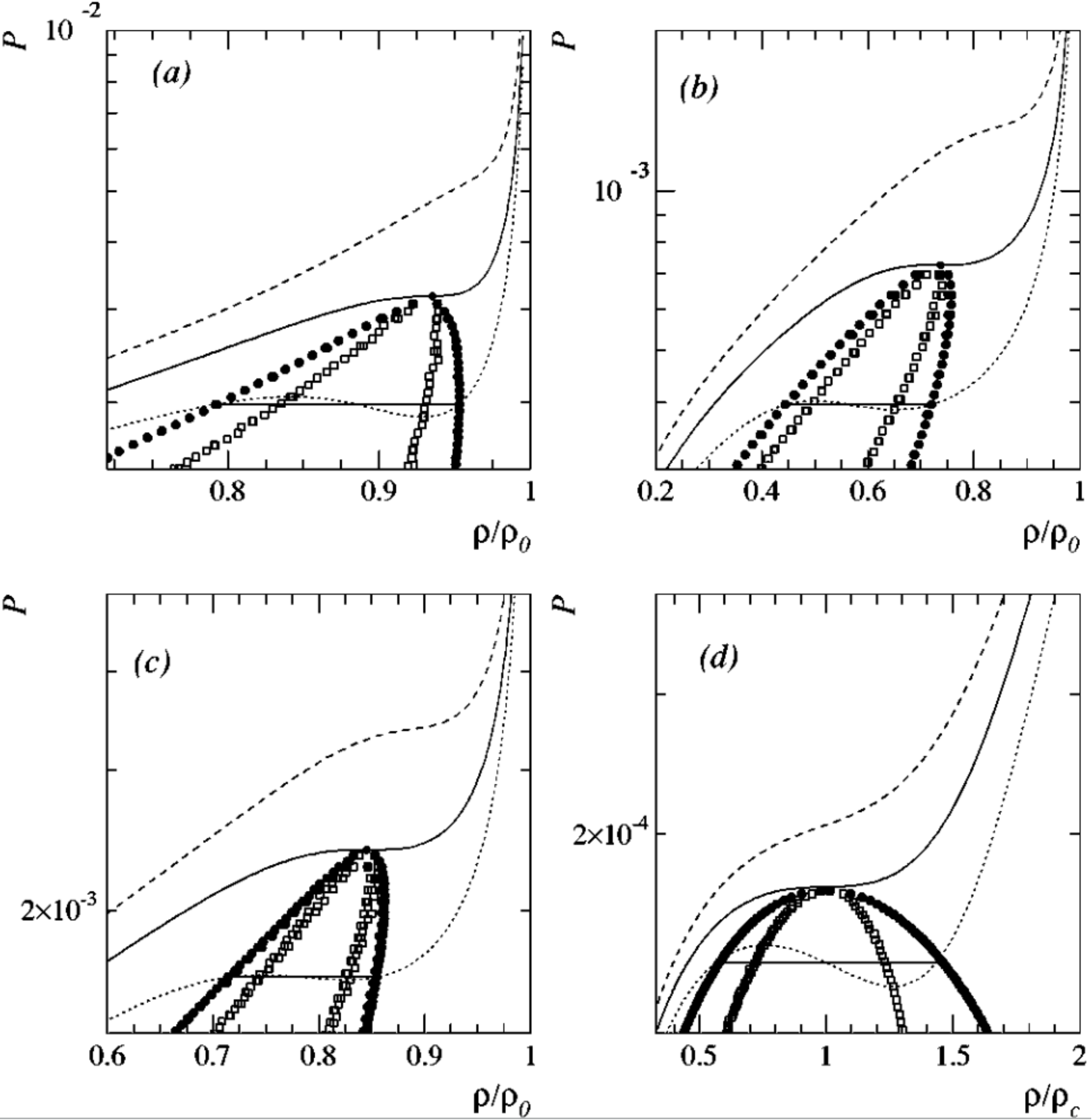}
\end{center}
\caption{Thermodynamics of the Mekjian model~\cite{Gup98},
from~\cite{ElHi00}. Isotherms of pressure
as a function of reduced density: (a) multifragmentation of charged
nuclei with $A=162$; (b) as in (a) but with no Coulomb; (c) as (b)
but with no temperature dependence of the surface free energies; (d)
a van der Waals fluid. Full symbols show the coexistence zone, open
symbols the spinodal zone. Solid curves correspond to the critical
isotherm for each case: (a) $T_{C}=7.6$ MeV; (b) $T_{C}=6.9$ MeV;
(c) $T_{C}=11.0$ MeV. Dotted and dashed curves are subcritical and
supercritical, respectively.}
\label{fig:thermoM}
\end{figure}
Note that very recently predictions for new signatures of phase transition for
hot nuclei to be confronted to data were
proposed~\cite{Mal17,Das18,DasM18}.

The thermodynamics of CTM/SMM were studied by Elliott and
Hirsch~\cite{ElHi00},
most notably the differences between charged or neutral matter, and
the influence of the surface energy temperature dependence. Calculated
pressure-density isotherms are presented in Fig.~\ref{fig:thermoM},
and in all cases coexistence and spinodal regions can be identified
up to some critical temperature. The effective critical temperature
of the model does not correspond to the value of the parameter $T_{C}=16$
MeV used for the calculations; indeed even without such temperature
dependence of the surface energy (Fig.~\ref{fig:thermoM}(c))
there is still a coexistence region delimited by a critical isotherm.
The effect of Coulomb on the critical temperature is surprisingly
small, of the order of 10\%. 

The density range covered by the coexistence and spinodal regions
changes most strongly according to the ingredients of the model. With
the standard Coulomb and surface energy terms (Fig.~\ref{fig:thermoM}(a))
the coexistence densities are surprisingly high, between 0.7 and 0.95$\rho_{0}$,
much higher than could be realized with a closest packing of normal
density nuclei as supposed in SMM and significantly higher than those
typically used to compare model predictions to data. With Coulomb
switched off (Fig~\ref{fig:thermoM}(b)) the densities
are more like those predicted by models for (neutral) nuclear matter,
of the order of $0.35-0.75\rho_{0}$.

To conclude one can also note that using a
classification scheme for phase transitions in finite systems
based on the Lee-Yang zeros in the complex temperature
plane~\cite{Yan52,Bor00}, it was shown that for this statistical model of
nuclear multifragmentation the predicted phase transition is 
of first-order~\cite{Mul01}.

\paragraph{Microcanonical models (MMMC and MMM)}

Historically, following the pioneering work of Randrup and Koonin~\cite{Ran81},
the Berlin group developed a microcanonical model~\cite{Zha87,Zhan87,Gro90} to better
understand mass distribution of fragments for hadron-nucleus
collisions at relativistic energies. In this rather simplified model 
the system of fragments is assumed to be stochastically expanded to a
freeze-out volume of 6$V_0$; the reason for this choice comes from the
difficulty to position the fragments in a smaller volume without
overlapping and consequently demanding a lot of CPU time. The model
only allows for sequential neutron evaporation from fragments. 
And no \emph{a priori} hypothesis is made concerning the internal energies
of excited fragments at freeze-out. This means that the vanishing of the
level density, which is expected to occur at high excitation energies 
is not taken into account. As a consequence no limiting temperature 
for fragments is introduced~\cite{Koo87}. This model which is known as
Microcanonical Metropolis Monte Carlo - MMMC illuminated qualitatively various
aspects of phase transition for hot nuclei.
A more complete
model called Microcanonical Multifragmentation Model - MMM
was developed ten years later.

Within a microcanonical ensemble, the statistical weight of a configuration
$C$, defined by the mass, charge and internal excitation energy
of each of the constituting $M_C$ fragments,
can be written as
\begin{eqnarray}
\nonumber
W_C(A,Z,E,V) = \frac1{M_C!} \chi V^{M_C} \prod_{n=1}^{M_C}\left( 
\frac{\rho_n(\epsilon_n)}{h^3}(mA_n)^{3/2}\right)
\\ 
\times
~ \frac{2\pi}{\Gamma(3/2(M_C-2))} ~ \frac{1}{\sqrt{({\rm det} I})}
~ \frac{(2 \pi K)^{3/2M_C-4}}{(mA)^{3/2}},
\label{eq:wc}
\end{eqnarray}
where $A$, $Z$, $E$ and $V$ are respectively the mass number,
the atomic number, the excitation energy and
the freeze-out volume of the system.
$E$ is used up in fragment formation, fragment internal
excitation, fragment-fragment Coulomb interaction and
kinetic energy $K$.
$I$ is the inertial tensor of the system whereas
$\chi V^{M_C}$ stands for the free volume or, equivalently, accounts for
inter-fragment interaction in the hard-core idealization. 
\begin{figure}
\begin{center}
\includegraphics[scale=0.80]{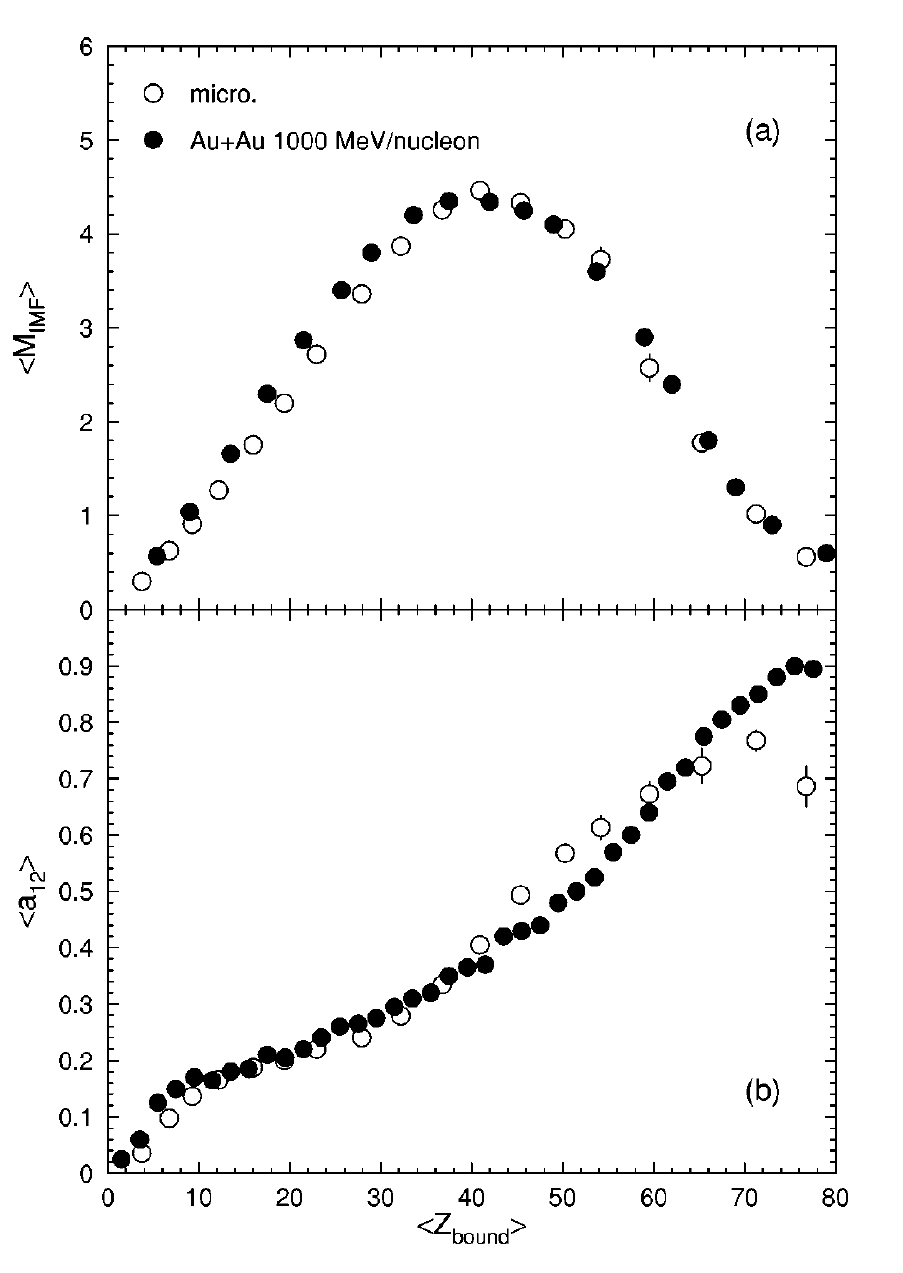}
\end{center}
\caption{Fragment multiplicity (top panel) and charge asymmetry of the
two heaviest fragments (bottom panel) of projectile spectators produced
in $^{197}$Au+$^{197}$Au collisions at 1000~MeV per nucleon
incident energy. Full points refer to data and open points to MMM
model. From~\protect\cite{Radu00}.} \label{mmmexp}
\end{figure}

In MMM~\cite{Rad97} the
statistical weights of each configuration and consequently the mean
value of any global observables can be expressed analytically.
Since the resulting formulas are not tractable, a statistical method
is proposed. The method, which is a generalization of Koonin and
Randrup's procedure~\cite{Koo87}, provides an exploration of
configuration space according to the detailed balance principle.
This method is then applied to describe the phenomenon of multifragmentation.
To obtain a realistic simulation, real binding energies of all the
elements with $A$ lying between 1 and 266 are used. Calculated level densities
include the excitation energy so as to describe the dependence of the factor
entering the Fermi-gas formula for the ($A,Z$)
nucleus with the binding energy $B$($A,Z$), limiting then temperature
for fragments. The freeze-out density or
volume is the only fitting parameter of the simulation. The model was
then refined~\cite{Radu00} by taking into account the experimental
discrete levels for fragments with $A<7$ and by including the stage of
sequential decays of primary excited fragments, thus allowing
quantitative comparisons with data.
As for SMM, radial expansion energies, fully
decoupled from thermal properties, were also added for a better
comparison with experiments.

Fig.~\ref{mmmexp} shows a comparison with data for projectile spectators
produced in $^{197}$Au+$^{197}$Au collisions at 1000~MeV per nucleon
incident energy, which was used to deduce the sequence of excitation
energy as a function of the projectile spectator $E^{*}(A)$.
$M_{IMF}$ is the multiplicity of fragments (2$<$ Z $<$31). $a_{12}$ is
the charge asymmetry of the two largests fragments 
$a_{12}$ = ($Z_{max} - Z_2$)/($Z_{max} + Z_2$) where $Z_{max}$ is the
heaviest fragment and $Z_2$ is the second heaviest fragment.
$Z_{bound}$ represents the charge bound in fragments.  

\begin{figure}
\begin{center}
\includegraphics[scale=0.80]{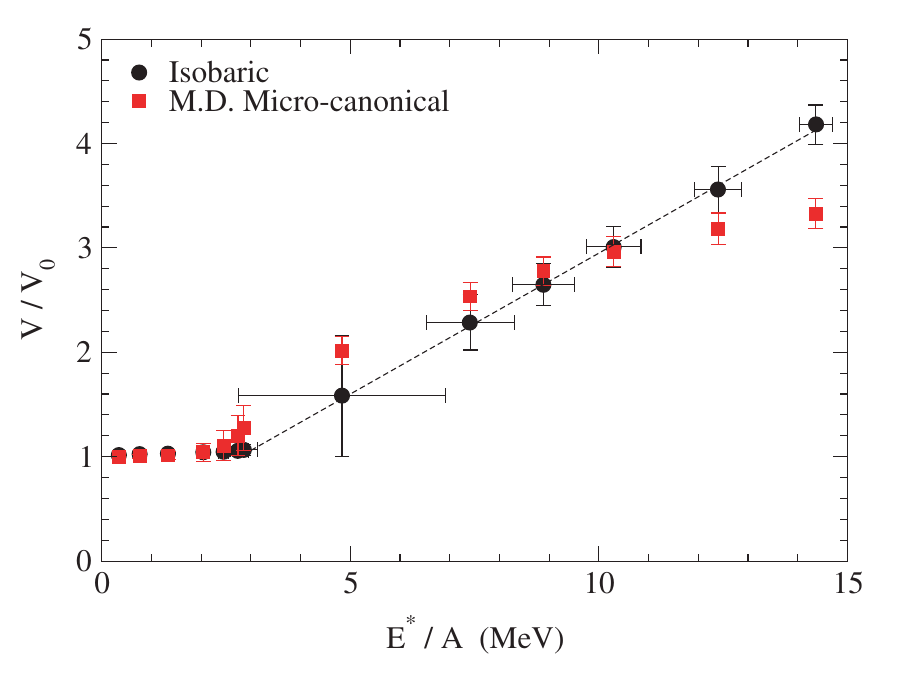}
\end{center}
\caption{Average break-up (freeze-out) volume as a function of the
excitation energy per nucleon predicted by the canonical isobaric
 and the Multiplicity Dependent microcanonical ensemble. Dashed line
emphasizes the fairly linear dependence of the break-up volume on the
excitation energy in the canonical isobaric ensemble.
From~\protect\cite{Agu06}.} \label{Vfocsmm}
\end{figure}

\paragraph{Nuclear multifragmentation: comparison of different
statistical ensembles}
The sensitivity of different ensembles to the underlying
statistical assumptions is a relevant information. Such a study was
investigated in~\cite{Agu06} by comparing microcanonical, canonical and 
canonical isobaric formulations within the SMM model. 
The work was carried out for the
nuclear system $A$=168 and $Z$=75. The same break-up temperature is
used in both canonical calculations and the break-up volume 3$V_0$ is the same
for both microcanonical and canonical ensembles. The one for which the
break-up volume is determined for each fragmentation mode is labelled 
``M.D. microcanonical'' (Multiplicity Dependent) to distinguish it
from the standard microcanonical version. The pressure, for the isobaric
ensemble, was fixed at P = 0.114~MeV/fm$^3$. The energy input used for
the microcanonical ensemble was the average excitation energy obtained
in the isobaric ensemble. We refer to~\cite{Agu06} for more details.
The main conclusions are the following: the microcanonical, canonical
and isobaric implementations predict very similar average physical
observables. Fig.~\ref{Vfocsmm} shows one example that concerns the
evolution of the average break-up volume as a function of the thermal
excitation energy: volumes obtained with the canonical isobaric and the M.D.
microcanonical implementations are very similar, which
indicates that the \textit{ad hoc} multiplicity dependence of SMM
is relevant.

\subsection{Dynamical models}\label{dynmod}

Beside statistical descriptions, there are microscopic frameworks that
directly treat the dynamics of colliding nuclei such as 
the family of semi-classical simulations based on
the nuclear Boltzmann equation (the Vlasov-Uehling-Uhlenbeck (VUU), 
Landau-Vlasov (LV),
Boltzmann-Uehling-Uhlenbeck (BUU) or Boltzmann-Nordheim-Vlasov (BNV)
codes~\cite{Kru85,Gre87,Ber88,Bon94}),
classical molecular dynamics (CMD)~\cite{Pra95,Str99,Cus02,Che04}, 
quantum molecular 
dynamics (QMD)~\cite{Pei89,Aic91,Luk93}, fermionic molecular dynamics
(FMD)~\cite{Fel90}, antisymmetrized molecular dynamics 
(AMD)~\cite{Ono93,Ono96,Sug99} and stochastic mean field approaches 
 related to simulations of the Boltzmann-Langevin equation
~\cite{Ayi88,Ayi90,Ran90,Rei92,Rein92,Reinh92}.
Boltzmann type simulations
follow the time evolution of the one body density. Neglecting higher
than residual two-body correlations, they ignore
fluctuations around the main trajectory of the
system (deterministic description), which becomes a severe
drawback if one wants to describe processes involving instabilities,
bifurcations or chaos expected to occur during the multifragmentation 
process. Such approaches are only appropriate during
the first stages of nuclear collisions, when the system is hot and 
possibly compressed and then expands to reach a uniform low density.
They become inadequate to correctly treat the fragment formation,
and for the description of multifragmentation it is
essential to include higher order correlations and fluctuations. 
This is done in molecular
dynamics methods and in stochastic mean field approaches.

\subsubsection{Quantum molecular dynamics: QMD and AMD
simulations}\label{QAMD}

QMD is essentially a quantal extension of the molecular dynamics approach
widely used in chemistry and astrophysics. Starting from the n-body
Schr\"odinger equation, the time evolution equation for the Wigner 
transform of the n-body density matrix is derived. Several approximations 
are made. 
QMD employs a product state of single-particle states where only the mean
positions and momenta are time-dependent. The width is fixed and is
the same for all wave packets. The resulting equations of motion are 
classical. Also the interpretation of mean position and momenta is purely
classical and the particles are considered distinguishable; this simplifies
the collision term which acts as a random force. 
All QMD versions use a collision term with Pauli blocking 
in addition to the classical dynamics.
Some versions consider spin and isospin and others do not distinguish
between protons and neutrons (all nucleons carry an average charge).
As with most dynamical models a statistical decay code must be coupled 
to describe the long time evolution (called an after-burner). 
However for the
code of Ref.~\cite{Aic91} there is no need to supplement the QMD
calculations by an additional evaporation model~\cite{Mul93}. 
It is important to emphasize here that QMD codes are certainly better
adapted for the higher incident energies.
\begin{figure}[htb]
\begin{center}
\includegraphics[scale=0.90]{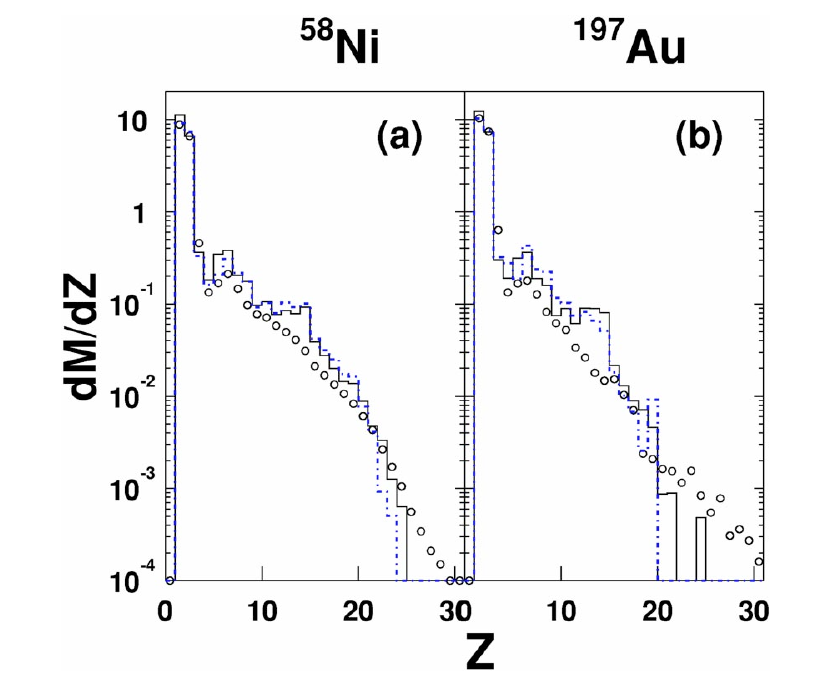}
\end{center}
\caption{Charge distribution of fragments produced in central collisions for
$^{64}$Zn on (a) $^{68}$Ni and on (b) $^{197}$Au at 47~MeV per nucleon incident 
energy. Experimental results are shown by circles and calculated 
results (AMD-V) correspond to dot-dashed lines (soft EOS) and to solid
lines (stiff EOS).  
From~\protect\cite{Wad04}.} \label{amdexp}
\end{figure}

An antisymmetrized version of
molecular dynamics (AMD) was constructed by incorporating the two-nucleon
collision process as the residual interaction into the fermionic molecular
dynamics (FMD). AMD describes the system with a Slater determinant of
Gaussian wave packets and therefore can describe quantum-mechanical features.
However, in the dynamics of nuclear reactions, there may be other phenomena
caused by the wave packet tail that are completely lost in AMD due to the
restriction of the single-particle states. So an improvement was realized
(called AMD-V) with the stochastic incorporation of the diffusion and the
deformation of wave packets which is calculated by the Vlasov equation
without any restriction on the one-body distribution~\cite{Ono96}. After
that the quantum branching process due to the wave packet diffusion
effect was treated as a random term in a Langevin-type equation of motion
whose numerical treatment is much easier. Moreover a new approximation 
formula was also introduced in order to evaluate the Hamiltonian in 
the equation of motion with much less computation time than the exact 
calculation, so that systems like Au+Au became tractable~\cite{Ono99}.
More recently a method was proposed to allow the possibility to form
particles of mass numbers $A$~= 2, 3 and 4. Details can be found
in~\cite{Ono13,Ike16}. As for most dynamical models
the stiffness of the effective interaction and the in-medium 
nucleon-nucleon cross-section are both important ingredients for determining the 
degree of agreement with experimental data. In order to test the 
sensitivity of the ingredients, a detailed study of reaction dynamics 
and multifragmentation was made in Ref.~\cite{Wad04} by comparing 
AMD-V calculations with data from heavy-ion reactions around the Fermi 
energy. Fig.~\ref{amdexp} presents the
charge distribution of reaction products from central collisions between
$^{64}$Zn projectiles and (a) $^{68}$Ni and (b) $^{197}$Au targets
at 47~MeV per nucleon incident 
energy. Lines correspond to AMD-V results obtained with
both stiff and soft EOS, which rather well agree with data. We refer
the reader to Ref.~\cite{Wad04} for more details. 

\subsubsection{Stochastic mean field approaches: SMF, BOB and BLOB
simulations}\label{BOBBLOB}
The stochastic mean field models used are semi-classical
nonrelativistic transport approaches. The time evolution of the
nuclear system is described in terms of the one-body distribution, as
ruled by the nuclear mean-field (plus Coulomb interaction for protons)
and hard two-body scattering, according to the so-called 
Boltzmann-Langevin equation (BLE),
\begin{equation}
\partial f / \partial t = \{h[f],f\} +  \overline{I}[f] + \delta I[f],  
\label{eq:BLE}
 \end{equation}
which was introduced for heavy-ion collisions in 
Ref.~\cite{Ayi88,Ayi90,Ran90}.
$f$ is the one-body phase space density. The first term on the r.h.s.
produces the collisionless propagation of $f$ due to the self-consistent 
mean field described by the effective Hamiltonian. The second term,
called collision term, represents the average effect of the
residual Pauli-suppressed two-body collisions; this is the term included 
in LV, BUU and BNV simulations. The third term is the Langevin term which
accounts for the fluctuating part of the two-body collisions.
Exact numerical solutions of the BLE are very difficult to obtain and
have only been calculated for schematic cases in one or two
dimensions~\cite{Cho91}. Therefore various approximate treatments of the BLE
have been developed and the test-particle method is used to solve
Eq.~(\ref{eq:BLE}). Fluctuations are introduced within the mean field
treatment according to various approaches. 

In the Stochastic Mean Field Model SMF~\cite{Bar02}, fluctuations 
are produced by agitating the spatial density
profile~\cite{Gua96,Colon98}. Once local thermal equilibrium is reached,
the density fluctuation amplitude $\sigma_{\rho}$ is evaluated by projecting on the
coordinate space the kinetic equilibrium value of a Fermi gas. Then,
in the cell of $\bm{r}$ space being considered, the density
fluctuation $\partial\delta_{\rho}$ is selected randomly according to
the Gaussian distribution
exp($-\partial\delta_{\rho}^{2}$/2$\sigma_{\rho}^{2}$). This determines 
the variation of the number of particles contained in the cell. A few
leftover particles are randomly distributed again to ensure
conservation of mass. Momenta of all particles are finally slightly shifted
to ensure momentum and energy conservation.
\begin{figure}[htb]
\begin{center}
\includegraphics[scale=0.90]{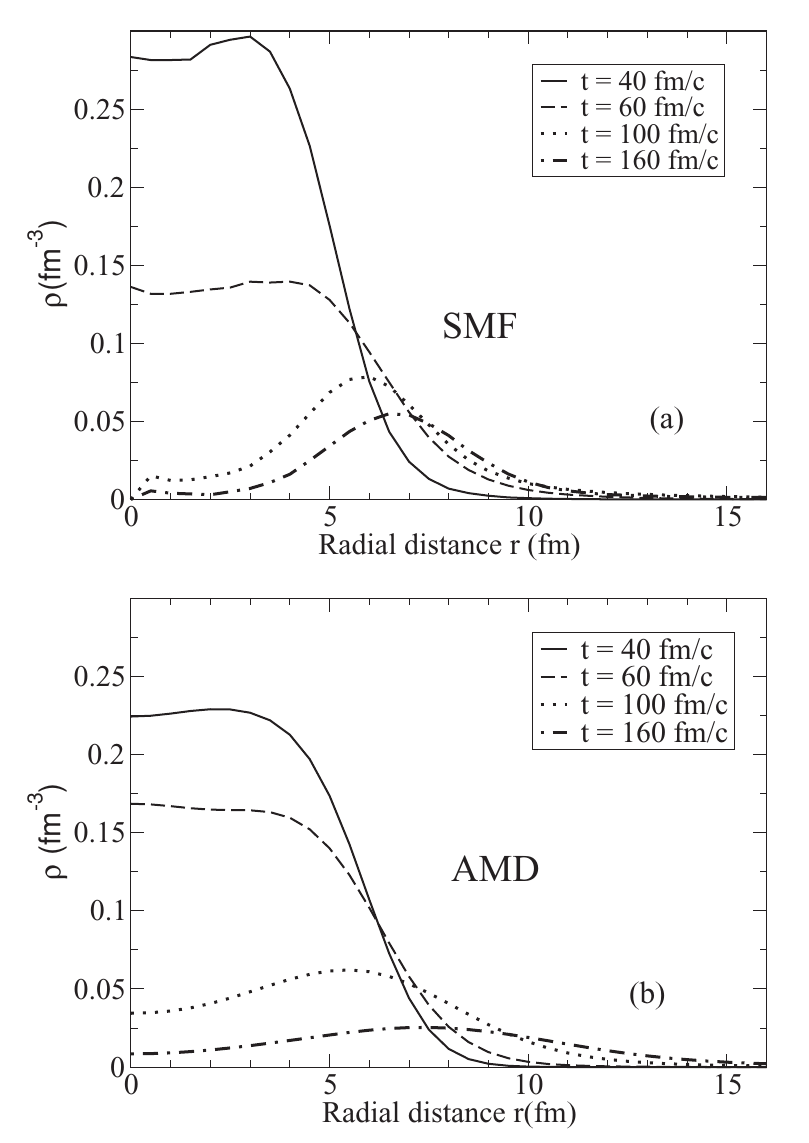}
\end{center}
\caption{Density profiles at several times obtained in SMF (a) and AMD
(b) models, for $^{112}$Sn + $^{112}$Sn central collisions
(b = 0.5~fm) at 50~MeV per nucleon incident energy. 
From~\protect\cite{Col10}.} \label{amdsmf}
\end{figure}
\begin{figure}[htb]
\begin{center}
\includegraphics[scale=0.70]{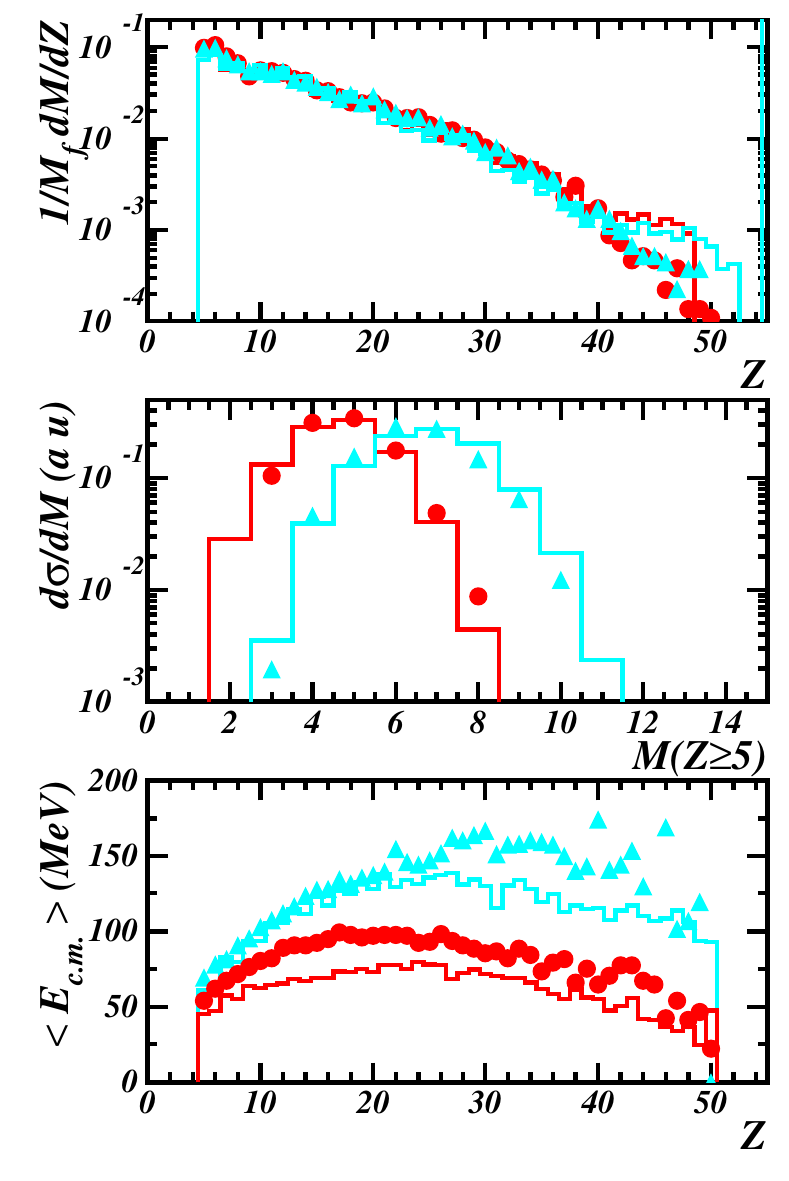}
\end{center}
\caption{Comparison of experimental data (quasifusion from central collisions: 
Gd+U - 36~MeV per nucleon and Xe+Sn - 32~MeV per nucleon incident energies) 
with BOB simulations for charge and multiplicity 
distributions of fragments (top and middle panels) and for their 
average kinetic energies (bottom panel). The symbols are for data and 
the lines for BOB simulations. Light grey lines and triangles stand 
for Gd+U and black lines and circles for Xe+Sn. 
Adapted from~\protect\cite{I29-Fra01}.}
\label{databob}
\end{figure}

A quantitative comparison of SMF and AMD models was made
in~\cite{Col10,Riz07}. They both predict fragment formation leading to
multifragmentation but with different mechanisms. For SMF, fragmentation
is linked to the spinodal decomposition mechanism (i.e. to
mean field instabilities) whereas in  AMD, many-body
correlations are sufficient to produce fragments.
Fig.~\ref{amdsmf} shows the time evolution of the density profile for
the two models and it concerns $^{112}$Sn + $^{112}$Sn central collisions
(b = 0.5~fm) at 50~MeV per nucleon incident energy. The qualitative
evolution of compression and expansion is similar but we notice that AMD
shows broader average density distribution than SMF as the system
expands, pointing to a faster expansion in AMD in which fragments have
already appeared and are distributed widely in space. We notice that
early fragment formation is observed for all models belonging to the
class of molecular dynamics. To conclude on SMF we also report to the
reader an exhaustive comparison of experimental data
with SMF model in~\cite{Bon14}.

In many domains of physics a diffusive behaviour is described by transport
theories which were originally developed for Brownian motion.
The effects of the disregarded degrees of freedom are simulated by a random
term in the dynamics of the retained variables. It is 
the basic idea of the Brownian One-Body dynamics model BOB~\cite{Cho94}.
The fluctuating term is replaced by
\[
\delta \tilde{I} [f] = - \delta \bm{F} [f] . \partial f / \partial \bm{p}
\]
where $\delta \bm{F} (\bm{r},t)$  is the associated Brownian force
($<\delta \bm{F}>=0$). Since the
resulting Brownian one-body dynamics mimics the BL evolution, the 
stochastic force is assumed to be local in space and time. The strength 
of the force is adjusted to reproduce the growth of the most unstable 
modes for infinite nuclear matter in the spinodal region (see~\ref{Spino}). 
Quantal fluctuations connected with collisional 
memory effects are also taken into account as calculated in~\cite{Ayi94}.\\
An extensive comparison data-BOB was made for two very heavy quasifusion 
systems produced in Xe+Sn and Gd+U  central collisions which undergo 
multifragmentation with about the same excitation energy 
($\sim$ 7~MeV per nucleon)~\cite{I29-Fra01,I40-Tab03,I57-Tab05}. Stochastic 
mean field simulations were performed for head-on collisions 
with a self-consistent mean field potential chosen to give a
soft EOS (K$_{\infty}$= 200 MeV). The
finite range of the nuclear interaction was taken into account using a
convolution with a Gaussian function with a width of 0.9 fm.
A term proportional to $\Delta \rho$  in the mean-field
potential was added; it allows to well reproduce the surface 
energy of ground-state
nuclei, which is essential in order to correctly describe the
expansion dynamics of the fused system.  In the collision term a constant
nucleon-nucleon cross-section value of 41 mb, without in-medium, energy, isospin or
angle dependence was used.
As a second step the spatial configuration of the primary fragments,
with all their characteristics as given by BOB, was taken as input in a
statistical code to follow the fragment deexcitation while preserving
space-time correlations. Finally the events were filtered
to account for the experimental device. 
These simulations well reproduce the observed charge and multiplicity
distributions of fragments (see Fig.~\ref{databob}).
Particularly experimentally observed independence of the charge
distribution against the mass 
of the system  was recovered~\cite{I12-Riv98}. 
More detailed comparisons of the charge distributions of the three 
heaviest fragments also show a good agreement~\cite{I29-Fra01}.
Kinetic properties of fragments are rather well reproduced
 for the Gd+U system, whereas 
for Xe+Sn the calculated energies fall $\sim$ 20\% below the measured 
values. 

At this stage it is important to stress that both statistical
(SMM) and dynamical (BOB) models are able to well reproduce
experimental data; it is illustrated by Fig.~\ref{fig:QFSMM} and
Fig.~\ref{databob} for quasifusion data from Xe+Sn at 32 MeV per
nucleon incident energy. It may at first seem surprising that the
results of a dynamical description are so close to those of a
statistical model. As thermodynamic equilibrium corresponds to an
unbiased population of the available phase space it is a strong
indication that the dynamics (first governed for BOB by mean field
instabilities) is effective in filling at least a large part of phase
space. The mixing of a selected set of events also contributes to the
large covering of phase space; it is also true for theoretical
simulations due to fluctuations in the number of nucleon-nucleon
collisions in the entrance channel. 

\begin{figure}
\begin{center}
\includegraphics[scale=0.80]{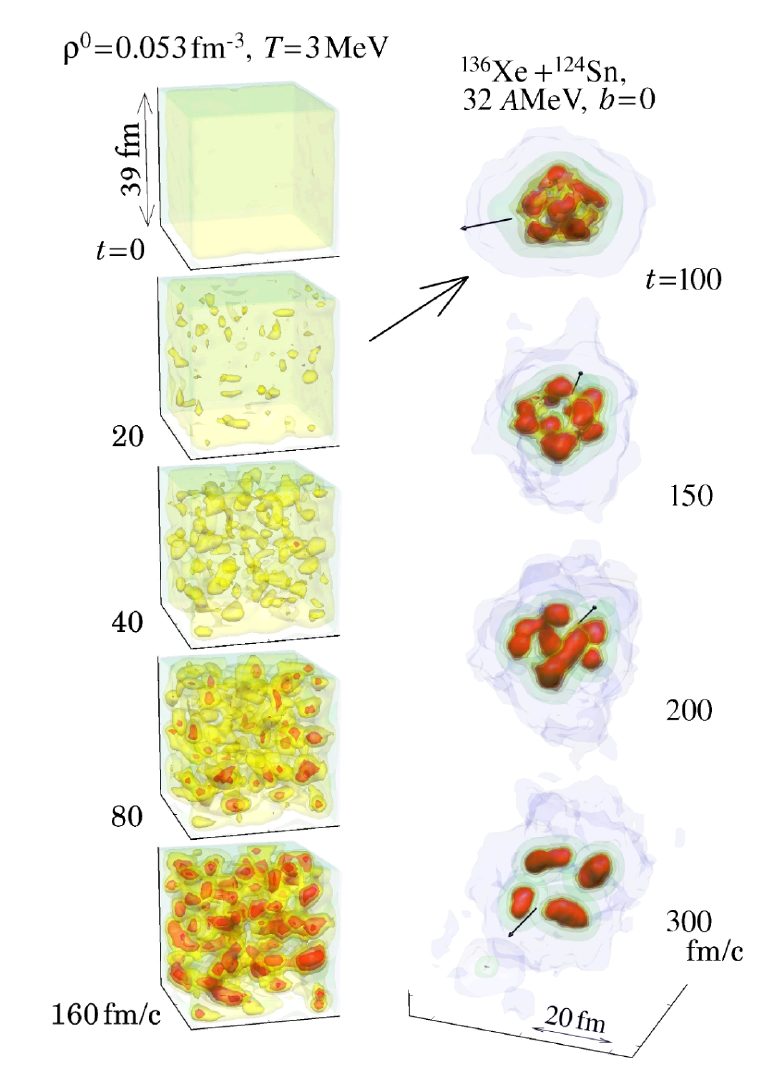}
\end{center}
\caption{Density landscape at several times for nuclear matter in a
periodic box (left panel) and  for a hot nuclear system (right panel) formed in a
head-on collision. Small arrows indicate the beam direction. For the
big arrow see text. From~\cite{Nap17}.} \label{BLOB}
\end{figure}
Recently a numerical treatment of Eq.~(\ref{eq:BLE})
in which fluctuations are introduced in full phase space from
induced nucleon-nucleon collisions has been proposed (BLOB simulation)~\cite{Nap17}.
This transport simulation based on a complete treatment of the 
Boltzmann-Langevin approach proves to be very efficient in 
building fluctuations for equilibrated systems and preserving
fluctuations of larger amplitudes, which leads to a more reliable
description of the onset of multifragmentation.  
Fig~\ref{BLOB} illustrates the correspondence between a portion of
nuclear matter ($T$ = 3~MeV and $\rho \sim \rho_{0}/3$) and a hot
nuclear system formed in a central collision between $^{136}$Xe and
$^{124}$Sn at 32~MeV per nucleon incident energy. We observe some
analogy (big arrow) between the early time when inhomogeneities emerge in nuclear
matter (20~fm/c) and when fragments start forming in a nuclear
system (around 100~fm/c) after accessing low-density spinodal conditions.
In both systems, a spinodal signal stands out by exhibiting equal-size
inhomogeneities in configuration space within a similar
timescale~\cite{Nap13} and it is smeared out later on (see~\ref{Spino}). 

To conclude on dynamical descriptions of multifragmentation and
fragment formation, one can make a few general comments.  
For AMD and stochastic mean-field simulations
at incident energies around 35-50~MeV per nucleon 
a maximum density of 1.2-1.6 $\rho_0$
is observed at 30-40~fm/$c$ after the beginning of central heavy-ion collisions 
and the normal density is recovered around 60 fm/c. Thermal equilibrium 
times  are found in the range 100-140 fm/c after the beginning 
of collisions, well before
freeze-out configurations (200-300 fm/$c$). Primary fragments exhibit
an equal or almost equal excitation energy per nucleon of 3-4~MeV in
good agreement with values deduced from
experiments~\cite{I39-Hud03,I66-Pia08}.
The mechanism of fragment production differs depending on model type.
 In molecular dynamics models fragments are
preformed  at early stages close to the normal nuclear density whereas
in stochastic mean field calculations fragment formation is
linked to spinodal instabilities; mononuclear systems at
low density ($\sim$0.4$\rho_0$) formed at around 100~fm/$c$ develop 
density fluctuations during about 100~fm/$c$ to
form fragments. More constrained observables related to the formation 
of fragments by spinodal instabilities will be discussed in~\ref{Spino}.
As a last point, dynamical calculations exhibit radial collective energies
for fragments with average values in the range 0.1-2.0~MeV per nucleon for 
heavy-ion collisions in the Fermi energy domain which agree fairly well 
with values derived from experiments (see~\cite{Bor08}).
    
\subsection{Information on thermodynamic variables}\label{thermovar}
Excitation energy can be derived from experiments through
calorimetry with some precautions. Indeed the formed
multifragmenting hot nuclei/nuclear systems  are accompanied by 
preequilibrium emission produced mainly by first and second chance
nucleon-nucleon collisions. These early emitted particles, mostly neutrons, 
$H$ and $He$, must not be included in the
calculation of the mass and of the excitation energy of the fragmenting 
system. In experimental analyses, preequilibrium particles are excluded either 
through angular and energetic properties of the observed 
products~\cite{I29-Fra01} or with the help of models.
For thermal energy, as mentioned before, the collective energy
to remove (generally a small part of total excitation energy) is evaluated
from models constrained by experimental data.
As far as temperature is concerned several methods
have been proposed and applied. However substantial differences
between thermometers have been observed when the excitation energy or
incident energy increases, which seems to indicate that thermometers
are not always measuring the same thing. We will shortly discuss this
problem. It is why temperature is often deduced from a statistical model that 
well reproduces the data.

\subsubsection{Calorimetry} \label{Calo}
All procedures for obtaining the excitation energy of a fragmenting 
source, observed with a 4$\pi$ array, are based on the determination 
of its velocity. For central 
collisions the reaction centre of mass velocity is most often chosen
whereas the quasi-projectile and the projectile spectator
velocities are either identified with 
that of the biggest fragment, or with that of the subsystem containing all 
the fragments (Z$\geq$3 or 5), forward emitted in the centre of mass.
The excitation energy, $E^*$, of the source is then calculated event 
by event with the relation
$E^* = \sum_{M_{cp}} E_{cp} + \sum_{M_n} E_{n} - Q$.
$E_{cp}$ and $E_{n}$ are respectively the kinetic energies of charged
products and neutrons emitted by the source, $Q$ is the mass difference
between the source and all final products.  
Energies are expressed in the source reference frame. $M_{cp}$ is in most
cases the detected multiplicity of charged products. The energy removed
by gamma rays is small and most often neglected in the calculation.
The way in which different charged products are attributed to the
sources differs with the experimental apparatus and the type of collision
under study. 
For central symmetric heavy-ion collisions, all fragments with Z$\geq$3
(or 5) are attributed to the source. Preequilibrium in that case is
mostly forward/backward emitted, and indeed the angular distributions of the
light products appear isotropic between 60 and 120$^o$. The charge,
mass and energy
contributions of these particles are doubled for the calculation of the
characteristics of the source (i). Another possibility, to account for the
detector inefficiency, is to calculate the charge, mass and energy of the
anisotropic part, and to remove it from those of the composite system (ii).
For quasi-projectiles the most important contamination comes from 
mid-rapidity products and several techniques are used for the
quasi-projectile
reconstruction. i) All fragments forward emitted in the reaction 
c.m. system are attributed to the quasi-projectile. Variants consist either
in putting a low velocity cut for the lighter fragments~\cite{Lle93,Pla01},
or in keeping only events with a compact fragment configuration in velocity
space~\cite{I61-Pic06,I69-Bon08}. Then twice the light 
elements in the quasi-projectile forward hemisphere are added.
ii) Fragments are treated as above, but particles are attributed 
a probability to come from the quasi-projectile emission, either using a 3-source 
fit~\cite{Ma05}, or by taking a well characterized subspace as 
reference~\cite{H5Vie06,Pia06}.
The velocity of the quasi-projectile is then recalculated by including all its components.
For projectile spectators, the highest energy deposits are obtained
with an intranuclear cascade model~\cite{Sum90}. They represent the sum
of the hole energies left behind by nucleons knocked-out from the
spectator and of the energies carried by struck nucleons captured into the
spectator. The projectile spectator reconstruction is generally made
including the measured abundances for Z $\geq$ 2 and the yields of
hydrogen isotopes are deduced by extrapolating to Z = 1 whereas free
neutrons are usually measured~\cite{Poc95,Tra98}.
Finally in hadron-induced collisions, products emitted from the
source are chosen from energetic considerations, by excluding those 
with an energy per nucleon above a given threshold either 
fixed~\cite{Hau96} or varying with Z~\cite{Vio06}.
All those procedures assume forward-backward 
symmetry of  particle emission in the source frame. For quasi-projectiles the
symmetry of the source emission may be questionable when highly excited 
quasi-projectiles and quasi-targets start emitting right after their
separation~\cite{Jan05,Hud05}: the close proximity of the partner
deforms phase space and emission is favoured between quasiprojectile and 
quasitarget. This possible effect is generally ignored.

Once all charged products have been attributed to the source, its charge is known. 
A first uncertainty is introduced in calculating the mass of the
source in the cases where the masses of all associated decay products
(especially heavy fragments) are not measured. A single mass can be attributed 
to all nuclei with a given atomic number, either that of the most stable 
species, or that derived from formulae existing in the literature 
(EPAX~\cite{Sum00} or EAL~\cite{Cha98}). 
At that point neutrons must be included. Except in
experiments using a neutron ball or a neutron wall in experiments with
relativistic spectators, neither their multiplicity not their
energy is known. The neutron number can then be estimated by assuming that the
source has the same N/Z ratio as the total system (central or hadron-induced 
collisions) or as the projectile. The average neutron energy is then taken equal
to the proton energy averaged over the event sample after subtraction of some
estimate of the Coulomb 
barrier. Note that with neutron balls only the neutron multiplicity is 
measured, at the price of a poor geometrical coverage for charged 
products. In that case corrections accounting for the undetected particles
and neutrons are made~\cite{Gal05}. 
\begin{figure}
\begin{center}
\includegraphics[width=0.6\textwidth]{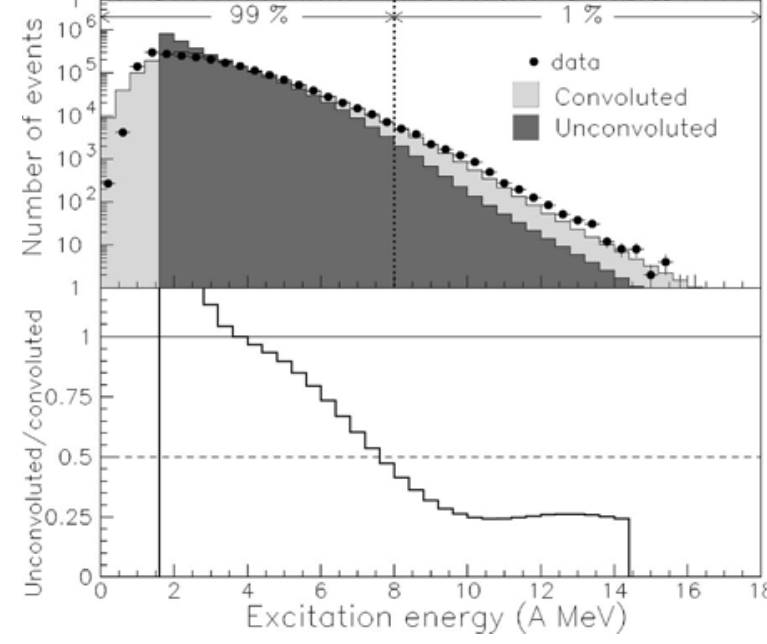}
\end{center}
\caption{{Top : Convoluted, unconvoluted and experimental excitation 
energy distributions obtained in $\pi$+Au reactions. Bottom :
Ratio of unconvoluted-to-convoluted distribution as a function of
the excitation energy per nucleon.  
Adapted from~\cite{Lef01}}\label{fig31.2}}
\end{figure}
In central heavy-ion reactions, the populated excitation energy domain is
narrow: $\sigma_{E^*} \sim$~0.7 - 1.25~MeV per nucleon; the width includes 
experimental effects (detector efficiency and resolution), calculation 
assumptions and physical effects (pre-equilibrium).
Conversely, in hadron-induced reactions as well as in quasi-projectile
and projectile spectator studies, a broad domain of excitation 
energy is populated, proportionally 
to the partial cross section, function of the impact parameter. However, 
due to on-line trigger effects very low energies are poorly sampled, 
for example due to the acquisition trigger based on a minimum
multiplicity of charged products; indeed neutron
emission - often not detected or with low efficiencies - is dominant in this region. 
At the other end of the distribution, the very high energies 
probably result from significant fluctuations. In
all cases the reliable domain extends from about 2 to 8~MeV per nucleon:
for example in Fig.~\ref{fig31.2} obtained in $\pi$+Au reactions, the
excitation energy distribution is unconvoluted assuming Gaussian 
fluctuations. More than half of the 1\% of events above the vertical 
dotted line have an energy overestimated by 1-2~MeV per nucleon~\cite{Lef01}.
\begin{figure}
\begin{center}
\includegraphics[width=0.6\textwidth]{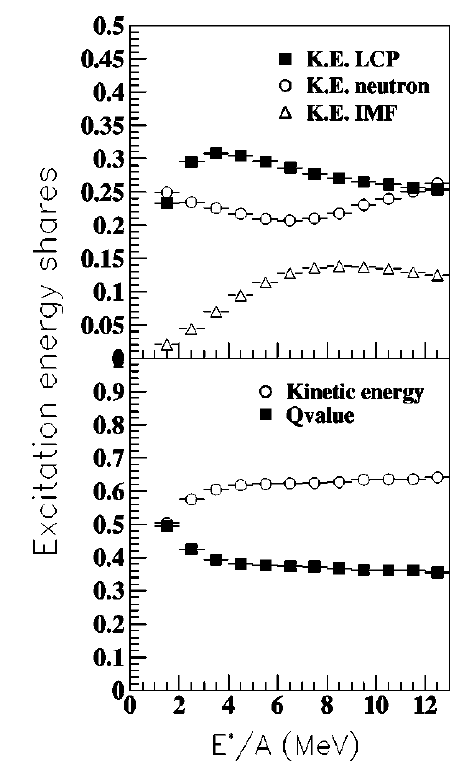}
\end{center}
\caption{{Relative share of excitation energy for various components of 
the reconstruction procedure as a function of the excitation energy per 
nucleon. From~\cite{Lef01}}\label{fig31.3}}
\end{figure}

How reliable are the energies so obtained ? 
Because of compensation of the errors on the mass and on the 
energy, the excitation energy per nucleon is a more robust experimental
observable than the total excitation energy.
By comparing values obtained by different methods for quasi-projectiles,
differences on $E^*/A$ smaller than 10\% were 
found~\cite{I27-Ste01,I69-Bon08}. From
simulations with an event generator, the reconstructed
values were found to differ from the true values by less than 
10\%, except for very peripheral collisions where the
discrepancies are much larger~\cite{H5Vie06}. In central collisions, 
excitation energies slightly smaller than the available
energies are generally found, which is what can reasonably be expected.
It was verified in the INDRA Xe+Sn data that the two procedures
for central collisions give the same results when a high
degree of completeness is required, e.g. that at least 90\% of the system 
charge be measured for each event in the considered sample. 
For lesser completeness (80\%), the difference between both types of
calculation increases with the incident energy, reaching 1~MeV per
nucleon (10\%) at 50~MeV per nucleon incident energy.
The main source of uncertainty in the calculation of $E^*$ comes from the
neutron terms. However, compensation occurs in the calorimetry equation
between the kinetic and the mass balance terms; indeed the weight of 
these two terms is similar for quasi-projectiles and in hadron-induced 
reactions 
($Q/E^* \sim$30-36\%~\cite{Lef01,T41Bon06} - see Fig.~\ref{fig31.3}); in 
central collisions the $Q$ term accounts for only $\sim$20\% of the 
excitation energy. 

\subsubsection{Temperature measurements} \label{Temp}
Two reviews extensively describe the methods most used for temperature 
measurements~(\cite{DasG01,Kel06} and references therein). A brief summary
will be given here. We will also draw attention to the inconsistency
of temperature measurements when using different thermometers.
The concept of temperature for a nucleus, which is a 
microscopic, isolated Fermionic charged system, is not \emph{a priori}
obvious. However we recall following our previous discussions
(see~section~\ref{pseudo}) that the intensive variables such as
temperature can always be defined for statistical ensembles constructed 
from homogeneous samples of selected hot nuclei. 
 As no probe can be used to measure the temperature of these small systems,
it has to be derived from the properties of particles that they emit
during their cooling phase. 
Three families of methods are used to "measure" temperatures.   
\begin{enumerate}

\item \textit{Kinetic temperatures.}
Historically, temperatures of compound nuclei were derived from the 
slopes of the kinetic energy spectra of the emitted neutrons or 
charged particles that they evaporate, as the spectra can be fitted with 
Maxwell-Boltzmann distributions~\cite{Wei37} (see Fig.~\ref{spectreE}). 
At higher energies, when long 
chains of particles are emitted, the obtained result is an average over
the deexcitation chain, and may differ from one particle to another,
depending on the emission sequence. To retrieve the initial 
temperature, it was proposed to subtract from the spectra those 
of particles coming from the same nucleus formed at lower
excitation energies~\cite{Gon90}. For multifragmenting systems, the 
slopes of light product spectra lead to very high
"temperatures", and probably do not reflect only the thermal 
properties of the system, but also the collective energies coming 
from the dynamics of collisions.

\item \textit{Excited state temperature.}
Thermometers are based upon the relative populations of excited states
of emitted particles. The underlying idea for this method is that the 
population of the excited
states of a system in statistical equilibrium is given by the temperature 
of the system and the energy spacing, $\Delta E$ = $E_1$ - $E_2$, between the levels.
\begin{equation}
T = \frac{E_{1} - E_{2}}{ln(a'Y_{1}/Y_{2})} 
\label{eq:Tde}
 \end{equation}
Here $a' = (2J_{2} + 1)/(2J_{1} + 1)$, $E_i$ the excitation energy,
 $Y_i$ is the measured yield and $J_i$ is the spin of the state $i$.
This definition in itself bears the limits of the method:  when the temperature
is higher than $\Delta E$, the ratio between the population of two states
saturates. Considering particle-unbound states is thus interesting as it
allows to measure higher temperatures, and the population ratio should
in that case be less influenced by secondary decays. Anyhow the considered
emitted particles should be present at freeze-out.
\begin{figure}[htb]
\begin{center}
 \includegraphics[scale=0.75]{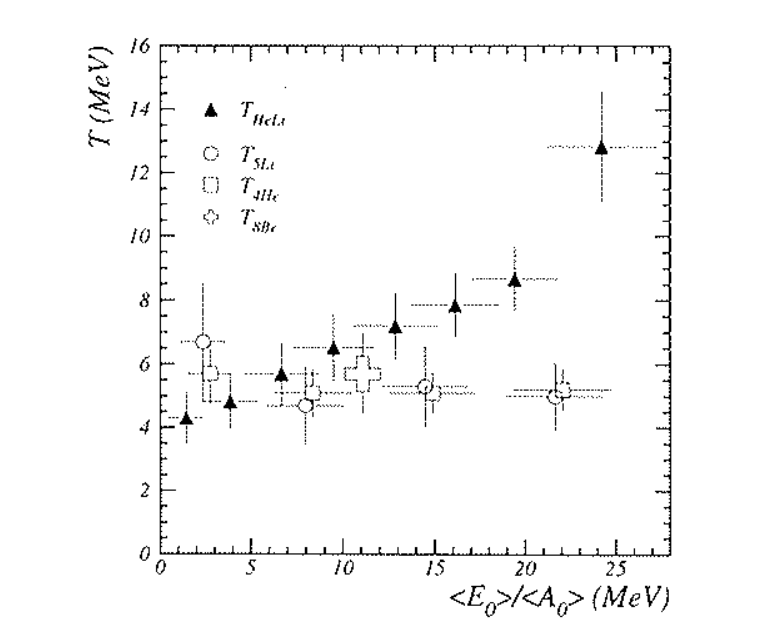}
 \includegraphics[scale=0.45]{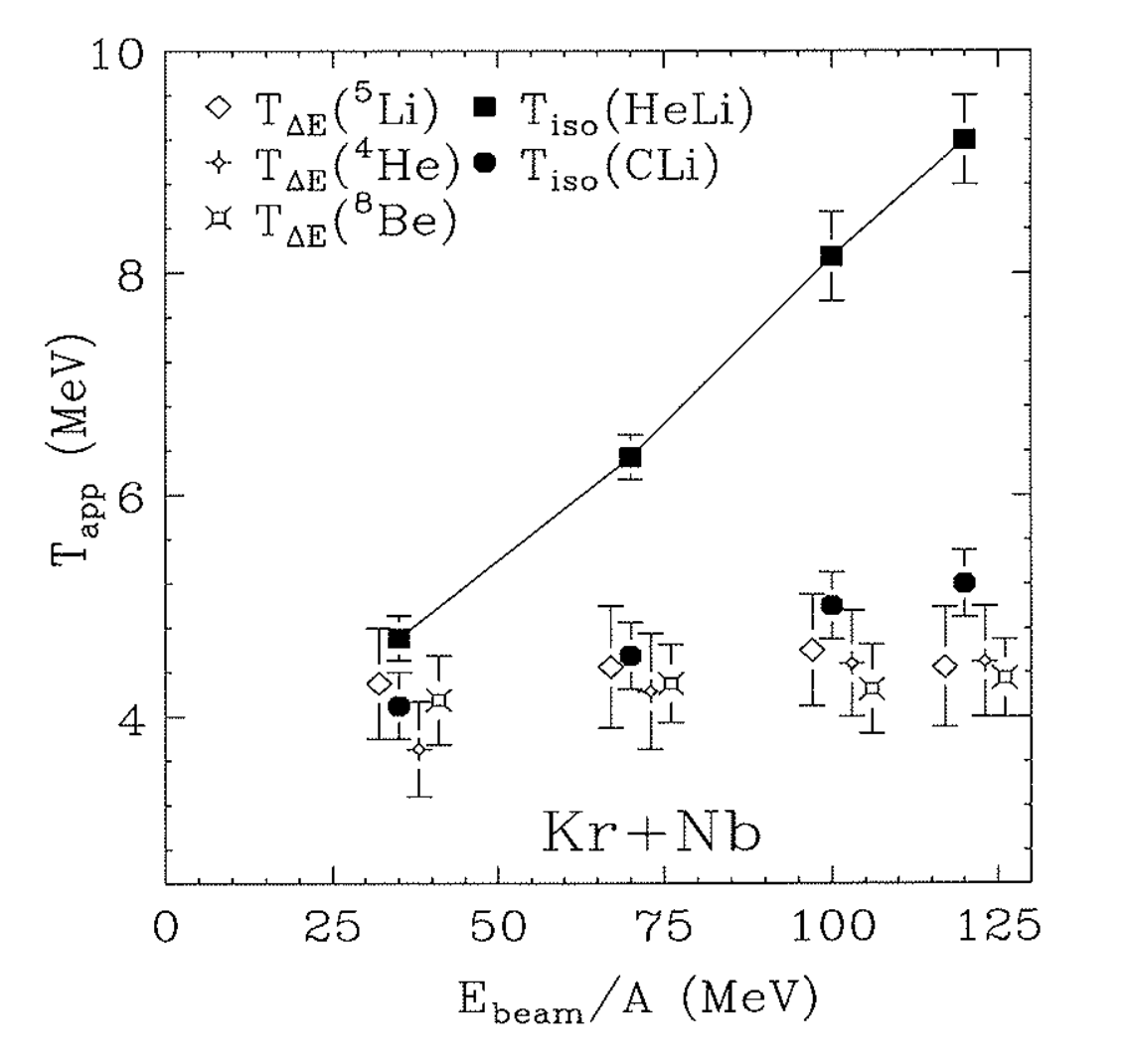}
\end{center}
\caption{Measured isotope temperatures and excited state temperatures;
(left) as a function of the experimental excitation energy per nucleon for
target spectators produced in Au on Au collisions at 1GeV per nucleon
incident energy; (right) as a function of incident energy for central
collisions between Kr and Nb. From~\protect\cite{Tra98} and 
adapted from~\cite{Xi98}.}
\label{fig:compT}
\end{figure}

\item \textit{Isotope temperature.}
This method uses the yields of different light isotopes produced by the
system. It was developed in the grandcanonical approach, and is valid for
systems at densities low enough to make fragment nuclear interaction 
negligible, thus the chemical composition of the system is frozen~\cite{Alb85}. 
The basic
assumption is that free nucleons and fragments are in thermal equilibrium
within an interaction volume V. The density of an isotope reads:
$\rho(A,Z)=N(A,Z)/V =A^{3/2} \omega(A,Z) \lambda_{T_N}^{-3} \exp (\mu(A,Z)/T)$,
where $\omega$ is the internal partition function of particle ($A,Z$), $\mu$
its chemical potential and $\lambda$ the thermal nucleon wavelength.
The condition of chemical equilibrium allows to define the chemical
potential of a species in terms of those of free neutrons and protons and of
its binding energy. Using two sets of two nuclei differing only by one
nucleon, the temperature is derived from the double yield ratio, the binding
energy differences $B$ and the partition functions only, the other terms 
disappear.
\begin{equation}
T = \frac{B}{ln(a(Y_{1}/Y_{2})/(Y_{3}/Y_{4}))} 
\label{eq:Tiso}
 \end{equation}
where $Y_1$, $Y_2$ are the yields of one isotope pair and $Y_3$, $Y_4$
is another isotope pair; $a$ contains the statistical weighting factor.
A first problem lies in the calculation of the binding energies which might
depend on density and temperature. Different corrections were proposed to
account for finite-size effects~\cite{Radu99} or secondary
decays~\cite{Tsa97,Rad00}. But again this equation assumes that the considered particles
should be present at freeze-out, and not  produced by secondary decays.
\end{enumerate}
This last point which appears for both thermometers can explain why 
they can give different temperatures. This is well illustrated by
Fig.~\ref{fig:compT}. Temperatures have been extracted from very different
experiments; for central Kr + Nb collisions (right) one can
roughly convert the beam energy scale saying that 50 MeV per nucleon
corresponds to 10 MeV per nucleon excitation energy. The same trend is
observed on both figures: the HeLi isotope temperature differs from
other measurements above around 10 MeV per nucleon excitation energy.
Without entering here in a deep discussion, it appears  \emph{a posteriori}
that the HeLi isotope temperature seems closer to the real temperature above
10 MeV per nucleon excitation energy. 

More recently another method for measuring the temperature of hot nuclei was
proposed~\cite{Wue10,Zhe11}. It is based on momentum fluctuations of emitted
particles, like protons, in the centre of mass frame of the fragmenting nuclei.
In a classical picture, assuming a Maxwell-Boltzmann distribution of the momentum
yields, the temperature $T$ is deduced from the quadrupole momentum fluctuations
defined in a direction transverse to the beam axis:\\
$\sigma^2$ = $<Q_{xy}^2>$ - $<Q_{xy}>^2$ = $4m^2T^2$\\
with $Q_{xy}$ = $p_{x}^2$ - $p_{y}^2$;
$m$ and $p$ are the mass and linear momentum of emitted particles.
If we now take into account the quantum nature of particles, a correction
$F_{QC}$
related to a Fermi-Dirac distribution was also proposed~\cite{Zhe11,Zhe12}.\\
In that case $\sigma^2$ = $4m^2T^2$ $F_{QC}$ where $F_{QC}$ =
$0.2(T/\epsilon_f)^{-1.71}$ + 1;\\ $\epsilon_f$ = 36 $(\rho/\rho_0)^{2/3}$
is the Fermi energy of nuclear matter at density $\rho$ and
$\rho_0$ corresponds to saturation density. Density can be estimated from
models. Again this method can be useful for relevant measurements by selecting 
particles emitted at freeze-out.

To conclude, one can say that
direct temperature measurements are questionable and we will see 
that, very often, temperatures are derived from comparisons of data
to statistical models or from simulations starting from data and able
to recover with a good level of confidence freeze-out properties.

\subsubsection{Break-up densities and freeze-out volumes}
\label{DenVol}
No sufficiently accurate method is available to determine precisely the spatial 
extension of hot nuclei/nuclear systems which undergo multifragmentation.
However we will see that derived estimates are rather coherent despite
the variety of methods used; three examples are chosen to illustrate
those determinations.

Break-up densities for projectile spectator fragmentation in $^{197}$Au+$^{197}$Au
collisions at 1000~MeV per nucleon incident energy were estimated by using 
selected particle-particle 
correlations (particles from secondary decays are excluded by imposing 
an energy threshold)~\cite{Frit99}. Assuming zero lifetime,
the volumes of spectator sources were extracted 
and densities calculated by dividing the number of spectator 
constituents by the source volume. The estimated average values slowly 
decrease from about 0.4 to 0.2~$\rho_0$ when excitation energies of 
spectators increase from 4 to 10~MeV per nucleon. 
 
\begin{figure}[htb]
\begin{center}
 \includegraphics[scale=0.90]{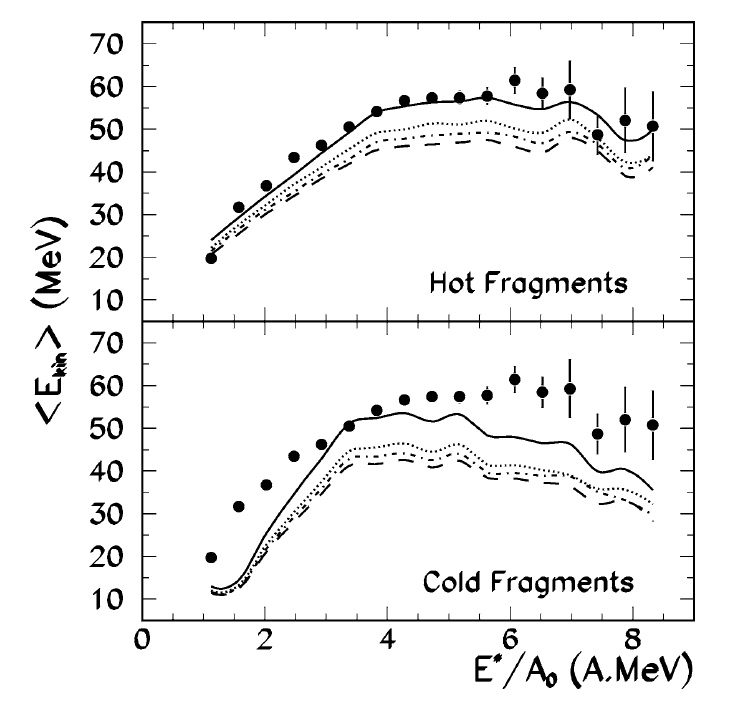}
\end{center}
\caption{Average fragment kinetic energy (corrected from collective
energy) as a function of the thermal
excitation energy for quasi-projectiles produced in
$^{197}$Au+$^{197}$Au mid-peripheral collisions at 35~MeV per nucleon incident energy.
Full points are corrected data and bars represent the statistical errors.
Lines correspond to mean many-body trajectory calculations for a volume
of 3$V_0$ (full), 4$V_0$ (dotted), 5$V_0$ (dashed dotted) and 6$V_0$
(long dashed). The bottom part refers to an extreme unphysical case. From~\cite{MDA02}.}
\label{fig:ekin}
\end{figure}

The average freeze-out volume can also be experimentally estimated
from the mean detected fragment kinetic energy ($<E_{kin}>$).
Concerning quasi-projectile data ($^{197}$Au+$^{197}$Au 
mid-peripheral collisions at
35~MeV per nucleon incident energy~\cite{MDA02}), after subtracting a small
contribution of collective energy (i.e. non thermal and non-Coulomb),
a many-body Coulomb trajectory calculation can be performed by
randomly placing the reconstructed primary fragments in a spherical
volume and letting them evolve in the Coulomb field.
Under the reasonable hypothesis that, on average, particles evaporated
from fragments do not affect fragment velocities, the superposition of
the average Coulomb and thermal motions provides an observable directly
comparable to $<E_{kin}>$ data corrected from collective energies. 
This comparison allows to estimate directly
from data a range of freeze-out volumes. The result is displayed in
Fig.~\ref{fig:ekin}. The full lines, which better agree with data, correspond to 
volumes of 3$V_0$ on average. Note that the hypothesis of cold
fragments (Fig.~\ref{fig:ekin} - bottom part), which is very extreme,
gives an idea of the sensitivity of the method. 

Another possibility is to make a simulation
for estimating the freeze-out properties in a 
fully consistent way.
Such a simulation was done for quasifusion hot nuclei
produced in Xe+Sn central collisions between 32 and 50~MeV per nucleon
incident energy~\cite{I58-Pia05,I66-Pia08}.
The method for reconstructing freeze-out
properties requires data with a very high degree
of completeness (measured fraction of the total available charge
$\geq$93\% in this study), crucial for a good estimate of Coulomb energy.
Quasifusion nuclei are reconstructed, event by event,
from all the available asymptotic experimental
information (charged particle spectra, average and standard deviation of 
fragment velocity spectra and calorimetry). 
Dressed excited fragments, which statistically deexcite, and particles 
at freeze-out are described by spheres
at normal density. Four free parameters are used to recover the data: 
the percentage of measured particles which were evaporated from primary 
fragments, the collective radial energy, a minimum distance between 
the surfaces of products at freeze-out and a
limiting temperature for fragments which was mandatory to
reproduce the observed widths of fragment velocity
spectra (see~\cite{I66-Pia08} for details). The agreement between 
experimental and simulated velocity/energy spectra for fragments, for the
different beam energies, is quite remarkable 
(see Fig. 3 of~\cite{I66-Pia08}).
Relative velocities between fragment pairs were also compared
through reduced relative velocity correlation 
functions~\cite{Kim92,I57-Tab05}
(see Fig. 4 of~\cite{I66-Pia08}).
Again a good agreement is obtained between experimental data and
simulations, which
indicates that the retained method (freeze-out topology built up
at random) and the deduced parameters are sufficiently relevant
to correctly describe the freeze-out configurations, including volumes.
Volumes were estimated from an envelope of all the nuclei. The average
values evolve from 3.9 to 5.7$V_0$ between 32 and 50~MeV per nucleon
incident energy whereas average thermal excitation energies of
quasifusion nuclei increase from 5.7 to 9.6~MeV per nucleon. 
A comparison with the results of a microcanonical statistical model (MMM)
was also performed to verify the overall physical coherence of the built
simulation~\cite{I66-Pia08}.

In conclusion we can say that working hypotheses
and approximations are used to give semi-quantitative information on
average break-up densities or freeze-out volumes. At large excitation
energy, around 10~MeV per nucleon, rather constant values 
around 0.2$\rho_0$ (5-6$V_0$) are found whereas 
values from 0.2 to 0.4$\rho_0$ are derived at lower excitation energies.

\section{Two well-identified phases}\label{phases}
Before discussing LG type phase transition signatures for hot nuclei,
we will recall some aspects of liquidlike behaviour of nuclei in 
their ground states or at low excitation energies and characterize
their vaporization at very high excitation energies.

\subsection{Liquid aspects of nuclei: binding energy, fission and evaporation} \label{liq}
For many purposes the nucleus can be 
viewed as a charged liquid drop. Three examples are chosen to
illustrate this vision of nuclei: the liquid drop model, the fission
shapes and the definition of the fissility parameter and the particle
evaporation from excited nuclei.

In the liquid drop model the nucleus is viewed as a charged spherical
liquid drop~\cite{Wei35,Bet36}.
In such a model, one would expect the nuclear binding energy to be
expressed as a bulk or volume term due to the attractive force between
nucleons, a surface correction due to the fact that surface nucleons
are surrounded on average by a smaller number of nucleons, a
second negative contribution due to repulsive Coulomb forces between
protons and a third one produced by an excess of protons or neutrons.
Thus
\begin{equation}
B(N,Z) = a_{V}A - a_{S}A^{2/3} - a_{C}\frac{Z^2}{A^{1/3}} -
a_{A}\frac{(N-Z)^2}{A}
\label{eq:BW}
 \end{equation}
where successive terms represent respectively bulk, surface, 
Coulomb (Coulomb energy of a charged sphere) and asymmetry contributions.
The parameter related to the symmetry energy of finite nuclei at saturation
density (see also~\ref{asymmat}) was deduced using formula (\ref{eq:BW}) 
in~\cite{Dan09} from a fit to over
3100 nuclei with mass number $A>10$ and gives $a_{A}$~=~22.5 MeV.
In this idealization of the nucleus no account is taken of shell effects
or residual interactions arising from independent particle motion in
the nucleus.

\begin{figure}
\begin{center}
\includegraphics[width=0.6\textwidth]{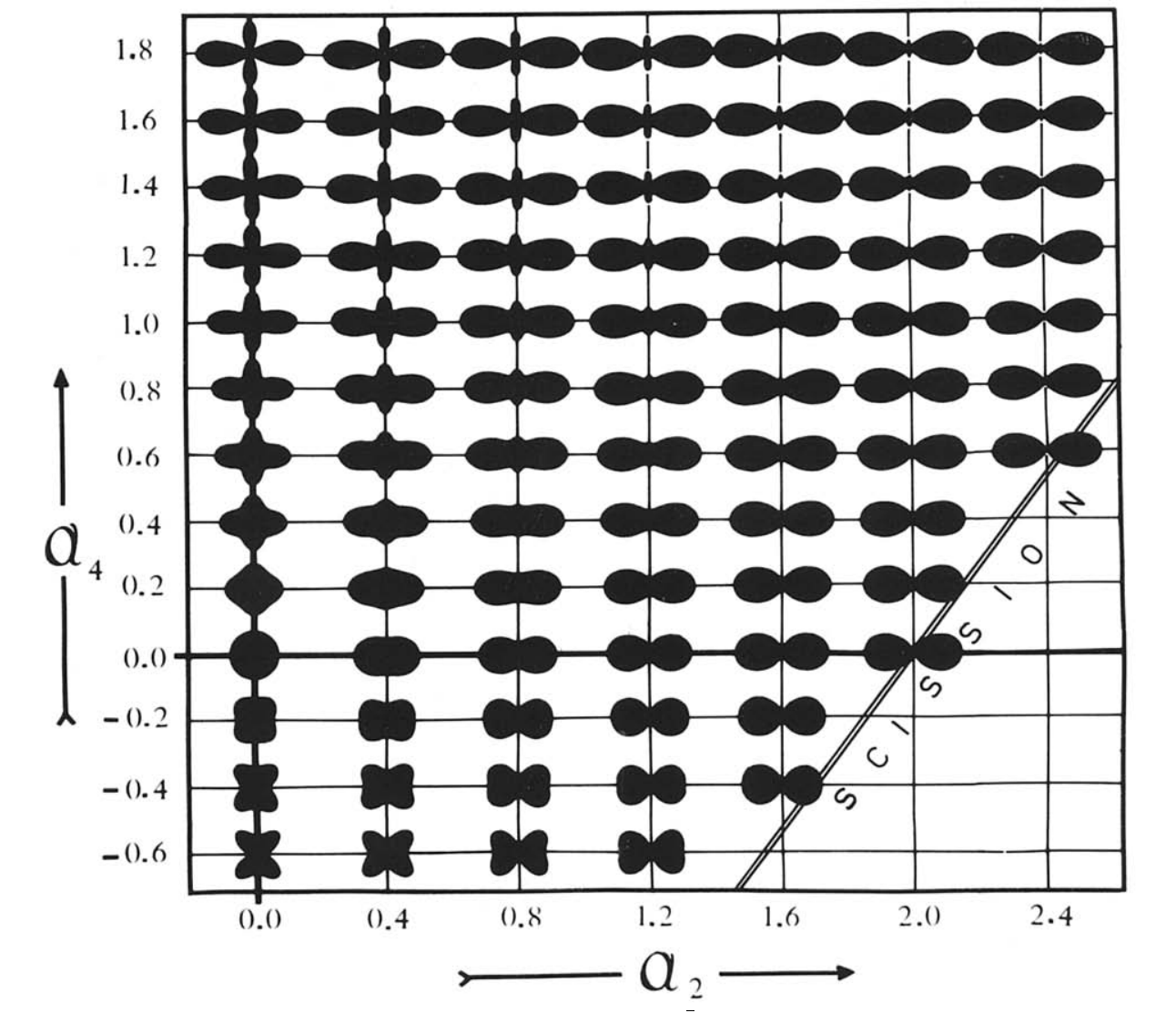}
\end{center}
\caption{Liquid drop shapes corresponding to various locations on
an $\alpha_2$-$\alpha_4$ map. The figures possess rotational symmetry
about the horizontal axis.
 From~\cite{Coh62}.}
\label{shapef}
\end{figure}
Considering fission, in order to accurately describe distorsions of a
sphere as large as are encountered at the top of the fission barrier, 
or saddle point, it is convenient to describe the drop shape in terms of
an expansion in Legendre polynomials: 
\begin{equation}
R(\theta) = (R_{0}/\lambda) [1 + \sum_{n} \alpha_{n}P_{n}(cos\theta)].
\label{eq:Legp}
\end{equation}
The parameter $\lambda$ is a scale factor required to ensure that the
volume remains constant at the value for the sphere of radius $R_0$.
Shapes associated with different combinations of
the leading coefficients $\alpha_2$ and $\alpha_4$ are given 
in Fig.~\ref{shapef}. The surface and Coulomb energies for small distorsions
are given by~\cite{Boh39} $E_S = E_S^0(1+2/5\alpha_2^2)$ and
$E_C = E_C^0(1-1/5\alpha_2^2)$ where $E_S^0$ and $E_C^0$ are the surface
and Coulomb energies of undistorted spheres. In order for the charged
liquid drop to be stable against small distortions, the decrease of
Coulomb energy (-1/5$\alpha_2^2) E_C^0$ must be smaller than the increase in
surface energy (2/5$\alpha_2^2) E_S^0$. The drop will become unstable when
$E_C^0$/$E_S^0$ = 1. Following Bohr and Wheeler~\cite{Boh39} the fissility
parameter $x$ is defined to be equal to this ratio $x$ = $E_C^0$/$E_S^0$.
\begin{figure}
\begin{center}
\includegraphics[width=0.6\textwidth]{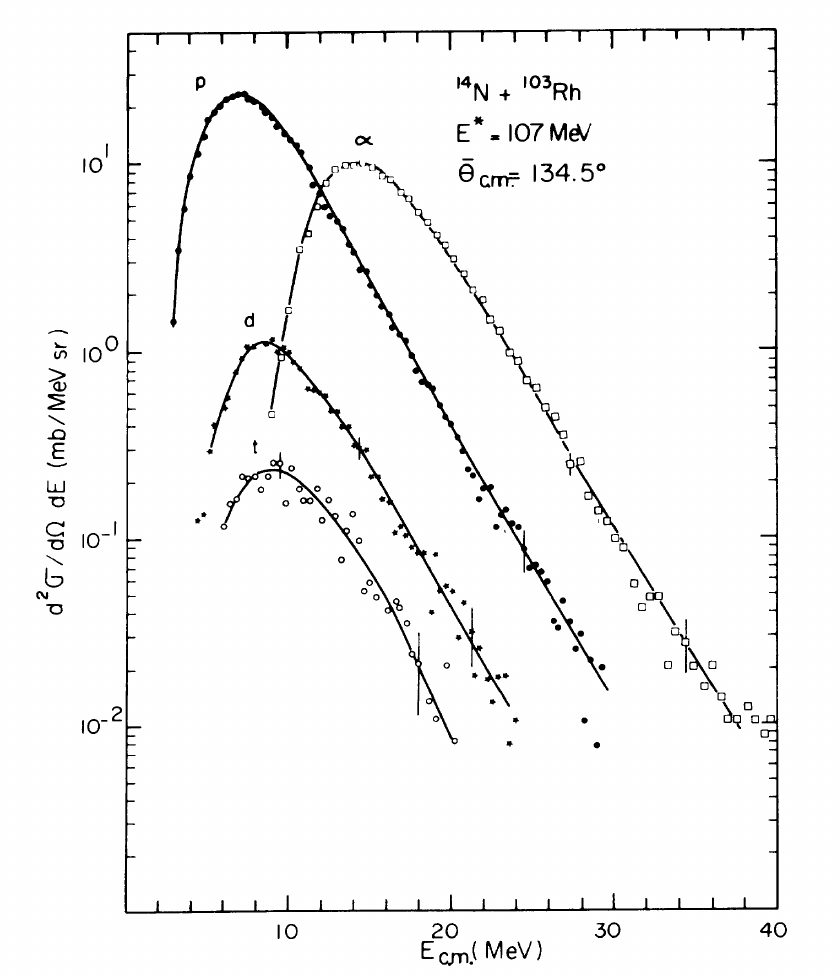}
\end{center}
\caption{Evaporated charged particle spectra emitted from the $^{117}$Te compound
nucleus formed around 1 MeV per nucleon excitation energy. From~\cite{Gal74}.}
\label{spectreE}
\end{figure}

For particle evaporation, the behaviour of nuclei at excitation 
energies around 1 MeV per nucleon has
been extensively studied and rather well understood using statistical
models~\cite{Wei37,Hau52,Col00}. At these energies there is a clear separation of
the timescales between compound nucleus formation, equilibration and
subsequent decay (see~section~\ref{pseudo}). 
The theoretical treatment of particle emission 
involves the estimation of microstate densities defined for equilibrium 
states. Excited nuclei are formed in fusion reactions below around 5
MeV per nucleon incident energies,
which produce a well-defined set of events and the canonical ensemble (fixed
number of particles) is the best suited to study deexcitation energy 
properties.
At such excitation
energies the density stays very close to normal density of cold nuclear
matter and the earliest evaporation model rests  on the basic idea:
an emitted particle can be considered as originally situated somewhere on 
the surface of the emitting nuclei at a given temperature and
with a randomly directed velocity, 
it is why we use the term evaporation. Moreover particles are emitted 
sequentially and independently without any correlation.
Fig.~\ref{spectreE} illustrates through Maxwellian spectra for protons,
deuterons, tritons and alpha-particles the evaporation process. Note
the similar high energy slope for all particles which indicates a given
temperature.

\subsection{Gas phase: onset and characterization} \label{gas}
\begin{figure}[htb]
\begin{minipage}[b]{0.49\textwidth}
 \includegraphics[width=\textwidth]{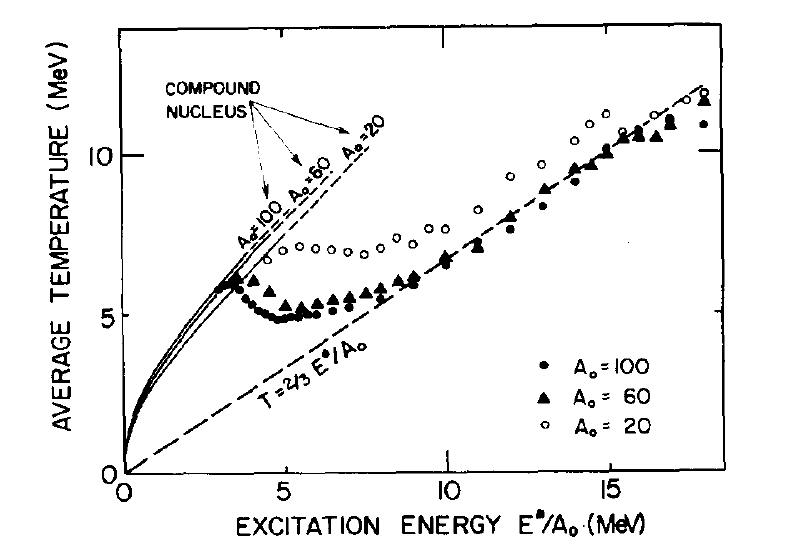}
\caption{Mean temperature $T$ as a function of the excitation energy
per nucleon for three different masses of nuclei. The dashed line
gives the free gas limit. From~\cite{Bon85}.}
\label{fig:ccbondorf}
\end{minipage}%
\hspace*{0.02\textwidth}
\begin{minipage}[b]{0.49\textwidth} 
\includegraphics[width=1.1\textwidth]{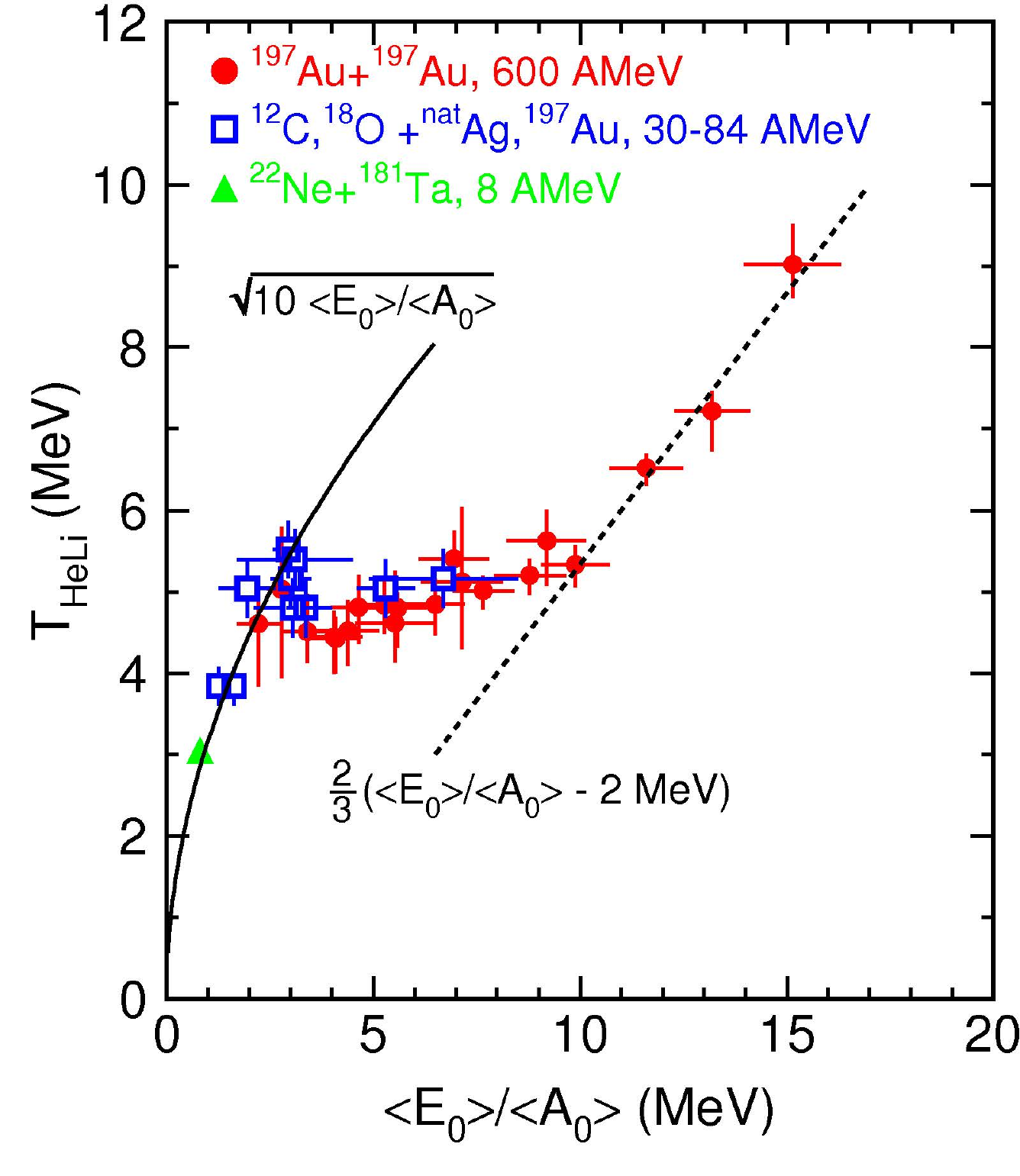}
\caption{Caloric curve of nuclei determined by the dependence of the
isotope temperature on the excitation energy per nucleon (see text).
From~\cite{Poc95}.}
\label{fig:ccGSI}
\end{minipage}
\end{figure}

First indications of excitation energy needed to enter the gas phase
were obtained from combined (and independent) determinations of
thermal excitation energy and of estimated temperatures of hot nuclei.
On the theoretical side, caloric curves were estimated using the
Monte Carlo method in a thermodynamical model based on a
finite-temperature liquid-drop description of nuclear
properties and a related canonical approximation~\cite{Bon85}.
Fig.~\ref{fig:ccbondorf} shows that at excitation energies above
around 10 MeV per nucleon the system behaves like a free gas if nuclei
are heavier than around $A$~= 50. Experimentally, the first caloric curve was derived by the
ALADIN collaboration
from a study of fragment distributions resulting from projectile
spectators produced in Au+Au
collisions at 600 MeV per nucleon incident energy. Excitation energies
per nucleon of primary fragments were determined from the measured
fragment and neutron distributions and temperatures estimated from
yield ratios of He and Li isotopes~\cite{Poc95}.
Fig.~\ref{fig:ccGSI} shows the isotope temperature as a function of
the total excitation energy per nucleon. Data for target residues
produced at lower incident energies between 30 and 84 MeV per nucleon
are also shown (open squares) together with one value below 2 MeV per nucleon
excitation energy provided by Ne+Ta fusion reaction at 8.1 MeV per
nucleon incident energy (triangle). Again, beyond an excitation energy
of 10 MeV per
nucleon, a steady rise of temperature with increasing excitation energy
is observed with a slope which alludes to a free gas; the offset of 2
MeV was interpreted as indicating a freeze-out density around 0.15 -
0.3$\rho_0$.
\begin{figure}
\begin{center}
\includegraphics[width=0.4\textwidth]{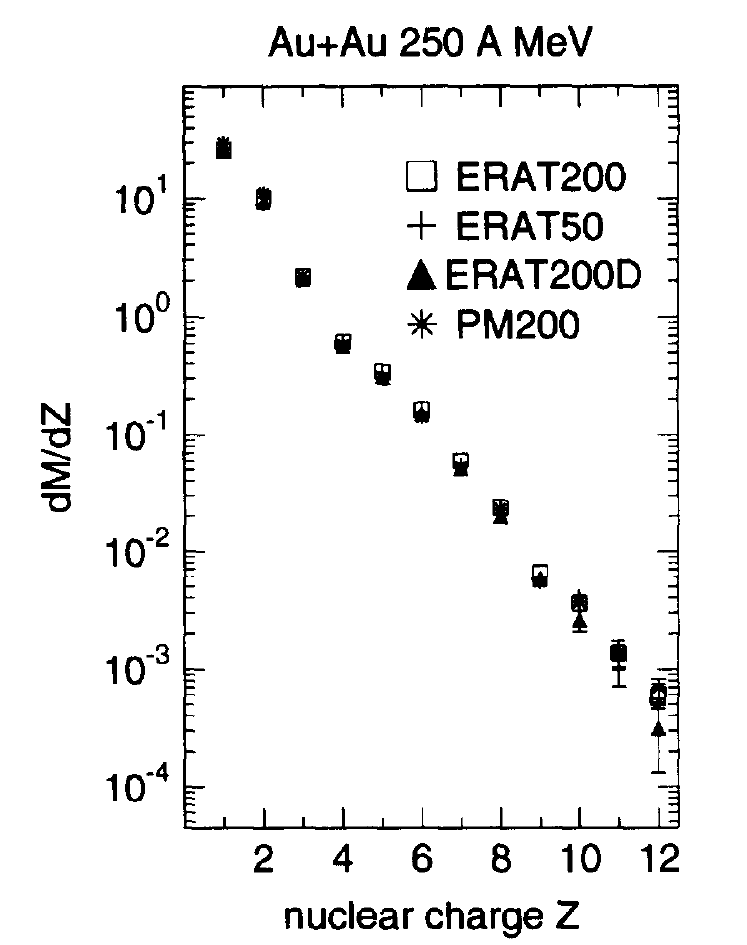}
\end{center}
\caption{Measured multiplicities of vaporized events produced in central
collisions as function of nuclear charge for various central event selections.
 From~\cite{Rei97}.}
\label{fig:Zgas}
\end{figure}

As theoretically predicted the nuclear gas phase is not only composed
of single nucleons, protons and neutrons. It was observed by the FOPI collaboration
studying central $^{197}$Au+$^{197}$Au collisions in the beam energy
range 150 - 400 MeV per
nucleon~\cite{Rei97}. Fig.~\ref{fig:Zgas} displays the composition of
the pure gas phase obtained at 250 MeV per nucleon incident energy
for charged products: we observe light nuclei up to Mg.
The measured complete composition is
the following: 45.3\% of neutrons, 51.2\% of Z = 1, 2 and 3.5\% of light
nuclei. The temperature of the gas was estimated at 26 $\pm$ 5 MeV. 

\begin{figure}[htb]
\begin{minipage}[t]{0.52\textwidth}
\includegraphics*[trim=0 0 52 0,width=\textwidth]{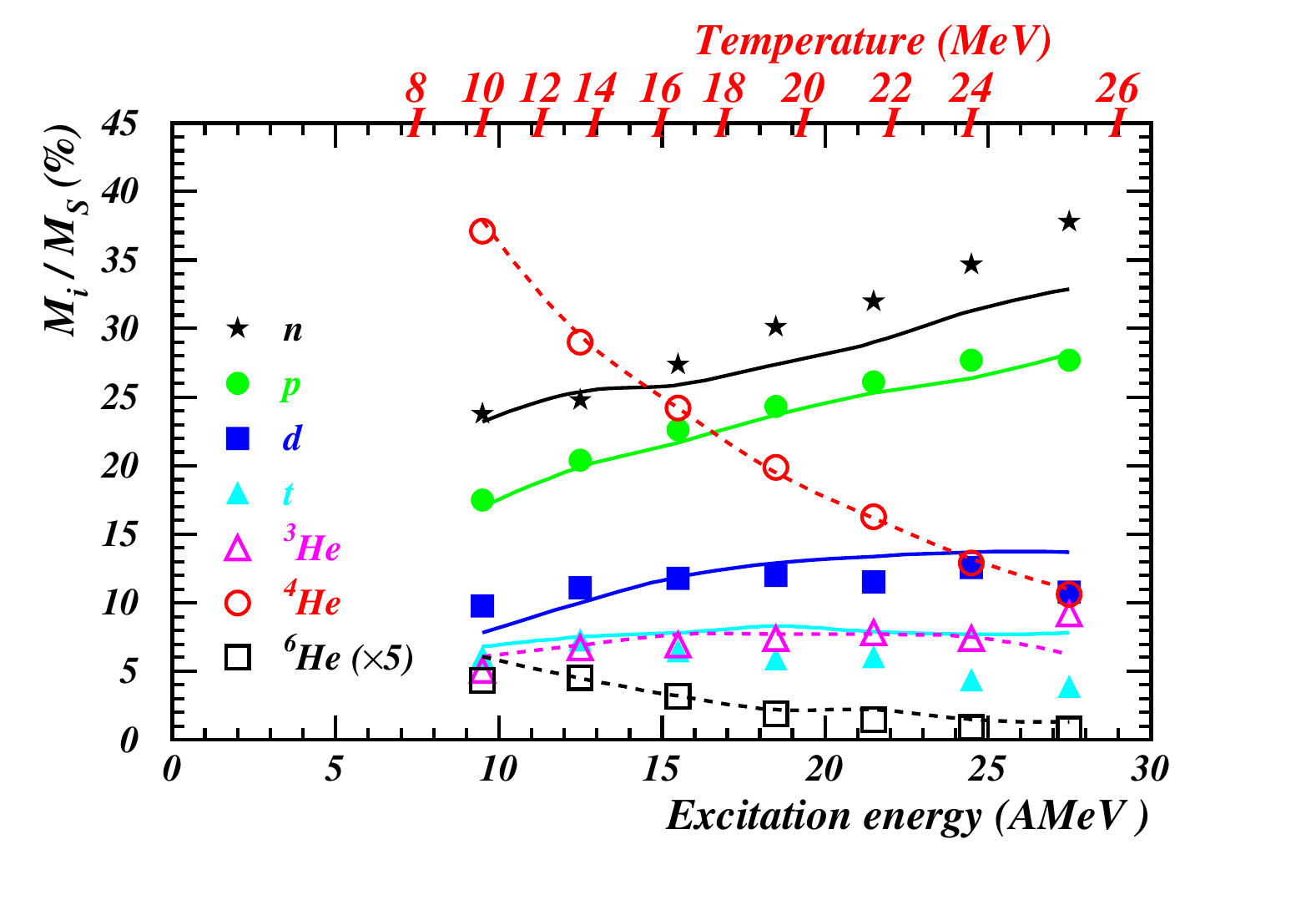}
\caption{Composition of vaporized quasi-projectiles (in percent), formed in 95 MeV
per nucleon 
$^{36}$Ar+$^{58}$Ni collisions, as a function of their excitation
energy per nucleon. Symbols are for
data while the lines (dashed for He isotopes) are the results of the model.
The temperature values used in the model are also given. 
From~\cite{I15-Bor99}.} \label{fig:b2}
\end{minipage}%
\hspace*{0.03\textwidth}
\begin{minipage}[t]{0.45\textwidth}
\includegraphics*[trim= 15 17 52 50,width=\textwidth]{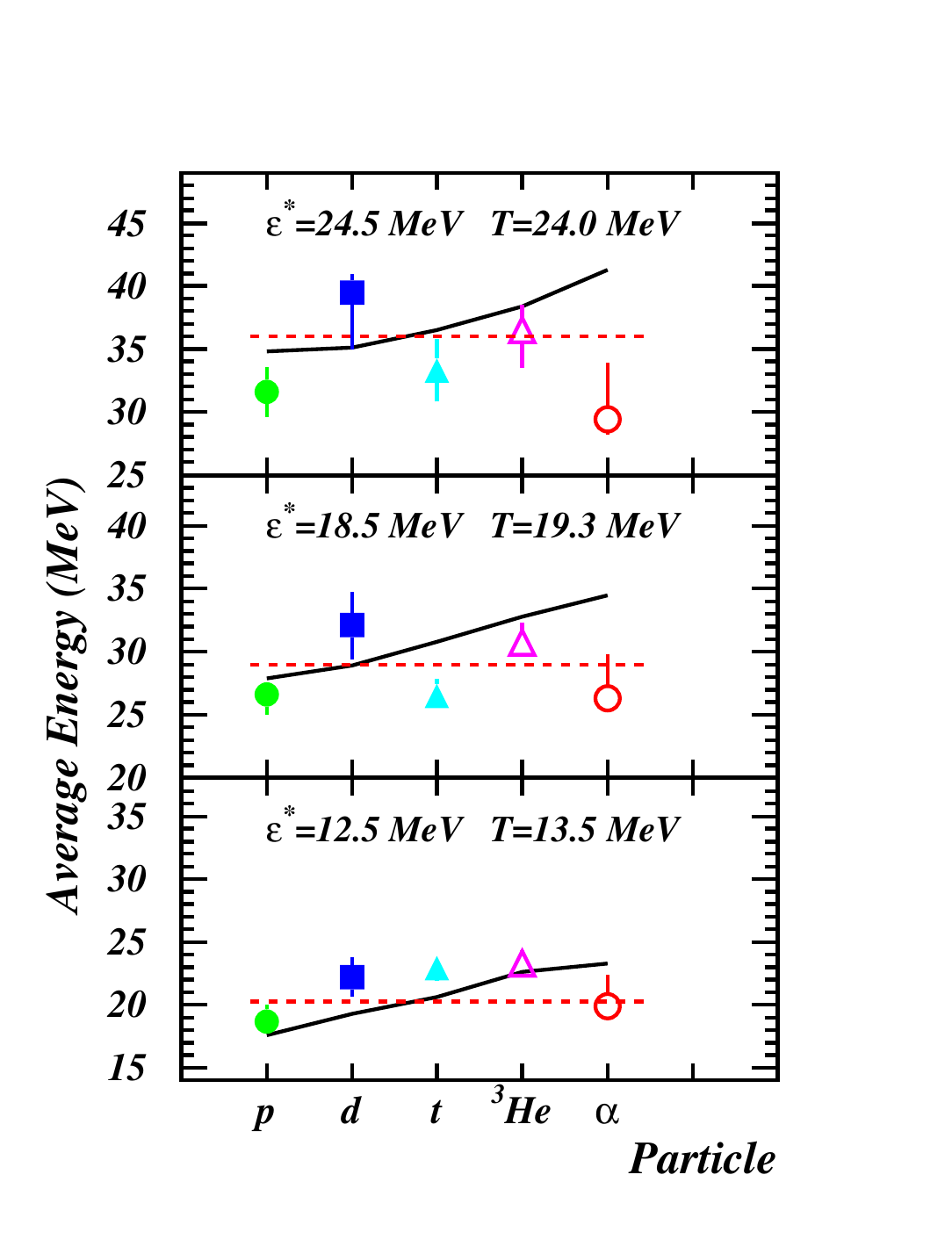}
\caption{Average kinetic energies of particles emitted by vaporized
quasi-projectiles at different excitation energies and formed in 95
MeV per nucleon
$^{36}$Ar+$^{58}$Ni collisions. Symbols are for data while full lines 
are the results of the model. The dashed lines refer to average kinetic 
energies of particles (3T/2) expected for an ideal gas at temperatures 
derived from the model. From~\cite{I15-Bor99}.} \label{fig:b3}
\end{minipage}%
\end{figure}
Following this, to characterize the gas phase, nuclei which deexcite
by emitting exclusively light particles (Z $\leq$ 2) 
were selected~\cite{I2-Bac95,I6-Riv96,Pie00}. By doing this, one excludes the
possible contamination from events of the phase coexistence region.
The gas phase was characterized by comparison with a model, by studying the
deexcitation properties of vaporized quasi-projectiles
produced in $^{36}$Ar+$^{58}$Ni reactions at 95 MeV per nucleon
incident energy~\cite{I15-Bor99}. Chemical composition
and average kinetic energies of the different
particles are well reproduced by a quantum statistical model 
(grandcanonical approach) describing a real gas of fermions and bosons
in thermal and chemical equilibrium. The
evolution with excitation energy per nucleon of the composition of vaporized
quasi-projectiles is shown in Fig.~\ref{fig:b2}. Nucleon production
increases with excitation energy whereas emission of alpha particles, 
dominant at lower excitation energies, strongly decreases. 
The regular behaviour observed is a strong 
indication that an abrupt change of
phase does not occur in the considered excitation energy range.
Note that an excluded volume correction due 
to finite particle size~\cite{Gul97} (van der Waals-like behaviour) was
found decisive to obtain the observed agreement. The consequence of
the excluded volume correction is to favour neutrons, protons and
alpha particles over the loosely bound particles like deuterons,
tritons, $^3$He and $^6$He. 
In the model, the experimental range in
excitation energy per nucleon of the source (9.5 to 27.5 MeV) was covered
by varying the temperature from 10 to 25 MeV and the only free parameter,
the excluded volume, was fixed at 3~$V_{0}$ in order to reproduce the
experimental ratio between the proton and alpha yields at 18.5 MeV per
nucleon excitation energy. $V_{0}$ is the volume of nuclei at normal density.
The average kinetic energies of the different
charged particles are also rather well reproduced over the whole excitation
energy range (Fig.~\ref{fig:b3}) but the model fails to accurately follow
the dependence on the different species especially for alphas. The dashed
lines in Fig.~\ref{fig:b3} indicate the average kinetic energies, 
3T/2, expected for a free gas, which appear as a rather good 
approximation. This is due to the low density, around 0.15 -
0.2$\rho_0$, of vaporized nuclei at freeze-out. 
We are in the presence of a quantum weakly-interacting gas.

\section{First-order phase transition in hot nuclei: from
predictions to observations}\label{transition}

This section is divided into five subsections. The first one 
presents the phase transition signatures related to the specific
consequence of the non-additivity inherent to finite hot nuclei, 
i.e. an abnormal curvature of entropy.
Then, consequences of finite size on scalings
and critical behaviours will be discussed from
fragment size distributions and fluctuations. In the third subsection
results using Landau free-energy approach will be presented. In
subsection four the most delicate point i.e. the nature of the dynamics
of the phase transition will be discussed. Finally in subsection five 
the coherence of observed signatures will be summarized.

\subsection{Phase transition signatures related to entropy
convexity}\label{Sconv}
In the physical situation encountered in the present studies concerning 
a first-order phase transition in isolated finite nuclear systems, extensive
variables like energy and entropy are no longer additive due to the 
important role played by the surfaces of particles and fragments
which are produced. As a consequence the number of states in the 
mixed/disordered region grows much faster with energy than the one
associated to an ordered phase and this creates a convex intruder in
the microcanonical entropy. In this context, \textit{coexistence 
does not refer to coexistence of two
distinct phases in static physical contact within a single system 
but rather to coexistence
of two phaselike forms among an ensemble of systems} (see subsection~\ref{statfs}).

\begin{figure}[htb]
\begin{center}
\includegraphics[width=0.6\textwidth]{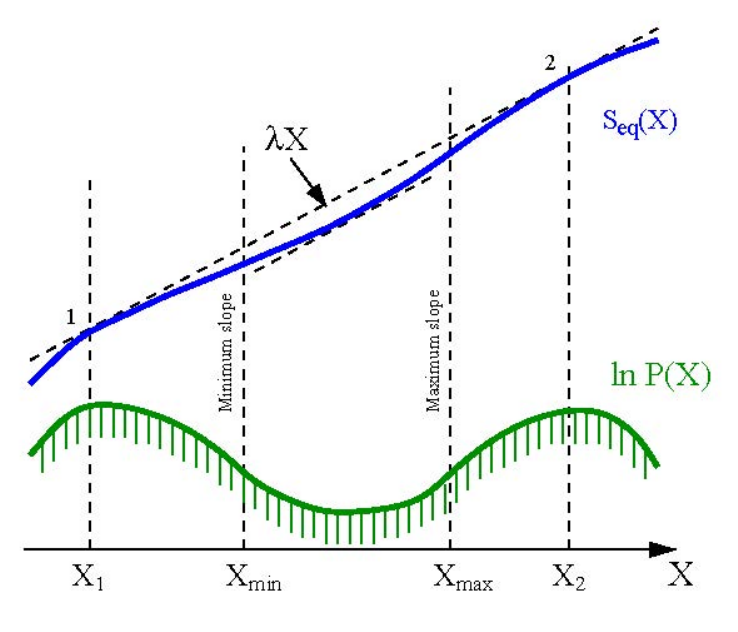}
\end{center}
\caption{
Canonical ensemble of finite systems. The bimodal equilibrium 
distribution of an extensive variable is given by $P(X)$ $\sim$
exp($S(X)$ - $\lambda X$). The figure 
shows the case when the Lagrange multiplier $\lambda$  is equal to the 
slope of the common tangent. From~\cite{Cho04}). \label{bimo}}
\end{figure}

The consequences of the curvature anomaly in the appropriate
thermodynamic potential are now discussed. The entire coexistence region
may be explored by varying the associated extensive variables.
The most direct phase transition signature corresponds to a situation
where the finite system is treated in the canonical ensemble,
the value of $X$ may fluctuate as the system
explores the phase space; the associated distribution at equilibrium
is $P(X)$ $\sim$ exp($S(X)$ - $\lambda X$) where $\lambda$ is the Lagrange
multiplier controlling the average. The distribution of $X$ 
acquires a bimodal character (see Fig.~\ref{bimo}).
That bimodality signature is the clearest indication that in finite
systems the LG-type phase transition reveals itself as a coexistence
between two types of events which manifest one of two phaselike forms. 
\begin{figure}[htb]
\begin{center}
\includegraphics[width=0.5\textwidth]{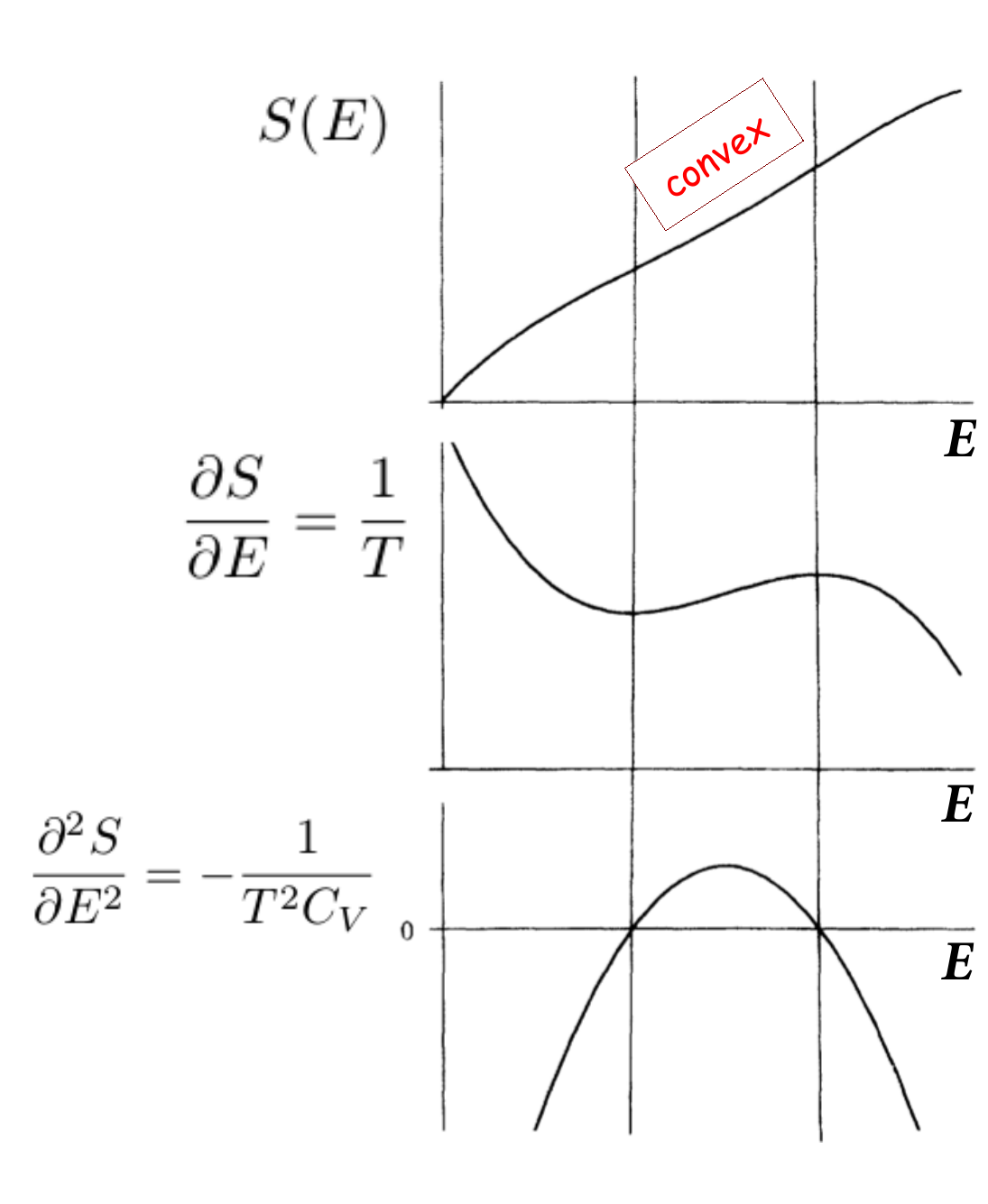}
\end{center}
\caption{ Microcanonical ensemble: evolution of entropy with
energy (top), its first derivative 1/T (middle) and second derivative
-~1/$T^2$$C_V$ (bottom), where T is the microcanonical temperature.
Adapted from~\cite{Wal94}.
\label{micro}}
\end{figure}

Another consequence of the entropy curvature anomaly appears by considering now
the microcanonical ensemble with energy as extensive variable and is
illustrated in Fig.~\ref{micro}:
the convex intruder implies a backbending in the temperature (first
derivative) and a negative branch for the heat capacity (second
derivative) between two divergences. A quantitative illustration 
using the 2-dimensional Potts model can be found in~\cite{Gro97}.
As we will see, the backbending of
caloric curves is not so
direct to observe since caloric curves are affected by the specific
dependence of the volume on the excitation energy. Indeed in the
microcanonical situation for the LG phase transition, one is forced to
introduce the volume which has to increase with energy to allow
the system to explore the partitions belonging to the disordered phase.
Conversely, negative heat capacity which always manifests by the
presence of abnormally large kinetic energy fluctuations in the
transition region should be easier to observe.

\subsubsection{Entropy convexity and bimodality}\label{Sbimo}
\begin{figure}[htb]
\begin{center}
\includegraphics[width=0.7\textwidth]{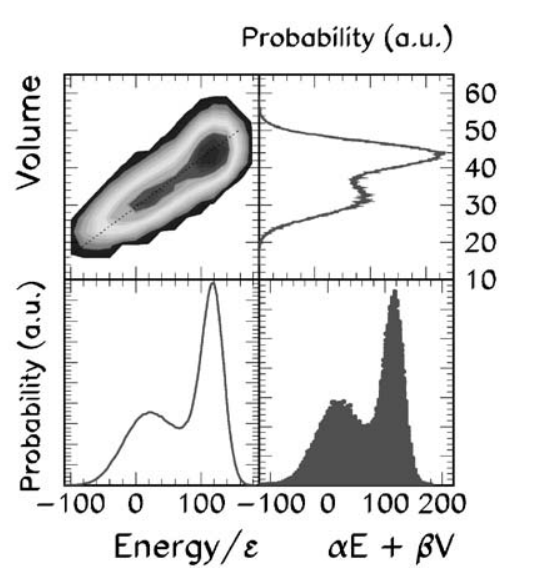}
\end{center}
\caption{ Lattice-Gas model in the isobar canonical ensemble.
Event distribution in the volume V versus energy E plane
in the first-order phase transition
region with three associated projections:
on the energy (bottom left), on the volume (top right) and on the line
which connects the two maxima and can be seen as the best order
parameter (bottom right). From~\cite{Cho01}.
\label{bimoLG}}
\end{figure}

For the macroscopic LG phase transition, the order parameter is the
difference in density between the liquid and the gas.
Since the density is related to both particle number and
volume, for finite systems one may consider an ensemble in which these two extensive
quantities are state variables. To further illustrate this point, one
can consider an ensemble of $N$-particle systems for which the volume is not
fixed but may fluctuate. In such an isobar canonical ensemble with
the Lagrange multiplier $\lambda = P/T$ associated to volume, from the 
lattice-gas model (see~\ref{PILP}) it is possible to define the 
statistical ensemble which contains the maximum of information on the
system properties. 
It was done for $N$ = 216~\cite{Cho01,Gul03}. Fig.~\ref{bimoLG} displays
volume and energy distribution and three associated projections.
The bimodal structure clearly emerges.
In the case of nuclear multifragmentation, related to the volume,
a natural order parameter can be the size of the heaviest fragment
emitted from highly excited nuclei~\cite{Cha07,Bru08} (see
also~\ref{Corlength}). 
Note that this observable provides an
order parameter for a large class of transitions or critical phenomena
involving complex clusters, from percolation to gelation, from reversible to
irreversible aggregation.
A priori that specific signature appears as robust and could be directly
observed if a large excitation energy range can be
covered in a single experiment.

A difficulty comes however from the sorting in the experimental data. 
The distribution of the energy deposit in collisions is
obviously not that of the canonical ensemble and the 
distribution of the charge, $Z_1$, of the heaviest 
fragment has no meaning in terms of statistical mechanics.
To cope with this problem, a simple procedure has been proposed in 
ref.~\cite{Gul07}. The bimodality in the canonical
two-dimensional probability distribution 
$p_\beta(E^*,Z_1)$ of a system of given size $Z_s$ at a first-order phase 
transition point reflects the convexity anomaly of the underlying 
density of states $W_{Z_s}(E^*,Z_1)$ \cite{Bin84,Lee96,I72-Bon09} 
according to: 
\begin{equation}
p_{\beta}(E^*,Z_1) = W_{Z_s}(E^*,Z_1) \exp (-\beta E^*) Z_{\beta}^{-1},
\label{distri_cano_2D}
\end{equation}
where $\beta$ = $1/T$ and $Z_{\beta}$ is the partition function. 
In an experimental sample, the energy distribution is not controlled 
through a Boltzmann factor, but it is given by a 
collision and detector dependent functional $g(E^*)$ according to:
\begin{equation}
p_{exp}(E^*) \propto \int dZ_1 W_{Z_s}(E^*,Z_1) g (E^*). 
\label{distri_exp}
\end{equation}
The convexity of the density of states can be directly inferred
from  the experimental distribution, by a simple weighting of the 
probabilities associated to each deposited energy:
\begin{equation}
p_{\omega}(E^*,Z_1)=\frac{p_{exp}(E^*,Z_1)}{p_{exp}(E^*)}=
\frac{p_{\beta}(E^*,Z_1)}{p_{\beta}(E^*)}=\frac{W_{Z_s}(E^*,Z_1)}{W_{Z_s}(E^*)}.
\label{reweighting}
\end{equation}
This procedure allows to get rid of the largely geometrical bias of
entrance channel impact parameter distribution that naturally favours the lower
part of the $E^*$ distribution. To produce a flat
$E^*$ distribution according to Eq.~(\ref{reweighting}), 
the $Z_1$ yield in each $E^*$ bin is weighted with a factor proportional
to the inverse of the bin statistics.
\begin{figure}[htb]
\begin{center}
\includegraphics[width=0.7\textwidth]{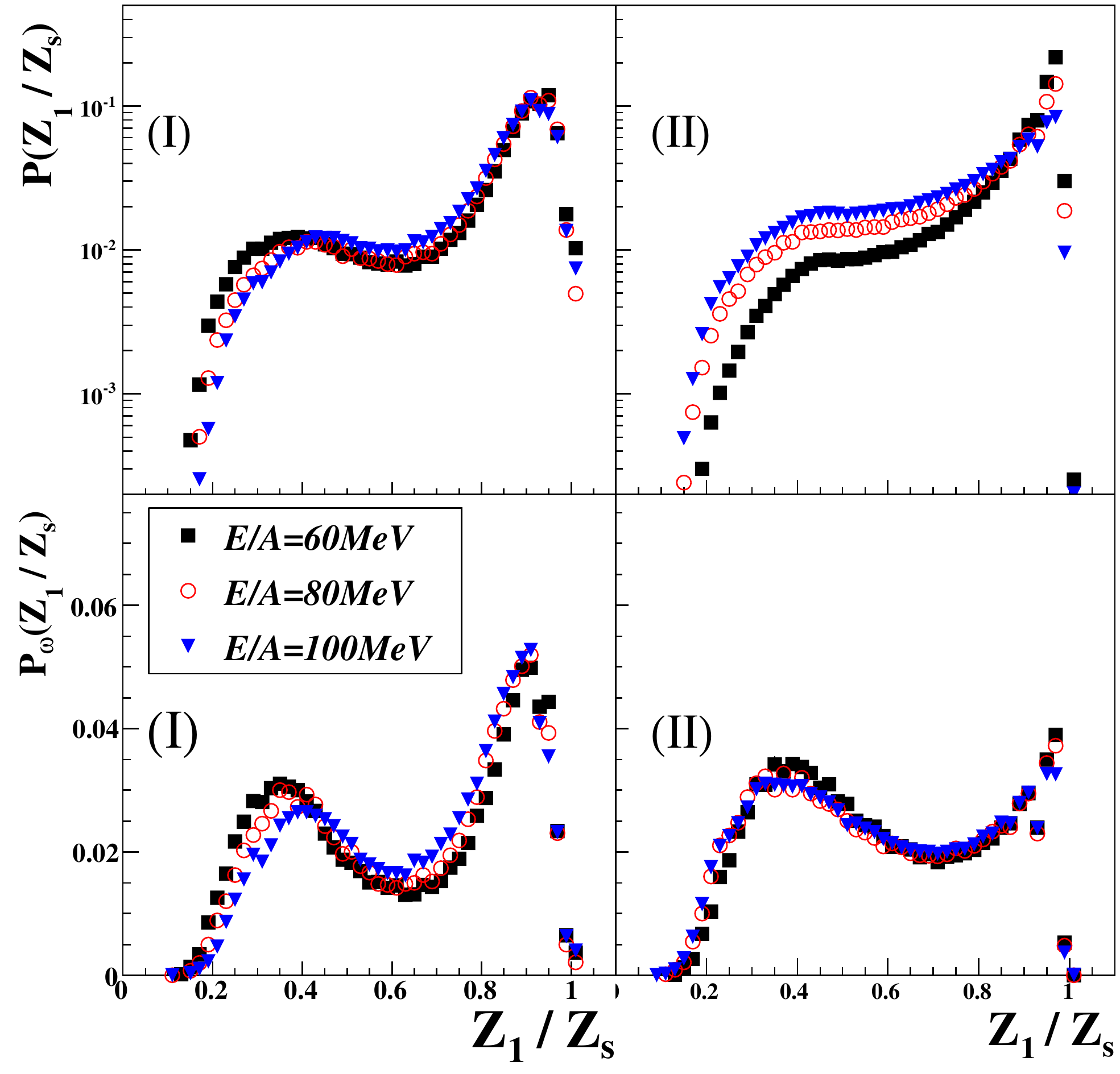}
\end{center}
\caption{ Upper part: measured distribution of the charge of the heaviest fragment
normalized to the charge of the quasi-projectile size detected in
$^{197}$Au+$^{197}$Au
collisions at three different bombarding energies. Lower part:
weighted distributions obtained considering the same statistics for
each excitation energy bin. (I) and (II) refer
to two different quasi-projectile selections.
From~\cite{I72-Bon09}. \label{bimoex}}
\end{figure}
\begin{figure}[htb]
\begin{center}
\includegraphics[width=0.7\textwidth]{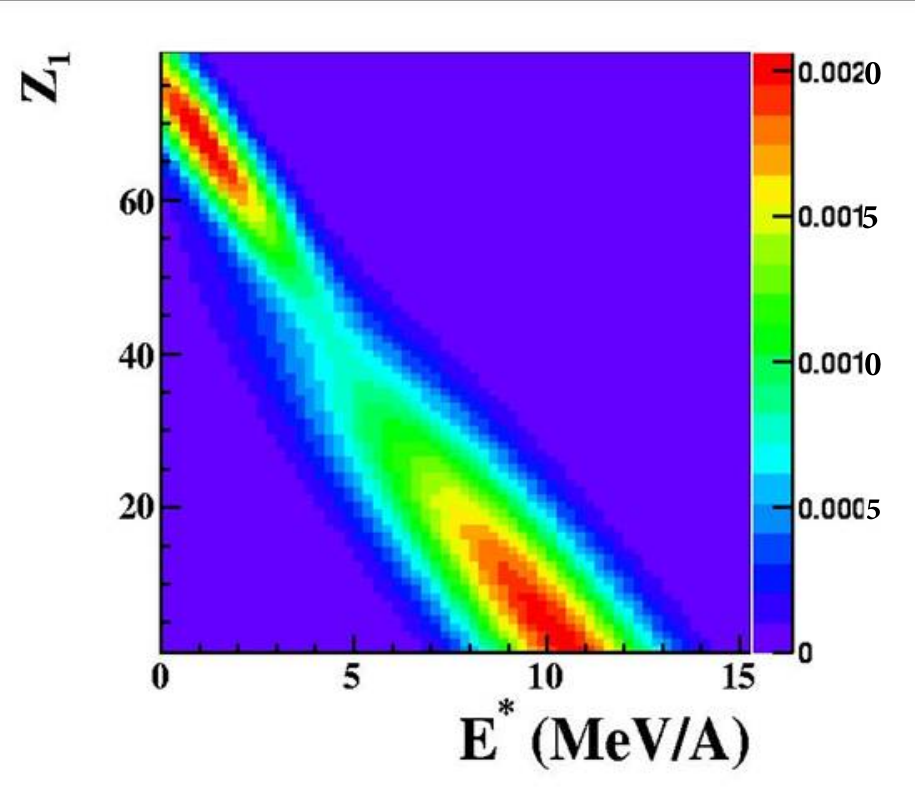}
\end{center}
\caption{Event distribution in the atomic number of the heaviest
fragment, $Z_1$, versus
excitation energy per nucleon plane. The picture is constructed using
the fit parameters extracted from the equivalent canonical
distribution for selection II. The distance between the two maxima,
``liquid'' and ``gas'' peaks, projected on the excitation energy axis
corresponds to the latent heat of the transition. From~\cite{I74-Bor10}.
\label{bimobi}}
\end{figure}

The results 
obtained with two different selection methods of quasi-projectiles
produced in 60-100 MeV per nucleon incident energies $^{197}$Au+$^{197}$Au
semi-peripheral collisions~\cite{I72-Bon09} are displayed in 
Fig~\ref{bimoex} (bottom). To take into account the small variations of the
source size, the charge of the heaviest fragment was normalized to the
source size, $Z_s$. A bimodal behaviour of the largest fragment
charge emerges with both selections whatever the bombarding energy and
for selection (II), results from different incident energies better
superimpose. 
Fig~\ref{bimoex} (top) also  displays the original measured distributions before
reweighting.
It is important to note that the two selection criteria, (I) and (II), 
produce similar but not identical distributions even after weighting, 
meaning that a residual bias on the density of states exists.
This is not surprising because any 
sorting procedure selects events according to observables correlated 
to the charge partitions. One may ask whether this inevitable bias 
prevents a sorting-independent extraction of the entropic properties 
of the system. 
To answer this question, comparison of information on the 
coexistence zone in the $(Z_1,E^*)$ plane (see Fig.~\ref{bimobi})
extracted from the two 
selection methods was done in~\cite{I72-Bon09}.
Eq.~(\ref{reweighting}) was solved 
for the canonical distribution $p_{\beta_t}(E^*,Z_1)$  at the transition 
temperature $\beta_t$ at which the two peaks of the energy distribution 
have the same height\cite{Gul03}.
This is easily obtained in a double saddle point
approximation~\cite{Gul07}: 
\begin{equation}
p_{\beta_t}(E^*,Z_1)= \sum_{i=l,g} N_i \frac{1}{\sqrt{det\Sigma_i}}
\exp\left ( -\frac{1}{2}\vec{x_i}\Sigma_i^{-1}\vec{x_i}\right).
\label{gcano_t}
\end{equation}
where $\vec{x_i}=\left (E-E_i,Z-Z_i\right )$, $\Sigma_l$ 
($\Sigma_g$) represents the variance-covariance matrix evaluated 
at the ``liquid'' $l$ (``gas'' $g$) solution, and $N_l$,$N_g$ 
are the proportions of the two phases.
The weighted experimental distribution was fitted with the function 
$p_w(E^*,Z_1)=p_{\beta_t}(E^*,Z_1)/p_{\beta_t}(E^*)$ which, using 
Eq.~(\ref{gcano_t}), is an analytic function.  
From the obtained parameter values and for selection (II), ``liquid''
and ``gas'' peaks have been respectively deduced; they are centred
at 1.05 and 10.3 MeV per nucleon excitation energy and values of the
largest fragment charge of 68.7 and 2.53. Those numbers seem
very reasonable after the indications
and comments that we were able to make on the appearance and
composition of the gas phase (see subsection~\ref{gas}).
The latent heat of the transition
of the hot nuclei studied (Z$\sim$70) was also estimated from all
available results~\cite{I72-Bon09}:

$\Delta E = E_g - E_l =  8.1(\pm 0.4)_{stat}(+1.2 -0.9)_{syst}$ MeV per
nucleon. 

Even though this study seems convincing, we must mention that
other physical scenarios have been invoked 
to interpret the observation of bimodality: Jacobi 
transition of highly deformed systems~\cite{lop05} or self-organized 
criticality induced by nucleon-nucleon collisions~\cite{Tra07,Lef08}.
Recently BUU simulations suggest that, depending on the bombarding
energy and impact parameter of the reaction, both entrance channel and
exit channel effects can be at the origin of the
bimodality~\cite{Mal18}:
fluctuations in the reaction mechanism induced by
fluctuations in the collision rate for central collisions, which agrees
with~\cite{Lef08}, as well as
thermal bimodality directly linked to the LG phase transition
for more peripheral collisions, which strongly supports
the results just presented.

\subsubsection{Entropy convexity and negative heat capacity
}\label{Cvneg}
\begin{figure}[htb]
\begin{center}
\includegraphics[width=0.75\textwidth]{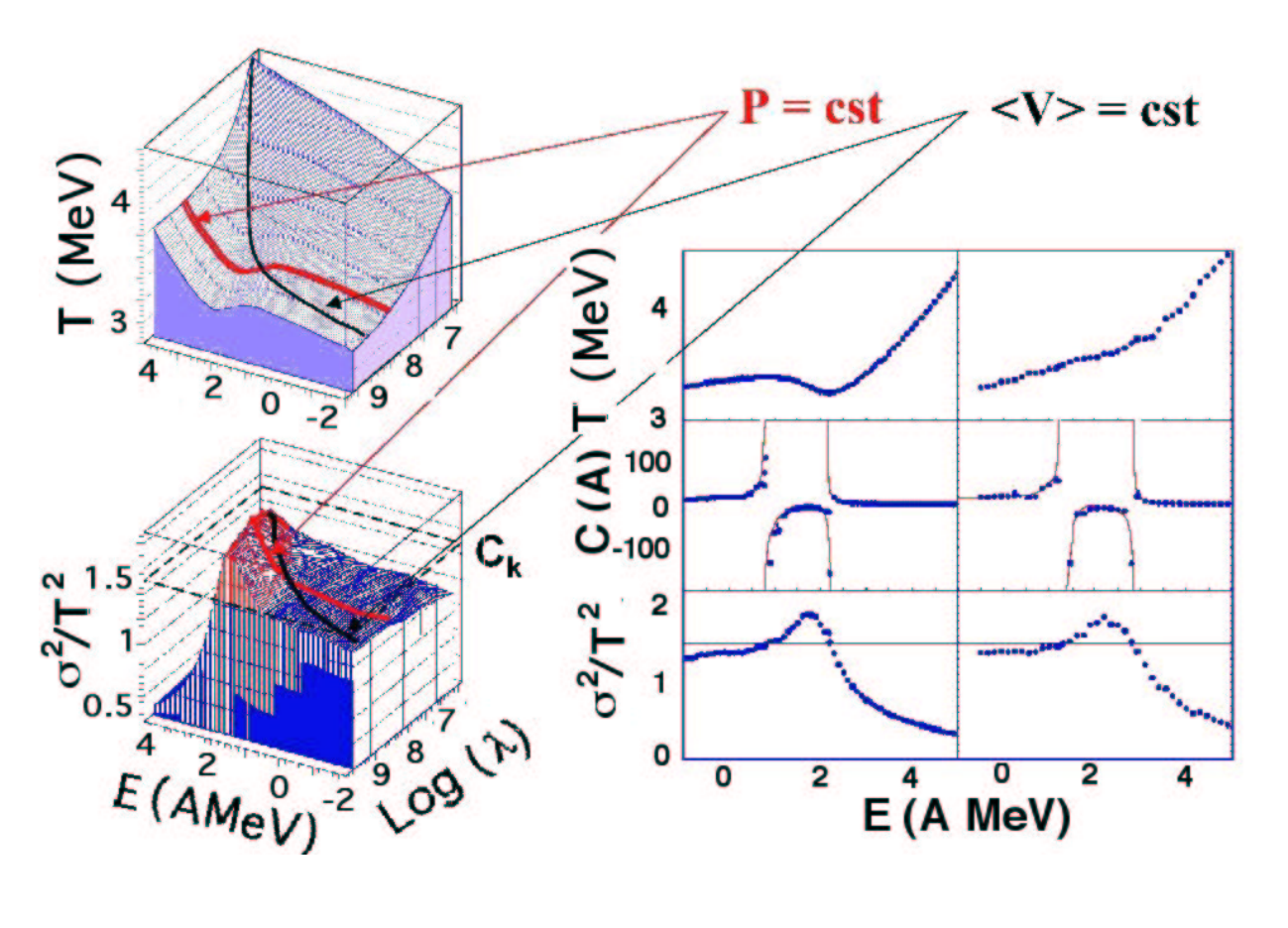}
\end{center}
\caption{Isobar lattice gas model. Left part (top/bottom): temperature/
normalized kinetic energy fluctuations as a
function of the energy per particle versus the Lagrange 
multiplier, $\lambda$ = $P/T$, associated to volume; the level
corresponding to the canonical expectation $C_k = \sigma_k^2 /T^2$ = 1.5 
is shown. Right part: thermodynamic quantities in the microcanonical
ensemble for a transformation at constant pressure and at constant 
 average volume; caloric curves are displayed in upper panels, 
normalized kinetic energy fluctuations are compared to the canonical 
expectation (lines) in lower panels and middle panels display microcanonical 
heat capacities (symbols) compared to 
the estimation (lines) from Eq.~(\ref{eq:CnegM3}).
From~\cite{Cho05}). \label{MLGM}}
\end{figure}

The observation of a plateau in nuclear caloric curves was experimentally 
proposed as a
direct signature of a first-order phase transition (see
Fig.~\ref{fig:ccGSI})~\cite{Poc95}. However,
from a theoretical point of view, a plateau-like shape cannot be an
unambiguous signature even if it is a strong indication of a physical 
change and
if its observation can help to better define the energy domain of interest
for the study of the phase coexistence.
As mentioned before, measured caloric curves can be
misleading because, depending on reactions involved
and impact parameter domain, different curve shapes can be
generated depending on the path
followed in the microcanonical equation of state landscape. 
As examples, calculated caloric curves and normalized kinetic energy
 fluctuations (microcanonical isobar lattice gas 
model - 216 particles - see~\ref{PILP})~\cite{Cho00} are displayed in Fig.~\ref{MLGM}.  
On the left hand side, landscapes of temperature and normalized kinetic
energy fluctuations are shown in the plane energy per particle -  Lagrange 
multiplier associated to volume $\lambda$ = $P/T$.
If for temperature very different curve shapes are obtained
depending on the path,
normalized kinetic energy fluctuations related to microcanonical heat capacity
are abnormally large in the coexistence region whatever the path.
To better appreciate the situation, the right hand side of the figure shows
the behaviours of caloric curves
(upper panel) at constant pressure or at constant average
volume in the subcritical region. At constant pressure a backbending
is clearly seen whereas at constant average volume a smooth behaviour
is observed with a slope change when entering the gas phase.
In experiments one does not explore a caloric curve at constant
pressure nor at constant volume, the different measured systems 
follow a path in the excitation energy - freeze-out volume plane and event by event
freeze-out properties must be deduced from performing simulations to
possibly derive relevant information. Constrained caloric curves will be
discussed in~\ref{CalC}.

Conversely the anomalously large fluctuation signal of kinetic
energy (lower panel) is always
seen, independently of the path, for systems undergoing a first-order
phase transition. From this theoretical observation 
a method was proposed for measuring microcanonical heat capacity using partial 
energy fluctuations~\cite{Cho99,Gul00,Cho00}.
The prescription is based on the fact that
for a given total energy, the
average partial energy stored in a part of the system is a good 
microcanonical thermometer, while the associated fluctuations can be
used to construct the heat capacity.
From experiments the most simple decomposition of the total energy 
is in a kinetic part, $E_{k}$, and a potential part, $E_{pot}$, 
(Coulomb energy + total mass excess).
However these quantities have to be determined at freeze-out  and
consequently it is necessary to trace back this
configuration on an event by event basis.
As discussed in  
section~\ref{how} the fragment properties entirely rely on the representation 
of the system at the freeze-out stage as non interacting fragments.
The true configuration needs  the knowledge of the freeze-out
volume and of all the particles evaporated from primary hot
fragments including the (undetected) neutrons. Consequently
some working hypotheses are used, possibly constrained by
specific experimental results (see for example~\cite{MDA02,Gul03}).
Then, the experimental correlation between the kinetic energy per nucleon
$E_{k}$/$A$ and
the total thermal excitation energy per nucleon $E^{*}$/$A$ of the considered
system can be obtained event by event 
as well as the variance of the kinetic energy
$\sigma_{k}^{2}$. Note that $E_{k}$ is calculated by subtracting 
the potential part $E_{pot}$ from the total energy $E^{*}$ and consequently
kinetic energy fluctuations at freeze-out reflect the configurational 
energy fluctuations.
In the microcanonical ensemble with total energy $E^{*}$ the total
degeneracy factor is simply given by the folding product of the
individual degeneracy factors $W_{k} = exp(S_{k}(E_{k}))$ and 
$W_{pot} = exp(S_{pot}(E_{pot}))$. One can then define for the total
system as well as for the two subsystems the microcanonical temperatures
and the associated heat capacities $C_k$ and $C_{pot}$.
If we consider now the kinetic energy distribution when the total
energy is $E^{*}$ we get
\begin{equation}
\noindent P_{E^*}(E_{k}) = exp(S_{k}(E_{k}) + S_{pot}(E^* -E_{k})).
\label{eq:Kindistri}
\end{equation}
Then the most probable kinetic energy $\overline{E_{k}}$ is defined by
the equality of the partial microcanonical temperatures
$T_{k}(\overline{E_{k}})$ = $T_{pot}(E^{*} - \overline{E_{k}})$ and  
$\overline{E_{k}}$ can be used as the microcanonical thermometer.
An estimator of the
microcanonical temperature of the system can be obtained by inverting the
kinetic equation of state:
$$ < E_{k} > = \langle \sum_{i=1}^{M} a_i \rangle T^2 +
\langle \frac{3}{2} (M-1) \rangle T $$
The brackets $\langle\rangle$
indicate the average on events with the same $E^{*}$,
$a_i$ is the level density parameter and M the multiplicity at
freeze-out. It may be noted that in this expression the same
temperature is associated with both internal excitation and thermal
motion of fragments. 
An estimate of the total microcanonical heat capacity 
is extracted using three equations.
\begin{equation}
\noindent C_k = \frac{\delta <E_k / A >}{\delta T},
\label{eq:CnegM1}
\end{equation}
is obtained by taking the derivative of $<E_k / A >$ with respect to
$T$ and is equal to 1.5 in the canonical ensemble.
Using a Gaussian approximation for $P_{E^*}(E_{k})$ the kinetic
energy variance can be calculated as\\
\begin{equation}
A\sigma_{k}^{2} \simeq T^2\frac{C_kC_{pot}}{C_k+C_{pot}};
\label{eq:CnegM2}
\end{equation}
Eq.~(\ref{eq:CnegM2}) can be inverted to extract, from the observed
fluctuations, an estimate of the microcanonical heat capacity:
\begin{equation}
\noindent (\frac{C}{A})_{micro} \simeq C_k+C_{pot} \simeq \frac{C_k^2}
 {C_k - \frac{A \sigma_k^2}{T^2}}.
\label{eq:CnegM3}
\end{equation} 
From Eq.~(\ref{eq:CnegM3}) we can see that the specific microcanonical
heat capacity $(C/A)_{micro}$ becomes negative if the normalized
kinetic energy fluctuations $A \sigma_k^{2}/T^2$ overcome $C_k$.
Fig.~\ref{MLGM} (middle panels of right hand side) illustrates the results of 
such a procedure in the framework of the microcanonical lattice gas model.
It is interesting to note that the constraint of energy conservation
leads in the phase transition region to larger fluctuations than in
the canonical case where the total energy is free to fluctuate. This
is because the kinetic energy part is forced to share the total
available energy with the potential part: when the potential part
presents a negative heat capacity the jump from ``liquid'' to ``gas''
induces strong fluctuations in the energy partitioning.
\begin{figure}[htb]
\begin{center}
\includegraphics[scale=0.44]{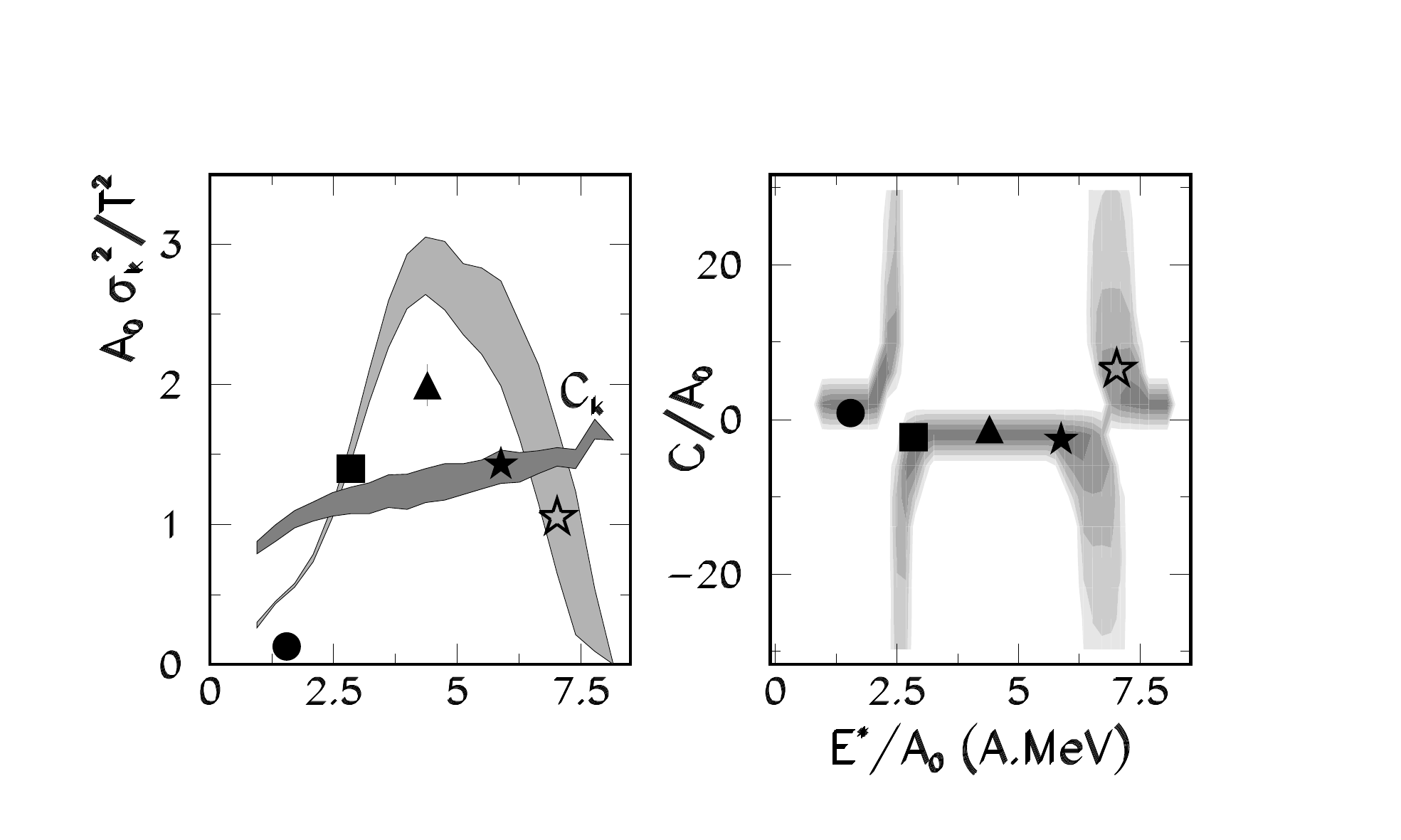}
\includegraphics[scale=0.22]{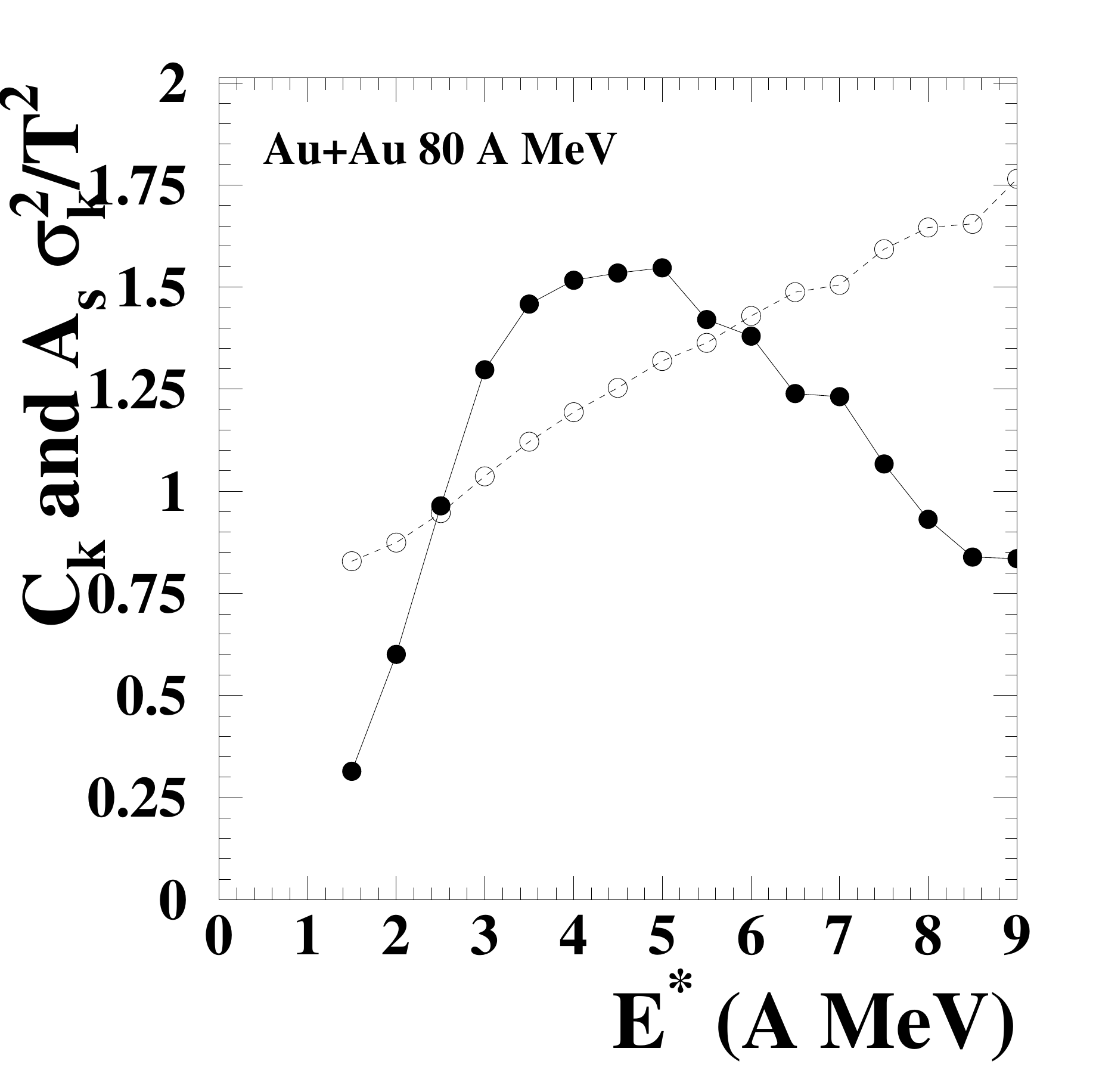}
\end{center}
\caption{Left panel: normalized kinetic energy fluctuations and estimated
$C_k$ values for quasi-projectile events produced 
in Au+Au collisions at 35~MeV per nucleon (grey zones) and for fused systems 
produced in central Au+C (black dots), Au+Cu (squares, triangles) 
and Au+Au reactions before (open stars) and after subtraction of 1~MeV
per nucleon 
radial flow (black  stars). Central panel: corresponding microcanonical heat 
capacities per nucleon. Grey zones indicate confidence regions. From~\cite{MDA04}.
Right panel: normalized kinetic energy fluctuations (filled circles)
and estimated $C_k$ values (open circles) for quasi-projectile events produced 
in Au+Au collisions at 80~MeV per nucleon. From~\cite{I63-NLN07}.
\label{fig:MULTINDRA}}
\end{figure}

That procedure was applied by the MULTICS and INDRA collaborations 
on quasi-projectiles and on quasifusion nuclei produced
respectively in semi-peripheral and central
collisions in the incident energy range 30-80 MeV per
nucleon~\cite{MDA00,NLNBorm00,I46-Bor02,MDA04,I63-NLN07}.
Fig.~\ref{fig:MULTINDRA} summarizes the results obtained by the MULTICS
collaboration. On the left hand side it is seen that normalized kinetic 
energy fluctuations overcome $C_k$; the middle part of the figure
illustrates the microcanonical 
negative heat capacities observed. On the right hand side of the
figure, as an example, one result of the INDRA collaboration is also presented.
We note that, with different selections of quasi-projectiles and
different bombarding energies for the reaction Au + Au, measurements
are compatible. 
These results provide a direct evidence of a first-order phase
transition. In relation with the reconstruction at freeze-out, they have to be
seen as semi-quantitative.

Exact microcanonical formulae, assuming  that a classical treatment 
of the motion of products emitted at freeze-out is appropriate, are also
proposed in~\cite{Radut02} to calculate heat capacity or alternatively
the second-order derivative of the system entropy versus energy.
They only depend on the total kinetic
energy and on the number of emitted products which have to be estimated event by
event at freeze-out. However up to now this method was not used to derive 
information from data.

Before concluding this part one can indicate that questions
are still under debate and concern the topology of the system at 
freeze-out. If the system is
still relatively dense at freeze-out, which seems improbable if we refer
to simulations from~\cite{I66-Pia08}, which reproduce remarkably well
experimental data and indicate freeze-out volumes in the range 3-6 times
the volume at normal density,
the fragment properties may be very different from the
ones asymptotically measured and the question arises whether the energetic
information measured on ground state properties can be taken for the freeze
out stage~\cite{Sat03}. Classical molecular dynamics calculations have shown
that the ground state Q-value is a very bad approximation of the interaction
energy of fragments in dense systems. This is due to the deformation of
fragments and to the interaction energy when fragment surfaces touch each
other. As a consequence, comparable kinetic energy fluctuations are obtained
in the subcritical and supercritical regions of the Lennard-Jones phase
diagram~\cite{Cam05}. On the other hand calculations with a similar model,
the Lattice Gas model, show that even in the supercritical region the correct
fluctuation behaviour can be obtained if both the total energy and the
interaction energy are consistently estimated with the same approximate
algorithm as it is done in the experimental data analysis~\cite{Gulm05}.
Concerning now the order of the transition, in certain theoretical
calculations it appears that a negative heat capacity is not always
incompatible with a continuous (second order) transition due to
finite-size effects, either in a generic case~\cite{Be06} or for
finite nuclei~\cite{De06}. 

\subsubsection{Constrained caloric curves}
\label{CalC}
In Ref.~\cite{I58-Pia05,I66-Pia08} freeze-out properties of
multifragmenting quasifusion nuclei produced in central $^{129}$Xe +
$^{nat}$Sn collisions at different beam energies (32, 39, 45 and 50 MeV
per nucleon) have been estimated by means of a simulation based on
all the available experimental information obtained with a
very high degree of completeness for events, which is crucial for a good estimate
of Coulomb energy. To check the overall physical coherence of this
approach, a detailed comparison with a microcanonical statistical model
(MMM - see section~\ref{multifragmod}) was also made.
Event by event, various quantities needed to build constrained caloric
curves were deduced, namely
the thermal excitation energy of quasifusion hot nuclei, $E^*$,
the freeze-out volume $V$
and
the total thermal kinetic energy at freeze-out $K$.
With regard to the pressure at freeze-out, it can be
derived within the microcanonical ensemble.
Taking into account that $S=\ln Z=\ln \sum_C W_C$ 
and that $\partial W_C/\partial V=\left(M_C/V\right) W_C$, 
where $W_C$ is the statistical weight of a configuration
, defined by the mass, charge and internal excitation energy
of each of the constituting $M_C$ products at freeze-out,
it comes out that
\begin{eqnarray}
\nonumber
P/T=
\left(\frac{\partial S}{\partial V}\right)&=&\frac1{\sum_C W_C} \sum_C 
\frac{\partial W_C}{\partial V}\\
&=&\frac1V \frac{\sum_C M_C W_C}{\sum_C  W_C}=\frac{\langle M_C\rangle}{V}.
\label{eq:Pmicro0}
\end{eqnarray}
The microcanonical temperature is also easily deduced from its statistical
definition~\cite{Radut02}:
\begin{eqnarray}
\nonumber
T=\left(\frac{\partial S}{\partial E}\right)^{-1}&=&\left(\frac1{\sum_C W_C} \sum_C 
W_C(3/2M_C-5/2)/K\right)^{-1}\\
&=&\langle(3/2M_C-5/2)/K\rangle^{-1}.
\label{eq:Pmicro1}
\end{eqnarray}
As $M_C$, the total multiplicity at freeze-out, is large,
\begin{equation}
\noindent
T\thickapprox
\frac{2}{3}\langle\frac{K}{M_C}\rangle
\label{eq:Pmicro2}
\end{equation}
and the pressure $P$ can
be approximated by
\begin{equation}
\noindent
P=T\frac{\langle M_C\rangle}{V}\thickapprox\frac{2}{3}\frac{\langle K\rangle}{V}.
\label{eq:Pmicro3}
\end{equation}
Knowing $\langle K \rangle$ and $V$ from simulations, pressure $P$ can be calculated for
events sorted in each $E^*$ bin. The temperature $T_{kin}$ that we obtain from
the simulations is identical to the microcanonical temperature of
Eq.~(\ref{eq:Pmicro2}). One can also note that the free Fermi gas pressure
exactly satisfies Eq.~(\ref{eq:Pmicro3}). 

In simulations, Maxwell-Boltzmann statistics
is used for particle velocity distributions at freeze-out
and consequently the deduced temperatures, $T$ = $T_{kin}$, are classical.
To build constrained caloric curves, authors of Ref.~\cite{I79-Bor13}
have used a thermometer based on momentum fluctuations of emitted
particles~\cite{Zhe11,Zhe12} for which, for the first time, the quantum
nature of particles is taken into account (see~\ref{Temp}). 
Momentum fluctuations of protons were used
and $\rho$ was estimated to
$\rho\sim0.4\rho_0$ from dynamical simulations,
which corresponds to $\epsilon_f\sim$ 20 MeV.
Systematic errors on $E^*$ and $T$ are discussed in~\cite{I79-Bor13}.
\begin{figure}[htb]
\begin{center}
\includegraphics[width=0.45\textwidth]{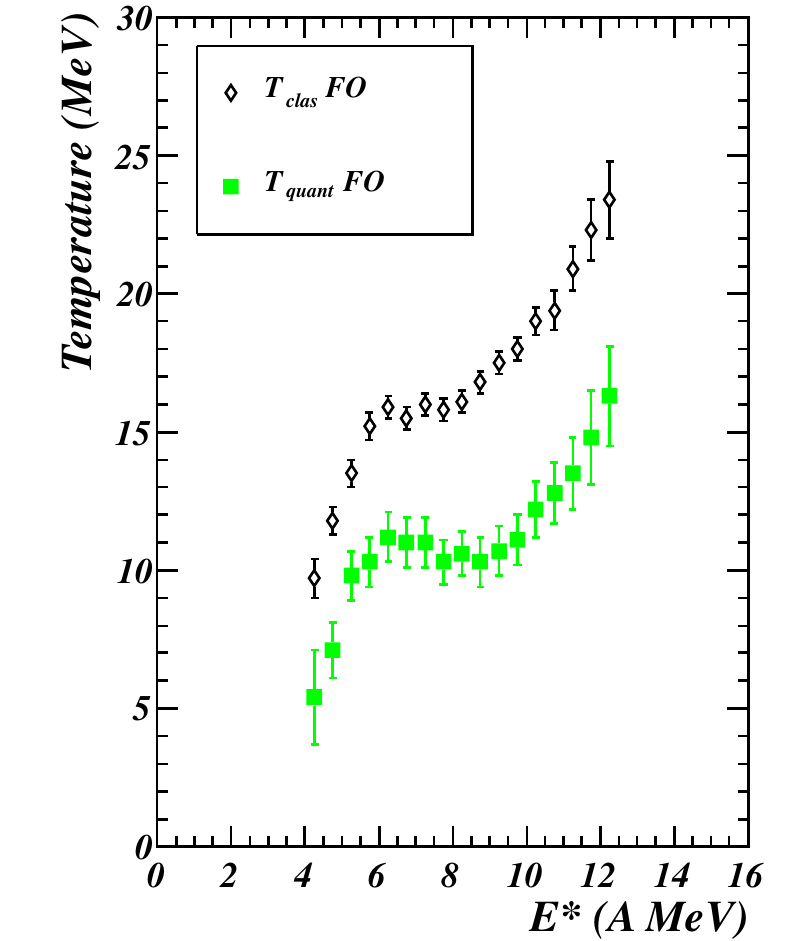}
\end{center}
\caption{ Caloric curves: classical
temperature (open diamonds)/
quantum corrected temperature (full squares) 
from proton momentum
fluctuations versus thermal excitation energy. Protons
are thermally emitted at freeze-out. Error bars include statistical and
systematic errors. From~\cite{I79-Bor13}).}
\label{fig:CC}
\end{figure}
Considering $E^*$ bins of 0.5 MeV per nucleon, to see the effect of
the quantum nature of particles, Fig.~\ref{fig:CC} shows the 
caloric curve with temperatures from quantum fluctuations (full squares)
compared to the one with classical temperatures derived from
the simulation (open diamonds). They both
exhibit a plateau with a significant difference for temperatures.
For the quantum corrected caloric curve the plateau is observed around a temperature
of 10-11 MeV on the $E^*$ range 5-10 AMeV.
Then constrained caloric curves, which correspond to correlated values of 
$E^*$ and quantum corrected temperatures have been determined.
$E^*$ values which are derived from experimental calorimetry have been corrected
\textit{a posteriori} using quantum temperatures instead of classical ones.
Pressure values were also corrected
using quantum temperatures in Eq.~(\ref{eq:Pmicro3}).
In Fig.~\ref{fig:CCon} (left hand side) constructed caloric curves for
two average freeze-out volumes are displayed.
As theoretically expected a monotonic behaviour of caloric curves is observed.
Fig.~\ref{fig:CCon} (right hand side) shows the caloric curves when pressure has
been constrained within two domains: (1.3-4.5) and (4.5-7.9)x $10^{-2}$ MeV fm$^{-3}$.
Again as theoretically expected, backbending is seen, especially for
the lower pressure range. 
For higher pressures the backbending of the caloric curve is reduced
and one can estimate its vanishing, indicating the critical temperature,
around 13 MeV for the selected finite systems.
So, constrained caloric curves confirm the previous signatures as far as a
first-order phase transition for hot nuclei is concerned. Note that
the caloric curve (quantum corrected temperature) of Fig.~\ref{fig:CC}
resembles the caloric curves constrained in pressure of
Fig.~\ref{fig:CCon}. This resemblance is not general and, in particular when
rather light nuclei are involved, the shape of caloric curves is
similar to the one of caloric curves constrained in average volume. 
\begin{figure}
\begin{center}
\includegraphics*[width=0.4\textwidth]
{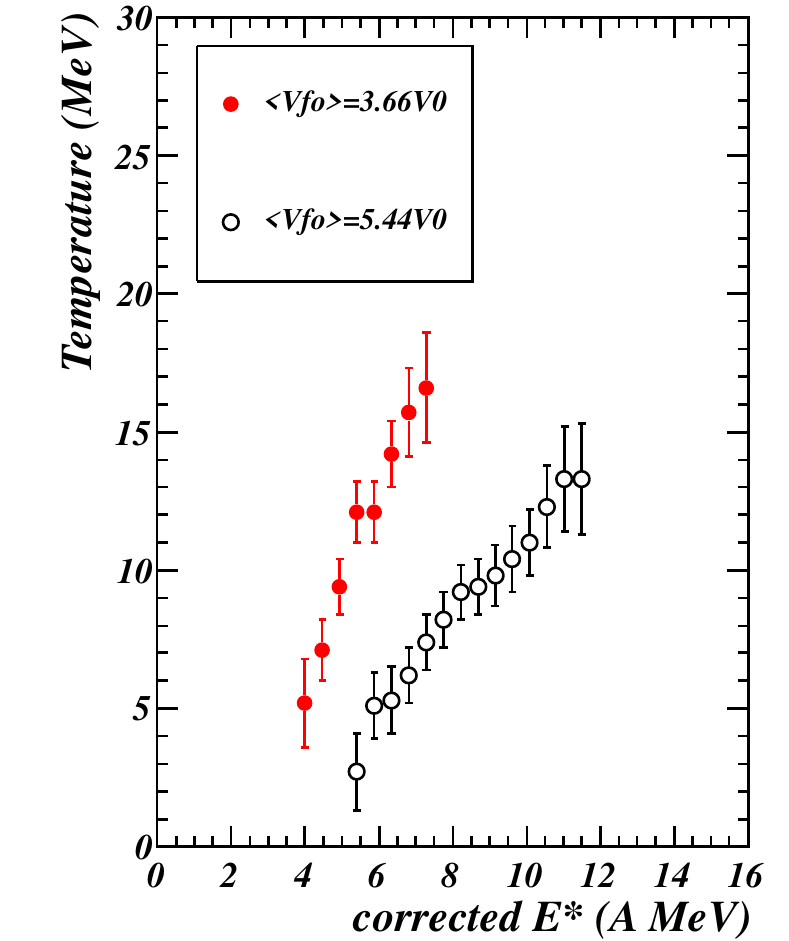}
\includegraphics*[width=0.4\textwidth]
{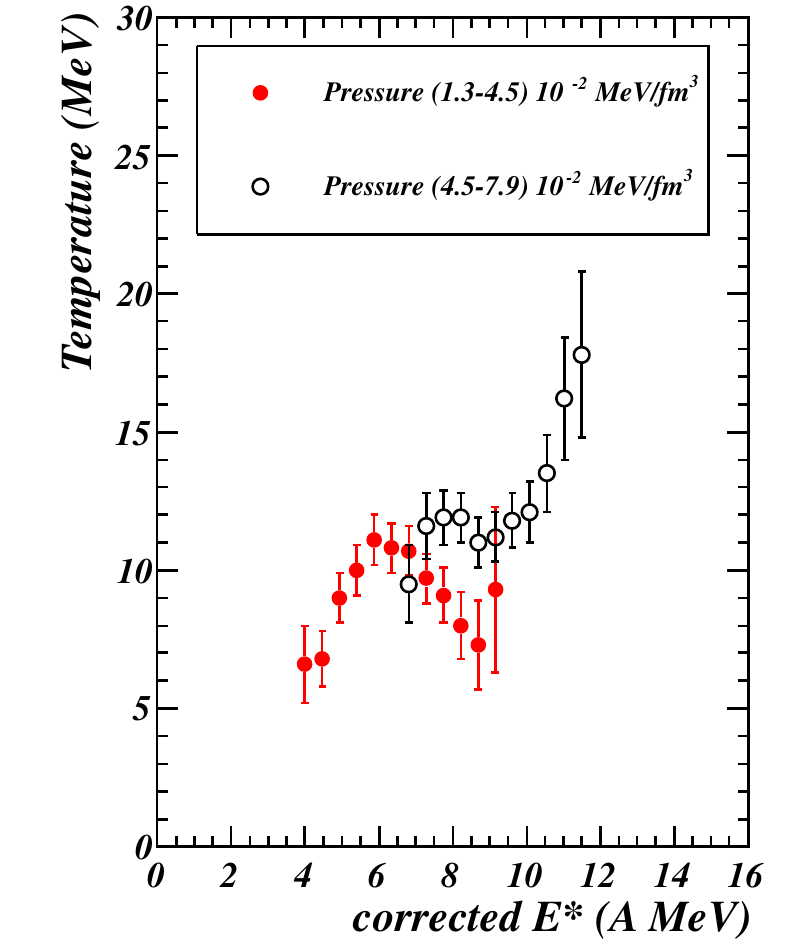}
\end{center}
\caption{(Caloric curves (quantum corrected temperature versus
corrected thermal
excitation energy) constrained at average volumes (left)
and for selected ranges of pressure (right).
Error bars include statistical and
systematic errors. From~\cite{I79-Bor13}).}
\label{fig:CCon}
\end{figure}

Finally one can say a few words about the effect of $N/Z$ content on caloric curves.
On the theoretical side the calculated temperature variation with
isospin is small~\cite{Ogu11,Bes89,Hoe07}. Experimentally,
semi-peripheral collisions for 600 MeV per nucleon
($^{124}$Sn, $^{124}$La, $^{107}$Sn) + $^{nat}$Sn~\cite{Sfi09} and
35 MeV per nucleon $^{78}$Kr + $^{58}$Ni and $^{86}$Kr + $^{64}$Ni~\cite{Wue10}
confirm a small isospin effect, with slightly higher temperatures for the 
neutron-richer systems. Conversely, in~\cite{McI13}, and for light 
quasi-projectiles of known A and Z formed in 35 MeV per nucleon
$^{70}$Zn + $^{70}$Zn, $^{64}$Zn + $^{64}$Zn and $^{64}$Ni + $^{64}$Ni
reactions, measurable effects, 
with lower temperatures for neutron-richer nuclei were observed. Note that,
unlike the ensemble of caloric curves presented in~\cite{Nato02}, none of those
derived in~\cite{Sfi09,Wue10,McI13} exhibits a plateau. In~\cite{McI13} 
the temperature linearly increases with energy between 2 and 8 MeV per
nucleon, reaching 12 MeV at an excitation energy of 8 MeV per nucleon.
Those measurements show again the necessity to constrain caloric
curves to derive relevant information.

\subsection{Criticality and correlation length}\label{Corlength}
In general, thermodynamic features of systems near a critical point
depend only on a small number of variables, mainly determined by the
dimensionality and symmetry properties of the system, but are insensitive
to details of the underlying microscopic 
properties~\cite{Ric01}.
They thus exhibit \emph{universality} and the way in which certain
physical properties (such as the difference between liquid and gas
density for fluids, or the susceptibility for magnetic systems) approach
their critical values are determined by\emph{ critical exponents} belonging
to different \emph{universality classes}. \emph{Scaling laws} are
frequently associated with critical systems, as the correlation length
diverges and fluctuations occur at all length scales, manifesting
a fractal, self-similar organization of matter. 

In the early 1980s an experiment at the Fermi National Accelerator
Laboratory led by a group from Purdue University was the first to
provide high statistics, high resolution measurements of the inclusive
mass distribution of fragments produced in high energy proton-nucleus
collisions \citep{Fin82,Hir84}.
The observation of a power-law,
\begin{equation}
\mathrm{yield}(A_{f})\propto A_{f}^{-\tau}
\label{eq:purdue-powerlaw}
\end{equation}
with exponent $\tau$ independent of the target mass strongly suggested
the interpretation of fragment production as a critical phenomenon,
by analogy with the well-known behaviour of real gases near the critical
point, which exhibit cluster distributions with $\tau$ between 2.1
and 2.3, as described by the Fisher droplet model~\cite{Fis67}
(see~\ref{Fishmod}).
Since then many works have been devoted to the extraction of critical
exponents from multifragmenting hot nuclei data, and results from different
experiments are consistent with the LG phase transition universality
class. Table \ref{tab:Critical-exponents-from} compiles some results
for the following critical exponents: $\beta$, which controls how
the difference between liquid and gas density goes to zero at the
critical point; $\gamma$, which describes the divergence of the isothermal
compressibility; $\tau$, the exponent of the mass-yield power law
at the critical point; and $\sigma$, an exponent used by Fisher to
relate cluster mass and surface energy (related to the dimensionality).
The scaling of fragment yields based on the droplet 
model of \cite{Fis67}
was even used to reconstruct the pseudo-coexistence curves at sub-critical
densities for the phase transition \citep{Ell02,Ell03}.

However, the question of criticality or the order of the phase transition
is far from unambiguous when dealing with finite systems. In the fragmentation
of small systems such as nuclei, critical behaviour has been shown
to be compatible with a first-order phase transition, due to finite
size effects \cite{Gul99}. There is
no contradiction between the scenario of nuclear fragmentation inside
the coexistence or the spinodal region associated with a first-order
phase transition and the observation of pseudo-critical signals in
fragment observables. The physical origin of the scaling behaviour
at subcritical densities lies in the finite size of the system: for
such small systems, correlations need only reach the same size as
the system itself in order to mimic critical fluctuations with infinite
correlation length.
\begin{table}
\caption{Theoretical values of critical exponents for the liquid-gas and percolation
phase transition universality classes along with experimental values.
For the experiments the critical excitation energy, $E_{C}*$, extracted
from fitting fragment yields with the Fisher droplet model \cite{Fis67},
is also given. QP and QF refer respectively to quasi-projectile and
quasifusion hot nuclei.}
\begin{centering}
\begin{tabular}{cccccc}
\hline
 & $E_{C}*$ {[}$A$MeV{]} & $\beta$ & $\gamma$ & $\tau$ & $\sigma$\tabularnewline
\hline
\hline
Liquid-gas & - & 0.33 & 1.23 & 2.21 & 0.64\tabularnewline
\hline
3D percolation & - & 0.41 & 1.8 & 2.18 & 0.45\tabularnewline
\hline
Au (1$A$GeV)+C \citep{Gil94,Ell96,Ell03} & 
4.6 $\pm$ 0.2 & 0.29 $\pm$ 0.02 & 1.4 $\pm$ 0.1 & 2.14 $\pm$ 0.06 & 
0.68 $\pm$ 0.05\tabularnewline
\hline
$\pi$(8 GeV/c) +Au \citep{Ell02} & 
3.8 $\pm$ 0.3 & 0.33 $\pm$ 0.25 & - & 2.18 $\pm$ 0.14 & 
0.54 $\pm$ 0.01\tabularnewline
\hline
Au (35 $A$MeV) + Au QP \citep{BON00} & 
4.5 $\pm$ 0.2 & 0.38 $\pm$ 0.02 & 1.4 $\pm$ 0.3 & 
2.2 $\pm$ 0.2 & \tabularnewline
\hline
Xe (32$A$MeV)+Sn QF \citep{I63-NLN07} & 
4.50 $\pm$ 0.03 & - & - & 2.09 $\pm$ 0.01 & 0.66 $\pm$ 0.01\tabularnewline
\hline
Au (80$A$MeV)+Au QP\citep{I63-NLN07} &
4.20 $\pm$ 0.03 & - & - & 2.56 $\pm$ 0.02 & 0.66 $\pm$ 0.01\tabularnewline
\end{tabular}
\par\end{centering} 
\label{tab:Critical-exponents-from}
\end{table}
\begin{figure}
\begin{centering}
\includegraphics[clip,width=0.9\textwidth]{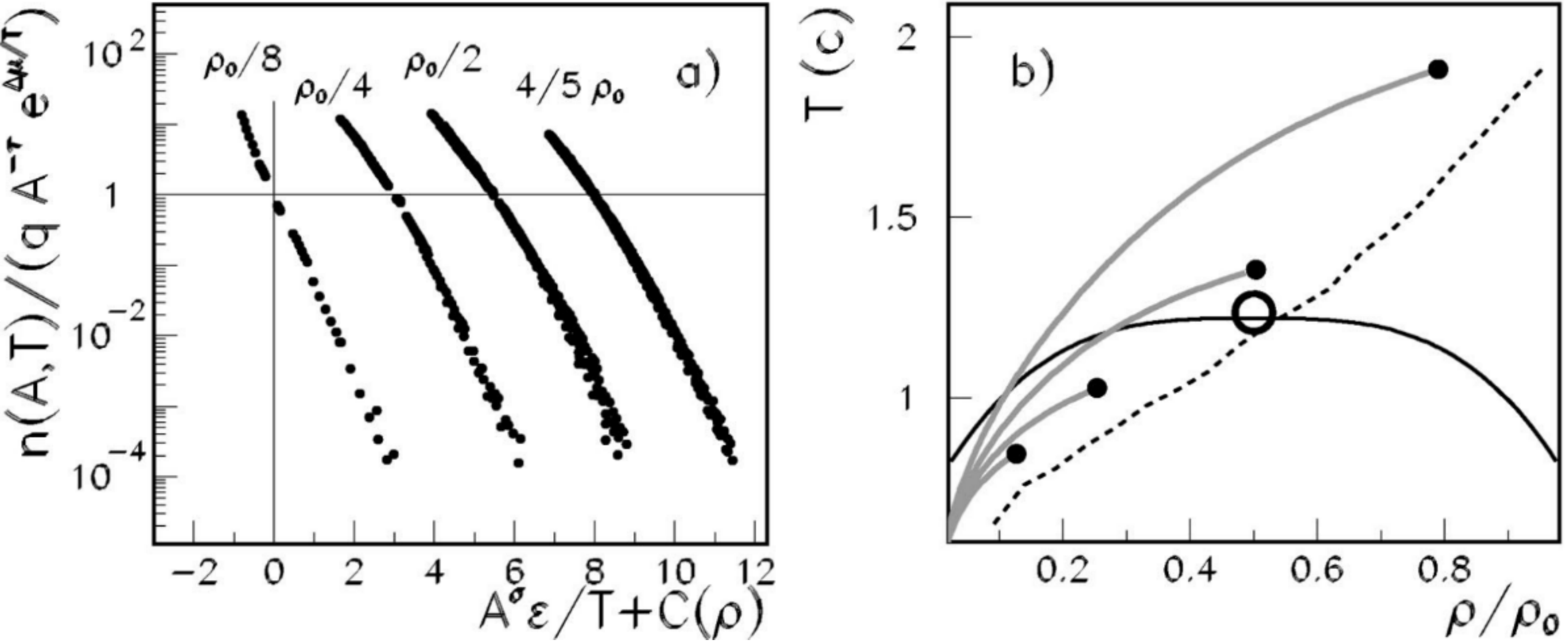}
\par\end{centering}
\caption{Application of the Fisher droplet model \cite{Fis67}
to a canonical version of the Lattice Gas model (see \cite{Gul02}
for details). (a) Cluster yields for different model densities scaled
according to the Fisher ansatz; for each density the points are offset
by a constant horizontal shift, $C(\rho)$, for clarity. (b) Thermodynamical
coexistence line (full line) and line of maximum cluster size fluctuations
leading to pseudocritical behaviour (dashed lines) from \citep{Gul99}.
Grey lines: coexistence line reconstructed from fragment partitions.
Adapted from \cite{Gul02}. 
\label{fig:fisher-lgm}}
\end{figure}
Some consequences of this are illustrated in Fig.~\ref{fig:fisher-lgm},
taken from \citep{Gul02}. The canonical
lattice gas model was used to calculate cluster
mass yields $n(A,T)$ for a wide range of temperatures and densities
both above and below the critical point. Then fits were performed
to the cluster yields using Eq.~(\ref{fisher}) of section~\ref{Fishmod}.
It can be seen in Fig.~\ref{fig:fisher-lgm}(a) that the Fisher
scaling ansatz works astonishingly well both below ($\rho<\rho_{0}/2$)
and above ($\rho>\rho_{0}/2$) the critical density of the model,
whereas the model is only strictly applicable to a subcritical system
\emph{i.e.} droplets surrounded by vapour. The extracted values for
the critical exponents are, however, compatible (within finite size
effects) with the universality class of the model. On the other hand,
Fig.~\ref{fig:fisher-lgm}(b) shows the sub-critical coexistence
curves (grey lines) deduced from the cluster yields using the same
methods as applied to data in \citep{Ell02,Ell03}.
For each density a different curve is obtained, terminating at (black
points) different pseudo-critical temperatures which increase with
the density of the system. Even for the calculations performed at
the critical density of $\rho=\rho_{0}/2$ the deduced coexistence
curve does not correspond to the thermodynamical one (black line)
and overestimates the critical temperature. As shown in \citep{Gul02}
the correct thermodynamics of the model are only retrieved for low
densities and high temperatures for which the clusters behave as an
ideal gas, as supposed by Fisher. When applied to data where no \emph{a
priori} knowledge of the density of the multifragmenting systems is
available, the interpretation of results obtained with Fisher's model
is therefore not so straightforward. 
\begin{figure}
\begin{centering}
\includegraphics[bb=0bp 0bp 643bp 616bp,clip,width=0.5\textwidth]
{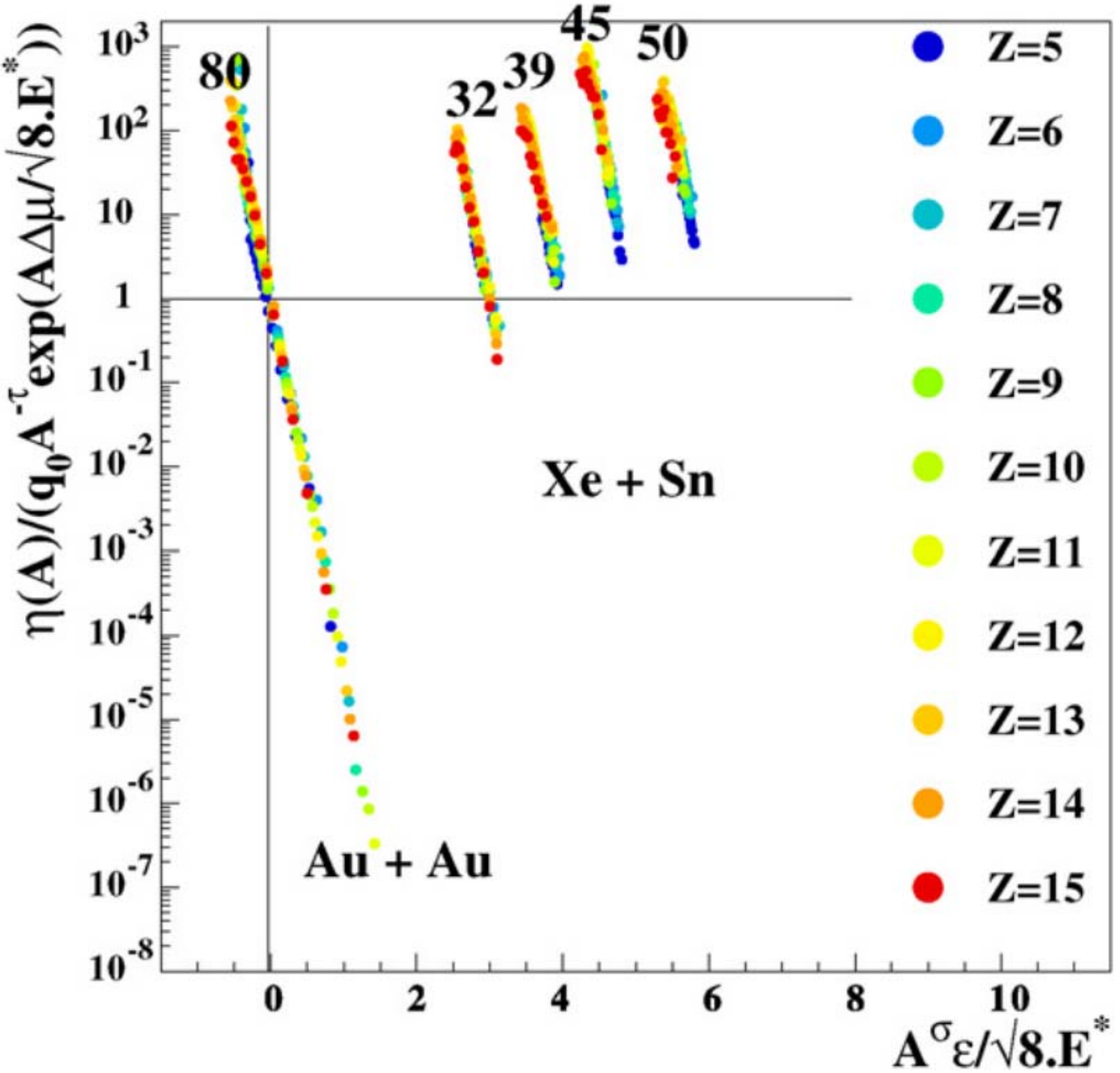}
\par\end{centering}
\caption{Fisher scaling of fragment yields applied to data for Au
quasi-projectiles (in the reaction Au+Au at 80 MeV per nucleon
incident energy) and for
quasifusion nuclei produced in Xe+Sn reactions from 32 to 50 MeV per
nucleon incident energies. Data for different
Xe+Sn reactions have been shifted horizontally for clarity.
From \citep{I63-NLN07}.} 
\label{fig:Fisher-scaling-INDRA}
\end{figure}

In \cite{I63-NLN07} a large body of data
was used to make a comprehensive survey of different signals of critical
behaviour in multifragmentation, including Fisher scaling. As shown
in Fig.~\ref{fig:Fisher-scaling-INDRA}, an excellent scaling is
observed for fragment yields in the break-up of Au quasi-projectiles
over a very wide range of excitation energies, which includes evaporative
decay at the lowest excitation energies (fission events were excluded).
A similar good scaling is also seen for multifragmenting hot nuclei
formed in central Xe+Sn reactions, albeit over a much
smaller range of excitation energies due to the reaction mechanisms
involved. These data were shown elsewhere to be consistent with predictions
based on the coexistence region of the nuclear phase diagram, not
the critical point, \emph{i.e.} spinodal decomposition
\cite{I29-Fra01,I31-Bor01,I40-Tab03,I89-Bor18},
negative heat capacity \cite{MDA02,I63-NLN07},
and, for the Au quasi-projectile data, bimodality of the order
parameter \cite{I72-Bon09}.
Therefore for experimental data as for models, behaviour such as Fisher
scaling of fragment yields, power laws, etc. is far from an unambiguous
signal of criticality or of the order of the transition. Similar conclusions
were reached in \cite{MDA03}.
What is interesting to note here is that, although Fisher's model
applies only to subcritical clustering, the yield scaling is of equally
good quality above the apparent critical temperature (\emph{i.e. }above
the horizontal line in Fig.~\ref{fig:Fisher-scaling-INDRA}). This
is especially true for the data on quasifusion nuclei from Xe+Sn
reactions for which almost all fragment yields are apparently compatible
with supercritical temperatures. Indeed, only the data from the 32 MeV
per nucleon incident energy
reaction exhibit a subcritical branch while, for the higher bombarding
energies, data move further and further away from the critical region.
The astonishing fact is that although these results are not consistent
in themselves with the Fisher model assumptions, they are consistent
with other analyses of the same data using very different approaches
and hypotheses. In \cite{I46-Bor02} the
negative branch of the heat capacity is clearly observed at 32 MeV per
nucleon incident energy,
less so at 39, and not significantly at all at 45 and 50 MeV  per
nucleon incident energies.
Moreover the so-called ``fossil signal'' of spinodal decomposition
which signs the systems' exploration of the mechanically unstable
region at the heart of the coexistence zone was shown to reach a maximum
for 32 - 39 MeV per nucleon before decreasing to almost nothing at 50
MeV per nucleon incident energy\cite{I40-Tab03,I89-Bor18}.

Another more general form of scaling law derived from the self-similar
nature of critical systems concerns the universality of order parameter
fluctuations, often referred to as $\Delta$-scaling \cite{Bot00,Bot02}.
Given an observable $m$, the $\Delta$-scaling is observed when its
probability distribution $P_{N}[m]$ for different system sizes $N$
can be reduced to a universal scaling function $\Phi(z_{(\Delta)})$
defined by
\begin{eqnarray}
<m>^{\Delta}P_{N}[m] & = & \Phi(z_{(\Delta)})\label{eq:delta-scaling}\\
z_{(\Delta)} & = & \frac{m-m*}{<m>^{\Delta}}\label{eq:scaling-var}
\end{eqnarray}
where $m*$ can be either the mode or the mean of the distribution.
In the trivial case of Poissonian fluctuations (such as for a non-critical
system or an observable $m$ not related to the order parameter),
distributions scale with $\Delta=1/2$. At or above the critical temperature
however, order parameter fluctuations scale like $\sigma_{m}\sim<m>$
and in this case the scaling law with $\Delta=1$ is expected (or
$1/2<\Delta<1$ if $m$ is linearly related to the order parameter).
A critical order parameter can therefore be identified by a change
in the scaling properties of its fluctuations (change of $\Delta$).
Moreover in the critical region the large-$z$ tail of the scaling
function should decrease like $\Phi(z_{(1)})\sim\exp-z_{(1)}^{\nu}$
with exponent $\nu>2$, \emph{i.e. }falling off faster than an exponential
or gaussian tail.

As well as providing an alternative to critical exponent analysis
in equilibrium systems, the authors of \cite{Bot00,Bot02}
claimed that $\Delta$-scaling analysis has the advantage of being
the only tool for the analysis of dynamical (non-equilibrium) systems
as the scaling laws are independent of whether one is dealing with
an equilibrium or a dynamical phase transition. They also stated that
the $\Delta$-scaling cannot be defined for systems which exhibit
a first-order phase transition, but it was later shown in \cite{Gul05}
that the same pseudo-critical behaviour as seen for Fisher scaling
also occurs for $\Delta$-scaling due to correlation lengths approaching
the (finite) system size inside the coexistence region along the line
of maximum fluctuations shown in Fig.~\ref{fig:fisher-lgm}. Other
model calculations \cite{Car02} indicate
that such a scaling may appear for first-order transitions only if
the system is sufficiently small, which is the case of all systems
which are experimentally accessible in nuclear physics.
A first application of the $\Delta$-scaling analysis to the multifragmenting
quasifused systems formed in $Xe+Sn$ reactions
was made in \cite{Bot01}. Both $\Delta=1/2$
and $\Delta=1$ scalings were observed in data. Fragment multiplicity
distributions at all incident energies collapse onto to a single gaussian
scaling function $\Phi(z_{(1/2)})$ with scaling $\Delta=1/2$ whereas
the size (charge) of the largest fragment of each event, $Z_{max}$,
exhibits the change of scaling expected for an observable closely
related with an order parameter of a critical phase transition, between
32 MeV and 39 MeV per nucleon incident energies. On the other hand the large-$Z_{max}$
tail of the scaling functions did not show any sign of proximity to
a critical region, falling off like $\exp(-z^{1.6})$.

The observed change of scaling law for $Z_{max}$ fluctuations, and
thus its identification with a critical order parameter for multifragmentation,
allows to establish in a model-independent way, \emph{i.e.} without
any \emph{a priori} knowledge of the microscopic processes involved,
that nuclear multifragmentation is more akin to the condensation of
vapour than it is to the shattering of glass \citep{Aic84}.
This is because all known critical phenomena with cluster degrees
of freedom fall into two generic families, for each of which the order
parameter is known:
\begin{itemize}
\item \emph{fragmentation scenarios} in which clusters result by breaking
the system into smaller pieces, rather like shattering glass or breaking
a dinner plate. In this case the order parameter is the mean multiplicity
of fragments; and
\item \emph{aggregation scenarios} in which clusters are built up from smaller
pieces; the order parameter is the mean size of the largest cluster.
This family includes the Fisher droplet model (and thus the LG phase transition), 
the Ising model and closely-related Lattice Gas
model, and the percolation model.
\end{itemize}

System mass (size) dependence of $\Delta$-scaling for central symmetric
collisions was studied in \citep{I54-Fra05}
where the behaviour of $Z_{max}$ as an order parameter was confirmed
for the lighter systems Ar+KCl and Ni+Ni. It was shown that
the bombarding energy at which the change of $\Delta$-scaling occurs
decreases with the size of the system; for the heaviest system studied
(Au+Au) only the $\Delta=1$ scaling is observed down to the lowest
studied beam energy of 40 MeV  per nucleon incident energy. 
For the first time the form of
the scaling distribution in the $\Delta=1$ scaling regime was clearly
identified as an extremal probability distribution, the Gumbel distribution
\begin{equation}
\Phi(z_{(1)})\sim\exp\left(-z-\exp\left(-z\right)\right).\label{eq:gumbel-dist}
\end{equation}
Whereas the central limit theorem leads to the gaussian distribution
for a sum of random variables, the Gumbel distribution belongs to
a family of distributions representing the extrema of a set of random
variables, and is characterized by an exponential tail on the extremal
side (for large $z$ if $z$ is a maximum; for small $z$ if a minimum).
In the $\Delta=1/2$ regime on the other hand the scaling function,
although clearly much more symmetric than Eq.~(\ref{eq:gumbel-dist}),
could only be qualified as ``quasi-gaussian''.
\begin{figure}
\includegraphics[bb=0bp 0bp 567bp 328bp,clip,width=0.45\textwidth]
{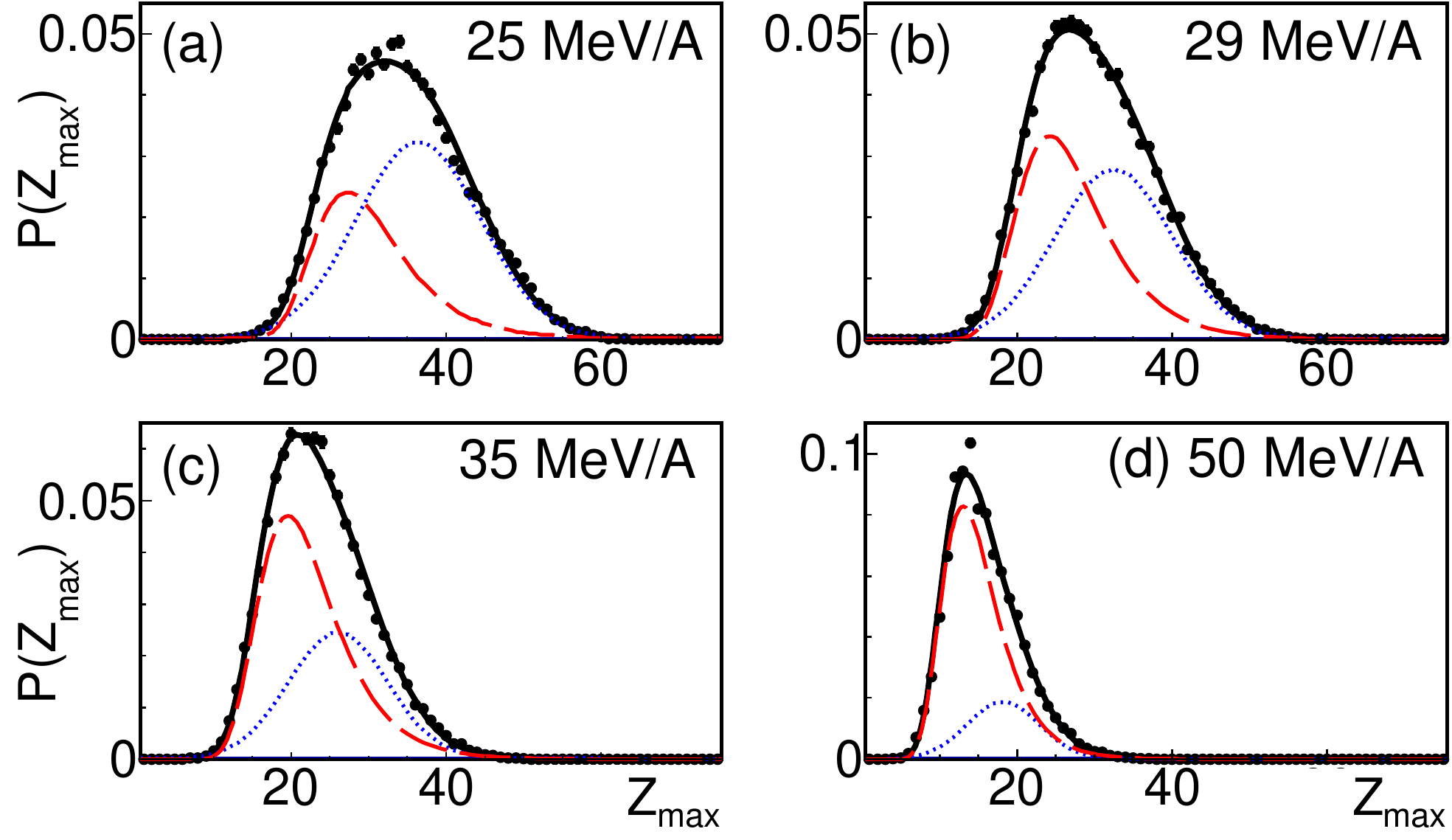}
\hfill{}
\includegraphics[bb=0bp 0bp 567bp 330bp,clip,width=0.45\textwidth]
{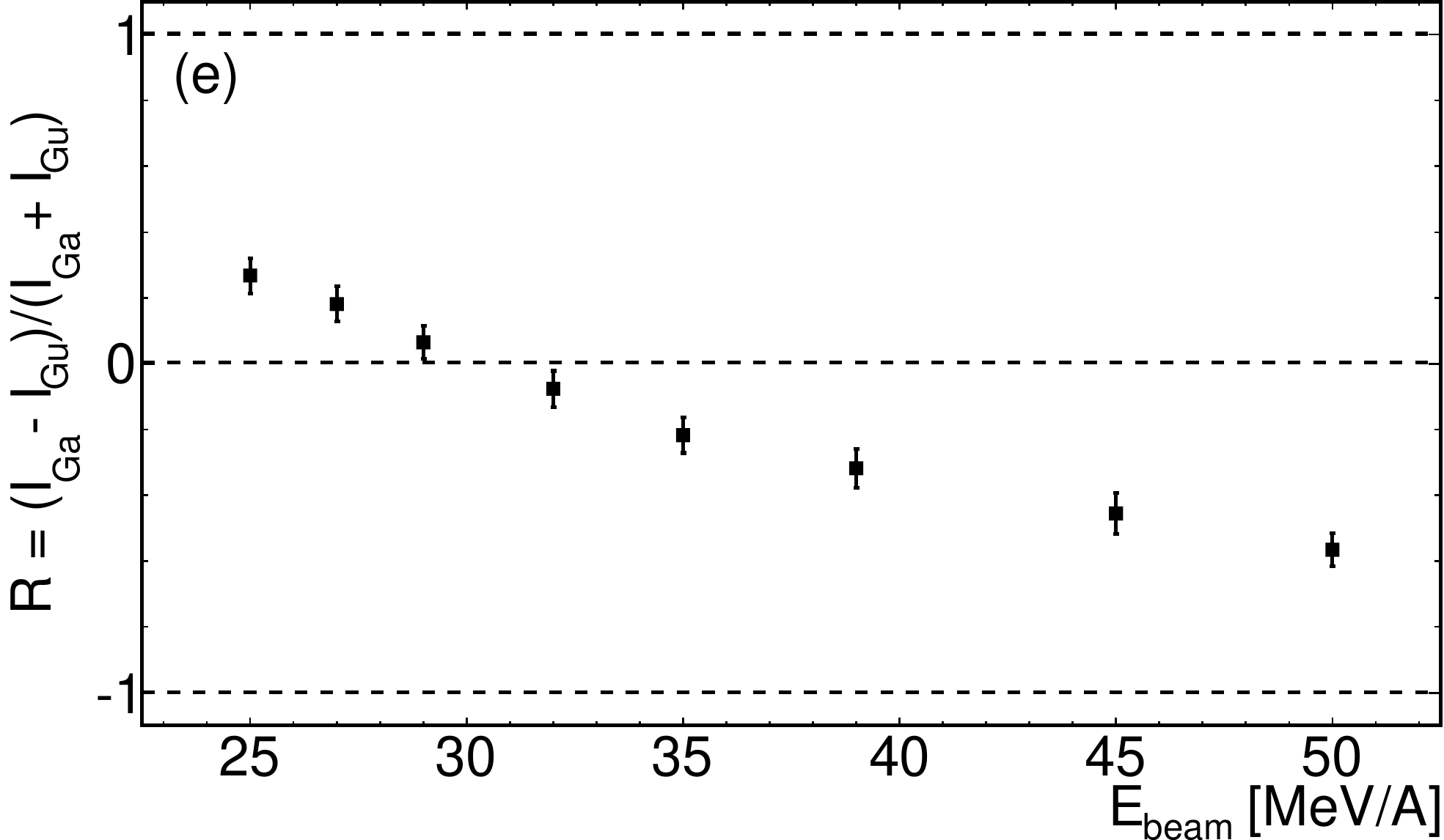}
\caption{(a)-(d): Charge distributions of the largest fragment of each event
$Z_{max}$ in central quasifusion collisions of Xe on Sn at the bombarding
energies shown. Data: points; black line:
fit of the distribution with an admixture of gaussian and gumbellian
distributions; red dashed line: gumbellian component; blue dotted
line: gaussian component. (e) Relative strengths of the two components,
$R$ (see Eq.~(\ref{eq:gauss-gum-ansatz})), as a function of the beam
energy. Adapted from \citep{I78-Gru13}.
\label{fig:delta-scaling-gauss-gumbel}}
\end{figure}
The reason for the mass-dependence of the $\Delta$-scaling change
and the exact form of the scaling function remained a mystery until
it was realized that in the irreversible aggregation model represented
by the Smoluchowski equations \citep{Don85}
the largest cluster size distribution $P(s_{max})$ can be reasonably
approximated by a sum of gaussian and gumbellian distributions \citep{Bot12},
\begin{equation}
P(s_{max})=\frac{R+1}{2}f_{\mathrm{Gauss}}(s_{max})+
\frac{1-R}{2}f_{\mathrm{Gumbel}}(s_{max}).\label{eq:gauss-gum-ansatz}
\end{equation}
In this model, the cluster mass distributions evolve over time according
to mass-dependent rates of aggregation or break-up, and after a time
$t_{C}$, called the critical gelation time, a large (infinite in
an infinite system) cluster appears corresponding to the sol-gel transition.
The relative strength of the two components in the distribution of
the largest cluster size, $R$, evolves with time in the model: at
short timescales ($t\ll t_{C}$) when little aggregation has occurred
$R\approx-1$, meaning that the Gumbel distribution dominates; at
longer timescales ($t\gg t_{C}$) $R\rightarrow1$ and the largest
cluster size distribution becomes gaussian; at the critical gelation
time $t_{C}$ (i.e. when fluctuations of the cluster size distribution
are maximal) $R\approx-0.45$. This evolution of the distribution
from gumbellian to gaussian means that the nature of the order parameter
$s_{max}$ changes from \emph{extremal} ($s_{max}$ corresponds to
the largest of a set of uncorrelated, random clusters) to \emph{additive}
(the largest cluster results from the aggregation of smaller clusters).
When applied to data on the largest fragment charge, the \emph{ansatz}
of Eq.~(\ref{eq:gauss-gum-ansatz}) provides a far better fit to the
$Z_{max}$ distributions for Xe+Sn central collisions than either
of the two asymptotic distributions 
alone (Fig.~\ref{fig:delta-scaling-gauss-gumbel}(a)-(d)).
Moreover the composition of the fit evolves in a regular way from
gaussian to gumbellian ($R$ goes from positive to negative values
in Fig.~\ref{fig:delta-scaling-gauss-gumbel}(e)) with increasing
bombarding energy. Values of $R\approx-0.45$ corresponding to maximum
fragment size fluctuations, analogous to the critical gelation time
in the irreversible aggregation, occur at beam energies close to 39
MeV per nucleon incident energy,
where the change of scaling law from $\Delta=1/2$ to $\Delta=1$
occurs.

By analogy with the irreversible aggregation model the authors of
\cite{I78-Gru13} therefore proposed that the
exact form of the charge distribution of the largest fragment in nuclear
multifragmentation, and hence its fluctuations, are also determined
by the timescale of fragment formation in such reactions. The onset
and increase of radial expansion at beam energies above 25 MeV per
nucleon
\cite{I69-Bon08}, essential to drive the system
towards the onset of spinodal instability and initiation of the break-up
into fragments \cite{Cho04}, will also effectively
shorten the time available for primary fragments to coalesce. Within
this framework, the mass-dependence of the energy at which the $\Delta$-scaling
law changes \citep{I54-Fra05} can
be ascribed to the entrance channel dependence of radial expansion
in central collisions: for light systems such as Ar+KCl or Ni+Ni
the bombarding energy required to achieve sufficient initial compression
for there to be significant radial expansion is higher than for the
heavier systems.

One can conclude this subsection by saying that now a coherent
comprehensive view is obtained on the criticality signal subject
for finite systems.
This view is moreover reenforced by the consistence with other
signatures, discussed up to now, of the coexistence region. 
As far as phase transition dynamics is concerned the 
aggregation scenario deduced  recalls microscopic approaches in which
fragments result from spinodal fluctuations occurring in the hot,
expanding nuclear matter formed in collisions.

\subsection{Landau free-energy approach}\label{Landau}
\begin{figure}
\begin{center}
\includegraphics[width=0.75\textwidth]{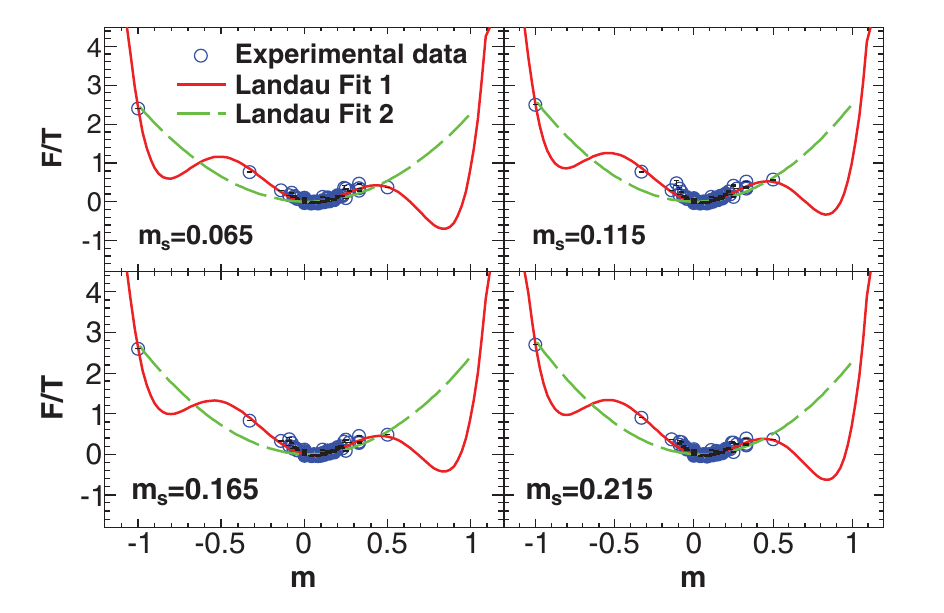}
\end{center}
\caption{
$F$/$T$ as a function of fragment neutron-proton asymmetry $m$ for an
excitation energy of 6.5 MeV per nucleon of quasi-projectiles.
Different panels
correspond to different neutron-proton asymmetry ($m_s$) bins of
quasi-projectiles.
Full lines are fits to the data using the complete Landau free 
energy (first order phase transition)
and dashed lines are fits which refer to single phase systems.
Statistical errors are smaller than the points. From~\cite{Mab13}.}
\label{fig:Landau}
\end{figure}
A new signature of first order phase transition was experimentally investigated
using the Landau free-energy approach~\cite{Mab13}.
Quasi-projectiles formed in  35 MeV per nucleon
$^{70}$Zn + $^{70}$Zn, $^{64}$Zn + $^{64}$Zn and $^{64}$Ni + $^{64}$Ni were
reconstructed and data sorted in quasi-projectile asymmetry
($m_s = (N_s-Z_s)/A_s)$ and excitation energy bins in the range
3 - 9 MeV per nucleon. According to the modified Fisher model to take into
account finite size effects, the free energy per nucleon of a
fragment of mass $A$ normalized to the temperature of the
quasi-projectile, $F/T$, can be
derived from the fragment yield $Y~=~y_0 A^{-\tau} e^{(-F/T)A}$;
$y_0$ is a constant and $\tau$ is a critical exponent. In the
Landau approach the free energy of a first order phase transition
is extended in a power series in the order
parameter $m$:
\begin{eqnarray}
\noindent
F/T=
\frac{1}{2}am^{2}+\frac{1}{4}bm^{4}+\frac{1}{6}cm^{6}-\frac{H}{T}m, 
\label{eq:Landau}
\end{eqnarray}
with $m = (N_f-Z_f)/A_f$; $N_f$, $Z_f$ and $A_f$ are the
neutron, proton and mass numbers of the fragment respectively.
The quantity $H$ is the conjugate variable of $m$ and $a$, $b$ and $c$
are fitting parameters which depend on the temperature, density or
pressure of the fragmenting system. In the absence of $H$, the free
energy $F/T$ is symmetric in the exchange of $m$ to $-m$ indicating
that nuclear forces are invariant when exchanging $N_f$ in $Z_f$.
In presence of $H$, which arises when the quasi-projectile is
asymmetric in the composition ($m_s$), the symmetry is violated.
More details can be found in~\cite{Mab13}.
In Fig.~\ref{fig:Landau} $F/T$ values as a function of $m$ are
displayed for quasi-projectiles at excitation energy of 6.5~MeV per nucleon and
for different asymmetry ($m_s$) bins of quasi-projectiles (different panels).
The value of $\tau$ = 2.3 $\pm$ 0.1 derived from previous works was used.
Dashed lines are fits to data using only the
first and last terms of Eq.~(\ref{eq:Landau}), a case corresponding  to single
phase systems. Better fits (full lines) to data are obtained with the complete form
of Eq.~(\ref{eq:Landau}), which is the signature that quasi-projectiles
are in the regime of a first-order phase transition. It is important
to note that statistical error bars are smaller than the points. 
       
\subsection{Phase transition dynamics}\label{Spino}
The knowledge of the nature of the dynamics involved during phase transition
in hot nuclei, i.e. fragment formation, is certainly the most delicate
point. Two mechanisms have been proposed. On one side, stochastic mean
field approaches predict the transition dynamics to follow the
spinodal fragmentation scenario proposed very early on, triggered by
phase-space fluctuations amplified in an unstable medium and, on the
other side, molecular dynamics models (QMD, AMD) in which many-body
correlations are sufficient to produce fragments at early times even
in absence of unstable conditions.
We have noticed in section~\ref{how} experimental evidence
for a radial extra energy boost (radial expansion energy) associated to
multifragmentation products. It can be attributed either to a dominant 
compression-expansion phase in central nucleus-nucleus
collisions or to thermal pressure for more gentle collisions:
hadron-nucleus or semi-peripheral nucleus-nucleus collisions.
The system might then reach densities and temperatures that
correspond to the unstable spinodal region where exponential
amplification of density fluctuations leads to clusterization: an
inhomogeneous mixture of fragments (normal density region), nucleons
and light fragments (low density region). This can be seen as an analogue
for phase separation in a finite system, for which spinodal
decomposition would be the microscopic mechanism.

One must first visit the major theoretical progress which
has been realized to understand and learn about spinodal fragmentation in the nuclear 
context especially for finite systems. A review can be found in Ref.~\cite{Cho04}.

What are the specificities of spinodal
decomposition as far as nuclear matter is concerned?
\begin{figure}[htb]
\begin{minipage}[t]{0.4\textwidth}
\includegraphics[width=\textwidth]{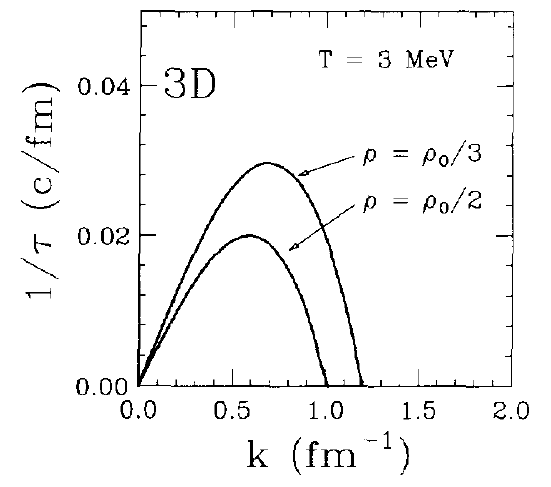}
\caption{Nuclear matter dispersion relation at 3~MeV temperature for
two different densities; $\rho_{0}$ is the normal density.
From~\cite{Colo97}.} \label{fig:BOB}
\end{minipage}%
\hspace*{0.05\textwidth}
\begin{minipage}[t]{0.4\textwidth}
\includegraphics[width=1.1\textwidth]{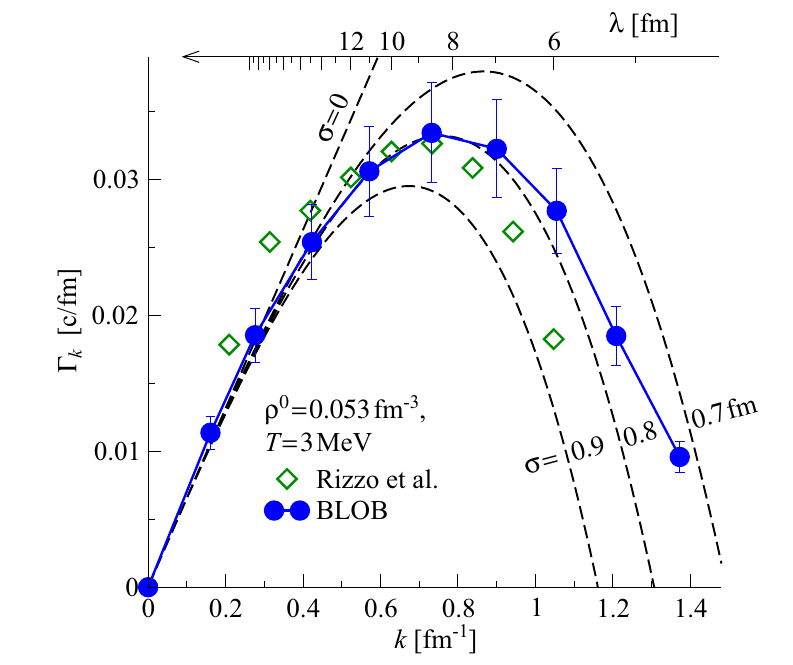}
\caption{Nuclear matter dispersion relation at 
3~MeV temperature for density around $\rho_{0}$/3 (full points).
The dashed lines correspond to analytical predictions and diamonds to
one dimension calculations from~\cite{Rizz08}.
From~\cite{Nap17}.} \label{fig:BLOB1}
\end{minipage}
\end{figure}
Associated to negative compressibility the mechanically unstable spinodal
region was investigated by studying the propagation of small density
fluctuations~\cite{Colo97,Ayi95}. By analogy with optics,
the nuclear dispersion relation can be calculated  for different 
conditions of temperature and density
by introducing the Boltzmann-Langevin equation with some stochasticity
in the evolution of the one body density (BOB, BLOB - see~\ref{BOBBLOB}).
Within the linear response theory framework, if one expands the solution
of the Boltzmann-Langevin equation as $ f = f_0 + \delta f$, where $f_0$ is
a solution of the Boltzmann equation and $\delta f$ the fluctuating part,
one finds the equation of motion
$\partial \delta f / \partial t = -i\mathcal{M} \delta f + \delta I[f_0]$
at the leading order in $\delta f$. The extended RPA matrix operator $\mathcal{M}$
represents the combined linearized action of the effective field and of the average
collision term.
Inside the spinodal region the eigenvalues of the matrix $\mathcal{M}$
become imaginary. Consequently the fluctuations associated with a given
eigenmode agitated by the source term $\delta I$ are exponentially 
amplified or suppressed, depending upon the sign of the imaginary 
part of the frequency. In the case of infinite nuclear matter the 
eigenmodes of the linearized
dynamics are plane waves, characterized by a wavenumber $k$ and
 an imaginary eigenfrequency, which is the inverse of the instability 
 growth time.  
Fig.~\ref{fig:BOB} presents an example
of nuclear dispersion relation at 3~MeV temperature for two different
densities $\rho_{0}/2$ and $\rho_{0}/3$.
Imaginary RPA frequencies are reported as a function 
of the wavenumber k of the considered perturbation. This dispersion 
relation exhibits a strong maximum at a given wavenumber followed by 
a cut-off at large k
values. This cut-off reflects the fact that fluctuations with wavelength
smaller than the range of the force cannot be amplified. The most 
unstable modes correspond to wavelengths lying around 
$\lambda \approx$~ 8-10 fm
and the associated characteristic times are almost identical, 
around 30-50~fm/$c$, depending on density ($\rho_{0}/2$ - $\rho_{0}/8)$ 
and temperature (0~-~9~MeV)~\cite{Colo97,Idi94}. Fig.~\ref{fig:BLOB1}
shows the results of the same study recently made in the framework of
a numerical treatment of the Boltzmann-Langevin
equation in which fluctuations are
introduced in full phase space from
inducing nucleon-nucleon collisions (BLOB simulation)~\cite{Nap17}.
Results obtained are very similar.
A direct consequence of the dispersion relation is
the production of ``primitive'' fragments with mass
$A\approx\rho\lambda^3$. For the leading wavelengths,
this corresponds to a distribution peaked around Ne nuclei. 
However this simple
 picture is expected to be largely blurred by several
effects.
The beating of different modes occurs. Coalescence effects due to the
nuclear interaction between fragments before the complete disassembly
are also expected~\cite{Colo97}. 

\begin{figure}
\begin{center}
\includegraphics[width=0.6\textwidth]{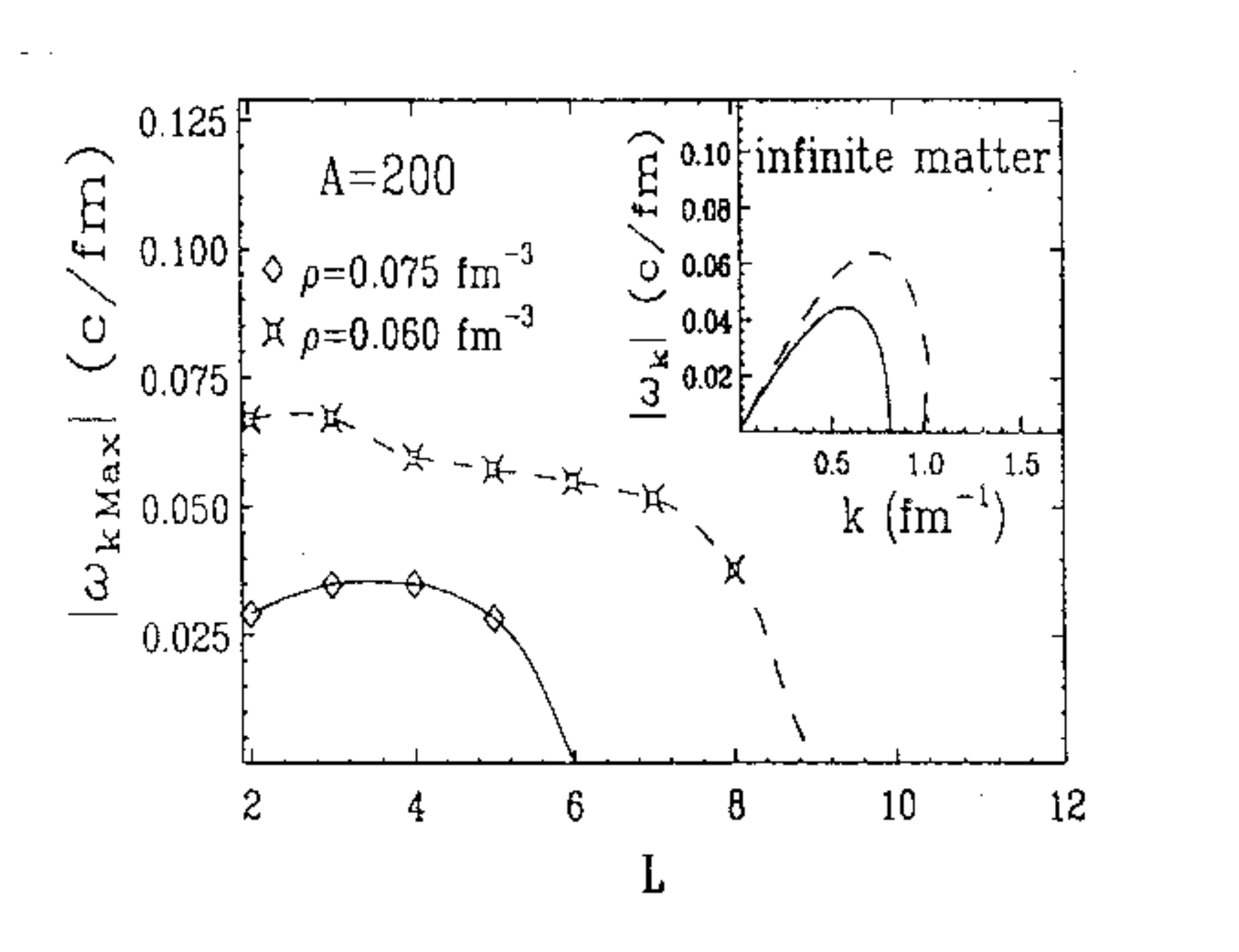}
\end{center}
\caption{Growth rates of the most unstable modes for a spherical source 
with 200 nucleons as a function of the multipolarity $L$ and for
two different central densities.
 From~\cite{Jac96}.}
\label{fig:dis_rel_fini}
\end{figure}

For finite systems the situation is even more
complicated. The presence of a surface introduces an explicit breaking
of the translational symmetry. Fig.~\ref{fig:dis_rel_fini} shows
the growth rates of the most unstable modes for a spherical nucleus
of A = 200 with a Fermi
shape profile and for two different central densities~\cite{Jac96}.
The growth rates are nearly the same for different multipolarities, $L$,
up to a maximum multipolarity $L_{max}$ (see also~\cite{Nor00}).
This result indicates that the unstable finite system breaks into different
channels depending on multipolarity~\cite{Jac96}.
Equal-sized ``primitive'' fragments
are then expected to be produced with sizes in the range
$A_F$/2 - $A_F$/$L_{max}$; $A_F$ being the part of the system leading
to fragments during the spinodal fragmentation.
Moreover the finite system produced during the nucleus-nucleus
collision has to stay or live long enough in the spinodal region ($\sim$
3 characteristic times - 100-150~fm/c - for symmetric matter) to allow
an important amplification of the initial density fluctuations.
And in addition, experimentally, fragments are detected after secondary decay,
which introduces a broadening of the fragment size distribution.
Taking into account the accumulation of all these effects on the final
extra production of equal-sized fragments, it is clear that any
signature that spinodal fragmentation is responsible for the phase
transition dynamics can only be what is called a fossil signature.
A full simulation of the spinodal decomposition of quasifused sources
using BOB simulations~\cite{I40-Tab03} already
testified to this fact, with less than 1\% of extra events with equal-sized
fragments. It is the reason why the signature is difficult to observe
experimentally.
One can also note that the Coulomb potential has a very small effect on the
growth rates of unstable collective modes except close to the border of the
spinodal zone where it stabilizes very long wavelength unstable
modes~\cite{JacT96}. On the other hand, for a finite system, Coulomb
interaction reduces
the freeze-out time and enhances the chance to keep a memory of the dynamical
instabilities; a similar comment can be made as far as collective radial expansion 
is concerned. Both effects push away the ``primitive'' fragments 
and reduce the time of their mutual interaction. So finally, even if
expected extremely
reduced, the presence of extra partitions with nearly equal-sized fragments is a
candidate to sign the role of spinodal instability in
multifragmentation.

Following early studies related to nearly equal-sized fragment 
partitions~\cite{Bru92}, an intra-event correlation method called higher order charge
correlations~\cite{Mor96} was proposed to enlighten any extra production 
of events with specific fragment partitions. The high sensitivity 
of the method makes it particularly appropriate to look for small numbers 
of events, as those
expected to have kept a memory of spinodal fragmentation properties.
All fragments of one event with fragment
multiplicity $M = \sum_Z n_Z$, where $n_Z$ is the number of fragments
with charge $Z$ in the partition,  are taken into account. By means of the
normalized first order:
\begin{equation}
\noindent
	\Zmoy = \frac{1}{M} \sum_Z n_Z Z
\label{equ:corrZ1}
\end{equation}
and second order:
\begin{equation}
        \sigma_Z^2 = \frac{1}{M} \sum_Z n_Z (Z - \Zmoy)^2
\label{equ:corrZ2}
\end{equation}
moments of the fragment charge distribution in the event, 
one may define the correlation function (CF):
\begin{equation}
\left. 1+R(\sigma_Z, \Zmoy)=\frac{Y(\sigma_Z, \Zmoy)}{Y'(\sigma_Z, \Zmoy)} 
\right| _{M}
\label{equ:corrZ3}
\end{equation}
Here, the numerator $Y(\sigma_Z, \Zmoy)$ is the yield of events with given
$\Zmoy$  and $\sigma_Z$ values. Because the measurement of the charge 
belonging to a given event is not subject to statistical fluctuations, 
expression  (\ref{equ:corrZ2}) can be used rather than the ``nonbiased estimator'' 
of the variance, $\frac{1}{M-1} \sum_Z n_Z (Z - \Zmoy)^2$, as proposed in
~\cite{Mor96}.
The exact identification of fragment charge up to at least 
$Z~\approx$~25 is mandatory to use such intra-event correlation method
for the proposed study. 
The denominator $Y'(\sigma_Z, \Zmoy)$ represents the uncorrelated yield
of pseudo-events and can be built in different ways.
It was built in~\cite{Mor96}, as for classical correlation  
methods, by taking fragments at random in different events of the selected  
sample of a certain fragment multiplicity. This Monte-Carlo generation of 
the denominator $Y'(\sigma_Z, \Zmoy)$ can be replaced by a fast algebraic  
calculation which is equivalent to the sampling of an infinite number of 
pseudo-events~\cite{Des02}. Its contribution to the statistical error of 
the correlation function is thus eliminated.  However, owing to the way the
denominator was constructed, only the  fragment charge distribution 
$\rd M/ \rd Z$ of the parent sample is reproduced  but the 
constraints imposed 
by charge conservation are not taken into account. This has, in particular, 
a strong effect on the total charge bound in fragments, 
which makes the denominator yield distributions as
a function of $\Zmoy$ wider and flatter than those of the
numerator~\cite{T28Tab00}.
Consequently, even in the absence of a physical correlation signal, the 
ratio (\ref{equ:corrZ3}) is not a constant equal to one. The correlations induced 
by the finite size of the system (charge conservation) distort the amplitude, 
or may even cancel other less trivial correlations. 
Therefore, a method for the evaluation of the denominator~\cite{Des02}, based on 
the ``intrinsic probability'' of emission of a given charge, was proposed.
It minimizes the effects just indicated and replicates all features of the partitions 
of the numerator, except the correlations due to other reasons 
than charge conservation. The principle of the method is to take into
account in a combinatorial way the trivial correlations due to charge
conservation.
The probability to observe 
a given partition (${\bf n}: (n_1,\ldots ,n_{Z_{\rm max}})$), at a given total
multiplicity, $m = \sum_Z n_Z$, is obtained by the multinomial 
formula. If the total charge is fixed ($Z_{\rm tot}=\sum_Z Z\,n_Z$), 
the partition probabilities are given by:
\begin{equation}
P({\bf n}|m) = \alpha \ m! \,
\prod_Z \frac{^{\rm intr}P_Z^{n_Z}} {n_Z!} \
\delta_{Z_{\rm tot},\sum_Z Z\,n_Z}\ ,
\label{equ:corrZ4}
\end{equation}
\noindent
where $\alpha$ is the normalization constant  (so that $\sum_n
P({\bf n}|m$) = 1) and $\delta$ is the Kronecker symbol. 
All the details can be found in~\cite{Des02,I40-Tab03}.
The intrinsic probability values, $^{\rm intr}P_Z$, are obtained by means of a recursive
procedure of minimization.
\begin{figure}
\begin{center}
\includegraphics*[width=0.65\textwidth]
{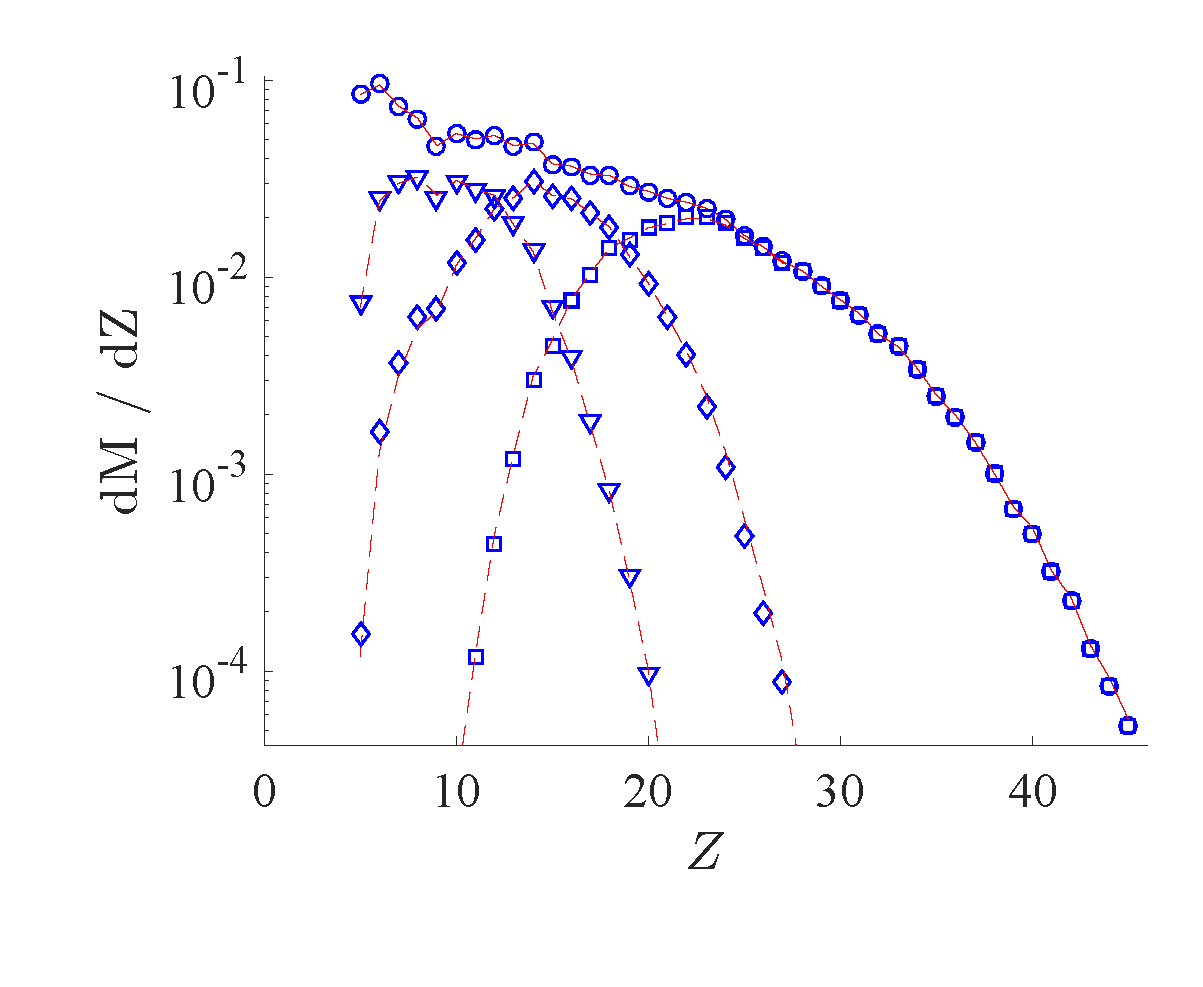}
\end{center}
\caption{Experimental differential charge multiplicity
distribution (circles) for quasifusion nuclei formed in central 32 MeV
per nucleon
$^{124}$Xe + $^{112}$Sn reaction with fragment multiplicity, $M$, equal
to 4. Experimental differential distributions for the first (squares), second (diamond)
and third (triangles) heaviest fragments of partitions are presented too. The full
and dashed lines, to be compared to data, correspond to the results of the intrinsic
probability method for the fragment probabilities. From~\cite{I89-Bor18}.}
\label{fig:spino1}
\end{figure}

From experiments, about fifteen years ago, there were indications that multifragmentation
may be induced by  spinodal instabilities but the confidence level of
the fossil signature was not sufficient (3 - 4~$\sigma$ at most), due to low statistics,
to allow drawing any definitive conclusion~\cite{I31-Bor01,I40-Tab03,Bor06}.
Only very recently, studies obtained from very high
statistics experiments ( a factor at least 10 to 15 higher as compared to
previous ones) were performed aiming to give a final answer.
At the same time, related isospin effects theoretically predicted were
investigated. If spinodal instabilities are at the origin of
multifragmentation, a reduction of 
instabilities for $N/Z$ asymmetric systems in relation with an increase of
the instability growth time is theoretically predicted~\cite{Col02}
(see Fig.~\ref{fig:spinonuc}).
Intra-event charge correlations were performed on fragments emitted
from multifragmenting quasifusion hot nuclei produced in central collisions between
$^{124,136}$Xe and $^{112,124}$Sn at two bombarding energies, 32 and
45 MeV per nucleon~\cite{I89-Bor18}. The considered event samples for the study were
those with fragment (Z~$\geqq$~5) multiplicities, $M$, from 3
to 6 which correspond to higher statistics.
Fig.~\ref{fig:spino1} shows one example on how the
experimental fragment charge distributions are
faithfully described by using the intrinsic probabilities,  
$^{\rm intr}P_Z$, which have been calculated independently
for each incident energy, for each reaction and for the different
fragment multiplicities. CF values greater than one were
observed at very low $\sigma_{Z}$ ($<$ 1)
but also for $\sigma_{Z}$ (1 - 2).
This observation was used to fix the
upper limit at 2  for the $\sigma_{Z}$ of events with nearly
equal-sized fragments.
\begin{figure}
\begin{center}
\includegraphics*[width=0.48\textwidth]
{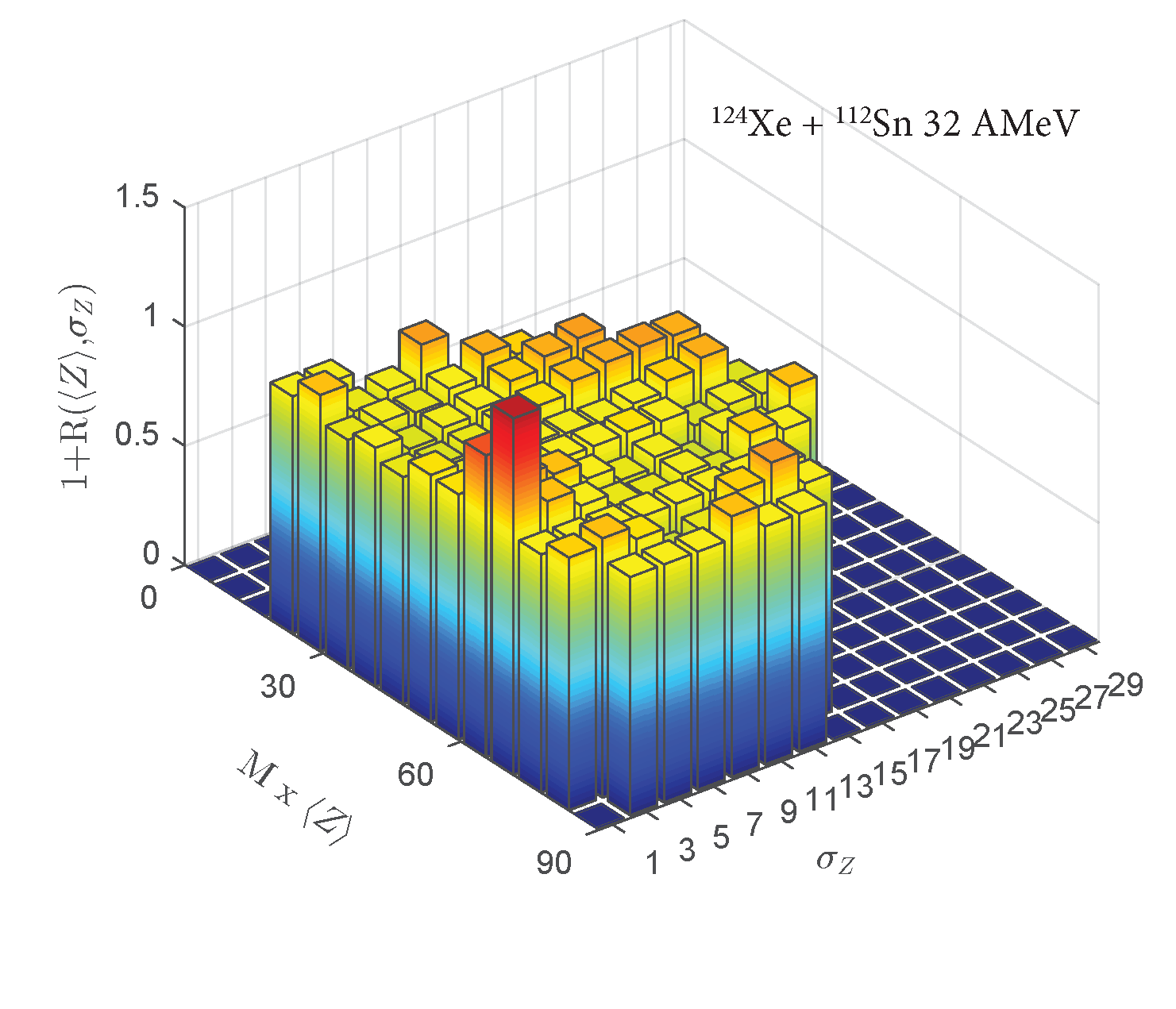}
\includegraphics*[width=0.48\textwidth]
{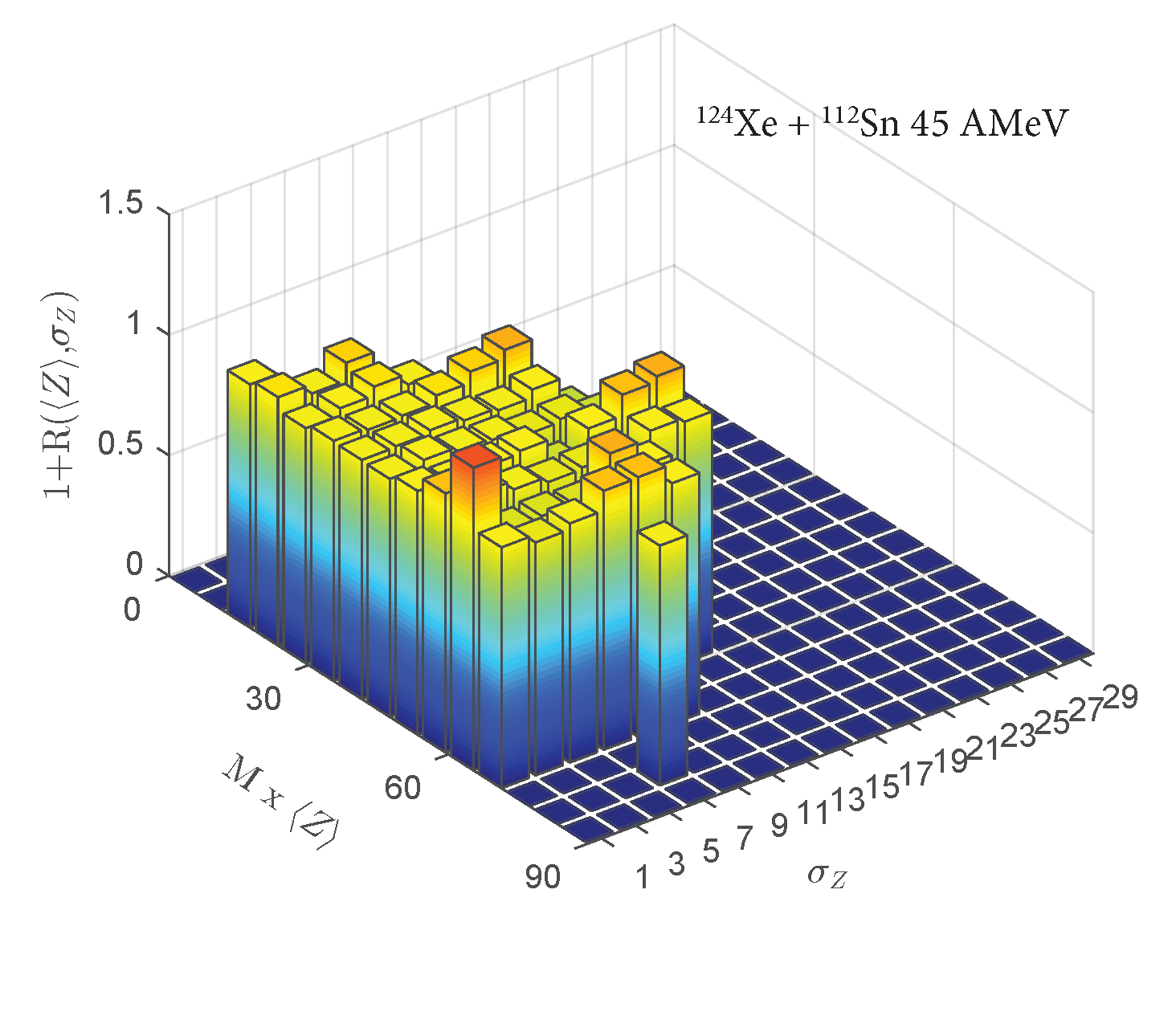}
\includegraphics*[width=0.48\textwidth]
{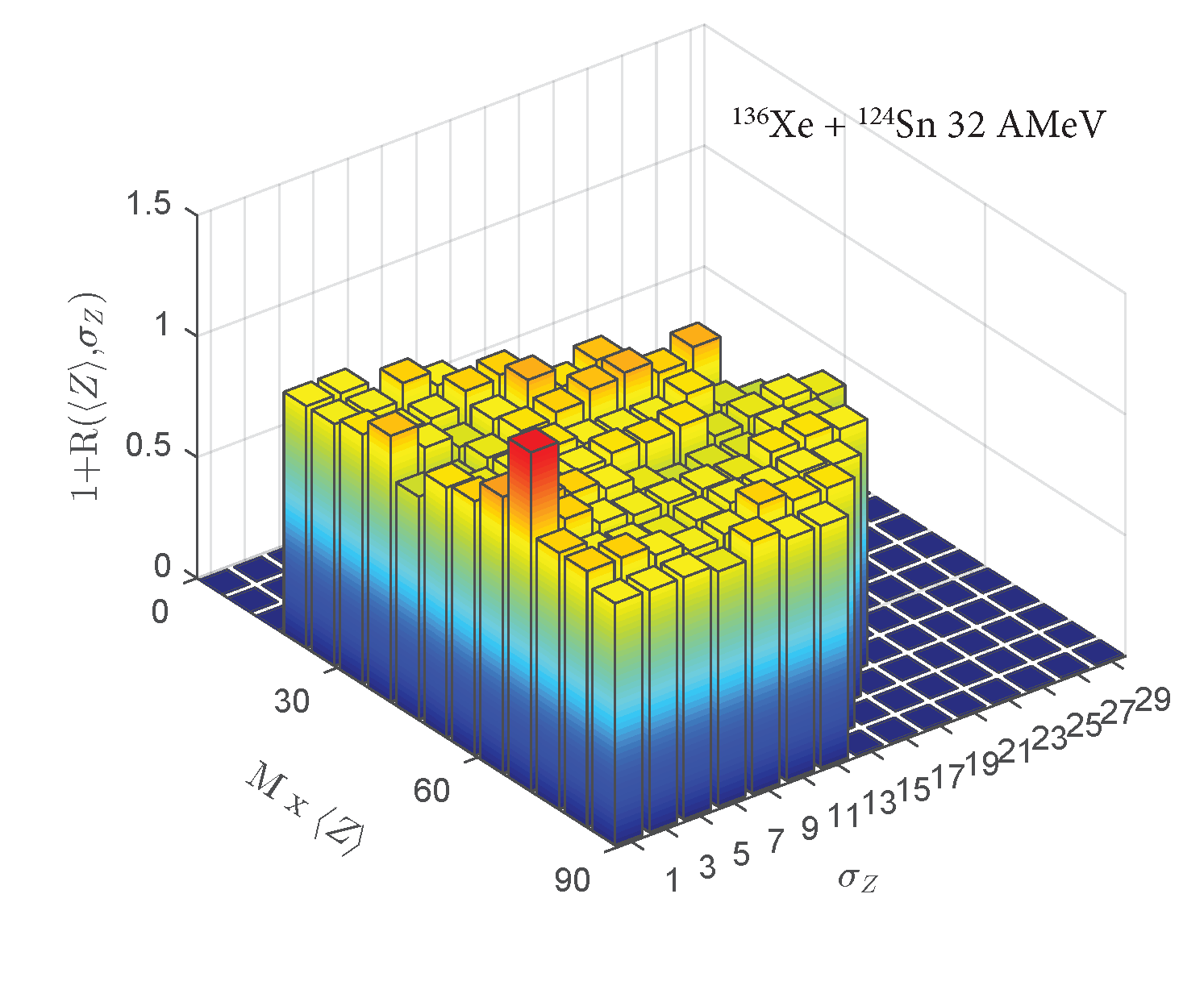}
\includegraphics*[width=0.48\textwidth]
{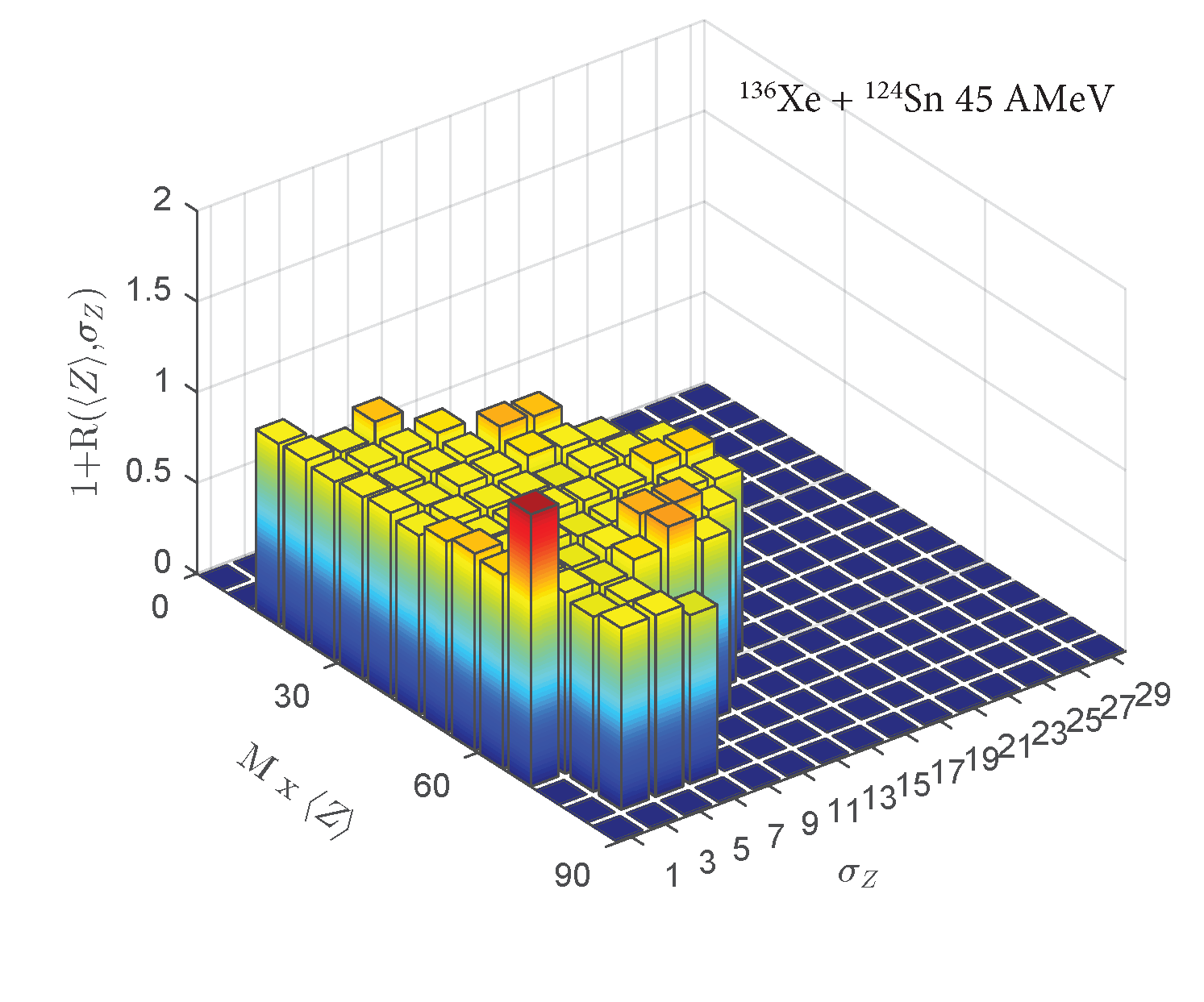}
\end{center}
\caption{Experimental correlation functions for selected
quasifusion events
formed in central $^{124,136}$Xe + $^{112,124}$Sn collisions.
Events with fragment multiplicities 3 to 6 are mixed. 
Correlation functions are calculated for a $\sigma_{Z}$ bin equal to 2
 and a $ M \times \Zmoy$ bin equal to 6. The colour/grey
scale for CFs has a maximum value for 1.6 which corresponds to dark
red/dark grey. From~\cite{I89-Bor18}.}
\label{fig:spino2}
\end{figure}
The complementary contribution ($\sigma_{Z}$ 1 - 2) comes from
the broadening of the
fragment $Z$ distribution introduced by the deexcitation of primitive
fragments (see~\cite{I89-Bor18} for details). For the
first time, the limited ranges of $\Zmoy$ contributing
to CF peaks were also clearly observed, which verifies what is theoretically
 expected for finite systems i.e. $M \times \Zmoy$
$\sim$ constant (see Table 1 of~\cite{I89-Bor18}). This observation also shows
that some spurious peaks at low $\sigma_{Z}$ were present in previous
experiments with low statistics.
To better visualize global results, CFs were built for all events of a reaction at a
given beam energy, whatever their multiplicity, by summing the correlated yields of
all $M$ and by replacing the variable $\Zmoy$ by  $ M \times \Zmoy$. 
Uncorrelated yields are then  constructed and weighted in proportion to real
events of each multiplicity.
Fig.~\ref{fig:spino2} summarizes the results.
For the four systems CF peaks are observed at low $\sigma_{Z}$.
At 32 MeV per nucleon incident energy the neutron poor system exhibits
two peaks with confidence levels greater around 6 - 7~$\sigma$ and the neutron
rich one peak above 6~$\sigma$ and one around 3~$\sigma$, which
definitively establishes the presence of spinodal fluctuations. 
At the
higher incident energy the two systems have one peak with confidence
level above 2~$\sigma$.
Covered  $ M \times \Zmoy$ domains  are the same (60 - 72) for both
reactions at 32 MeV per nucleon incident energy whereas at higher incident energy
the neutron rich system covers a range a little bit higher (66 - 72)
than the neutron poor one (54 - 66).
 Finally the percentages of events
($Y(\sigma_Z, \Zmoy)$/total number of events)
and extra events (($Y(\sigma_Z, \Zmoy)$ - $Y'(\sigma_Z, \Zmoy)$)
/total number of events) are reported in
Table~\ref{tab:spino}.
\begin{table}
\caption{Numbers and percentages of events and extra events with $\sigma_{Z} < 2$
for the different incident energies and reactions.
Calculated errors are statistical. From~\cite{I89-Bor18}.}
\begin{center}
\begin{tabular}{lccccc}
   \hline
     E (AMeV) &reaction &events &(\%) &extra events &(\%) \\
     \hline
32  &\nuc{124}{Xe}+\nuc{112}{Sn} & 1313 & 0.27 & 336 & 0.068 $\pm$ 0.004\\
32  &\nuc{136}{Xe}+\nuc{124}{Sn} & 1077 & 0.32 & 217 & 0.064 $\pm$ 0.004\\
45  &\nuc{124}{Xe}+\nuc{112}{Sn} & 1073 & 0.34 & 77 & 0.025 $\pm$ 0.003\\
45  &\nuc{136}{Xe}+\nuc{124}{Sn} & 68 & 0.030 & 15 & 0.0065 $\pm$ 0.0017 \\
\hline
\end{tabular} \\
\end{center}
\label{tab:spino}
\end{table}
Within error bars, extra event percentages are
similar for both systems at the lower incident energy. At higher
incident energy we observe a strong reduction of percentages.
The observed reduction  for the more symmetric system is in good agreement
with the negative heat capacity signatures observed experimentally
(see~\ref{Cvneg}) which fixed the upper limit of the coexistence zone 
(spinodal region) in the incident energy range 45-50 MeV per
nucleon~\cite{I46-Bor02}.
The large reduction of the signal observed for the neutron
rich system, a factor ten between 32 and 45~MeV per nucleon, can be
understood in terms of the expected $N/Z$ influence.
Indeed, if spinodal instabilities are at the origin of the dynamics
of multifragmentation, as said before, theoretical
calculations predict a reduction of 
instabilities for asymmetric systems in relation with an increase of
the instability growth time. More precisely, it is shown
in~\cite{Col02} (see also Fig.~\ref{fig:spinonuc}) that, 
for $Sn$ isotopes, the most unstable modes
associated to shorter instability growth times
($\simeq$~50 fm/c) disappear when $N/Z$ changes from 1.40 to 1.64. 
If we consider that quasifusion systems produced by the collisions, with
$N/Z$ varying from 1.27 to 1.50, have to stay
long enough in the spinodal region ($\sim$~3 characteristic times) to allow
important amplification of the initial fluctuations, one can
qualitatively 
understand the large extra reduction of the correlation signal for the
neutron rich system at high incident energy as coming from
insufficient reaction time. Such a situation also favours coalescence
of primary fragments. Finally, note that the set of reaction trajectories
in the density - temperature plane, close to the border of the
spinodal region at 45 MeV per nucleon can be slightly different for the two
reactions.

To summarize on these experimental results one can say that, 
using charge correlations, the fossil signature of
spinodal instabilities i.e. the abnormal presence of nearly equal-sized
fragments, even if very low as expected, was definitively established at a
confidence level of around 6 - 7 $\sigma$. 
Associated to this weak signal, it is important to
underline again the dominating role of chaotic/stochastic dynamics 
driven by spinodal instabilities for
fragment formation, especially for finite systems. It has to do
with beating of modes, breaking
of translational symmetry and coalescence during the final step of fragment 
formation. In a more mathematical language, for fragment formation one
can say that at an early stage the unstable modes are independent and their 
amplitudes evolve exponentially. Then, the modes become progressively
coupled and the evolution grows more and more complicated as the
non-linearities gain importance. Moreover, after the formation of
nascent fragments which are in mutual nuclear interaction the system
seeks to organize itself, it is the final coalescence phase.
On the other hand it could be that spinodal fluctuations, even if
present, do not play the dominating role which then would come from
fluctuations due to many-body correlations (molecular dynamics models
- see~\ref{QAMD}).

Considering the
intrinsic weakness of the fossil signal, it would be valuable to have
another signature. 
At the present time one can  consider new studies with the 
advent of future accelerators
which will provide beams covering a broad range in $N/Z$ ratios.
Considering the LG phase transition for asymmetric nuclear matter
analysed in a mean field approach, two different mechanisms of phase
separation have been compared: equilibrium
related to the highly chaotic character of collisions involved to
produce hot nuclei and spinodal instabilities~\cite{Duc07}.
\begin{figure}
\begin{center}
\includegraphics*[width=0.65\textwidth]{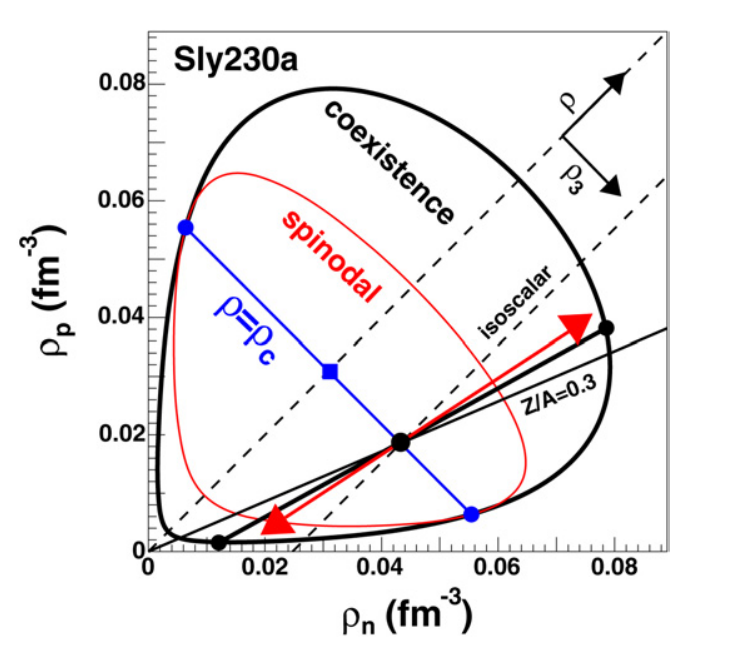}
\caption{Coexistence and spinodal regions in the
proton-neutron density plane at $T$ = 10 MeV. Illustration
of phase separation inside the instability region for matter with a proton
fraction $Z$/$A$ = 0.3 (see text). From~\cite{Duc07}.} \label{fig:EqSpino}
\end{center}	
\end{figure}
The isospin properties of
the phases are deduced from the free-energy curvature, which contains
information on the average isospin of the phases and on the system
fluctuations. The results are illustrated in Fig.~\ref{fig:EqSpino}
for neutron rich matter with $Z$/$A$ = 0.3 and a temperature $T$ of
10 MeV. If equilibrium is the origin of phase separation, the system
will undergo separation according to Gibbs construction. The two
phases, represented as black dots on the coexistence border do not belong
to the line of constant proton fraction ($Z$/$A$ = 0.3). The liquid
fraction is closer to symmetric matter than the gas phase. It is a
consequence of the symmetry energy minimization in the dense phase.
This unequal repartition of isospin between the two phases is the
well-known phenomenon of isospin fractionation (see also~\ref{asymmat}).
One can now study isospin fractionation if phase separation is driven
by spinodal instabilities. This is the spinodal fragmentation and the
local properties of the constrained entropy curvature determine the
development into phase separation. The double arrow in Fig.~\ref{fig:EqSpino}
shows the results; the spinodal fragmentation leads to a more pronounced
fractionation than equilibrium, the dense phase getting closer to the
symmetric matter. This fact can be a possible new signature of the dynamics of the
phase transition. However it appears as a very challenging task.
For experimentalists large $Z$/$A$ values are required to have enough
sensitivity. A robust reconstruction of primary fragments is also
mandatory. Moreover future $A$ and $Z$ identification arrays, like for example
FAZIA, are absolutely needed for such studies~\cite{Bou14,Sal16}.
On the theoretical side more realistic calculations involving
collisions between nuclei are also essential.

\subsection{Coherence of observed signals}\label{Coherence}
To conclude this section
one can say that it is now well-established that phase transitions can
be identified in finite systems such as hot nuclei. However their
manifestation is radically different to the behaviour expected in the
thermodynamic limit. A first-order phase transition manifests itself
in hot nuclei without real phase coexistence, these systems being
too small to contain different homogeneous phases and the interface
between them~\cite{Lab90,Wal94}. Rather they exhibit characteristic
behaviours determined by the same topological features of the
microcanonical entropy (local convexities) which lead to phase
separation in the thermodynamic limit.
\begin{figure}[htb]
\begin{center}
\includegraphics[width=0.6\textwidth]{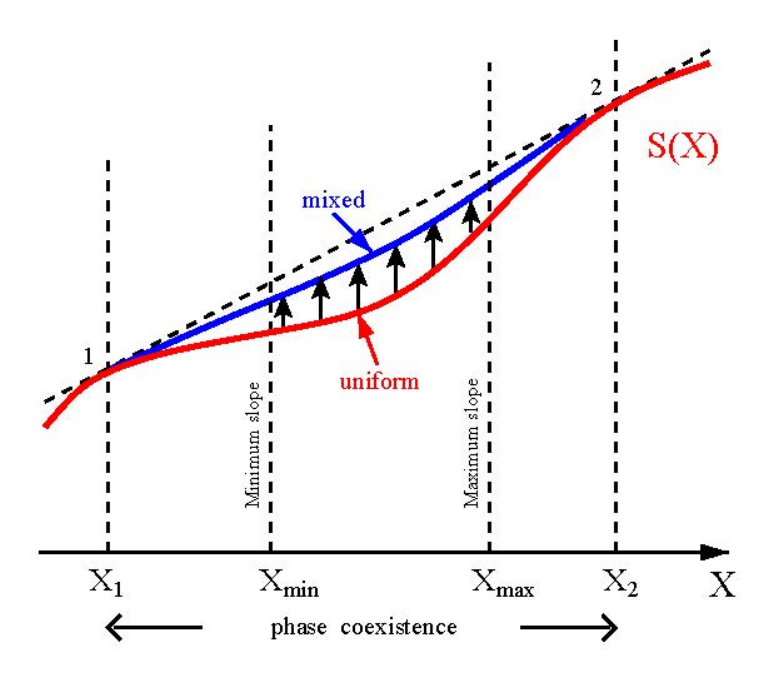}
\end{center}
\caption{
Isolated finite systems: the entropy function for a uniform unstable 
system (lower curve) has a local convexity region and the 
resulting equilibrium entropy function (upper curve) will always lie 
below the common tangent (dashed line). From~\cite{Cho04}. 
\label{convex}}
\end{figure}

For signal coherence, one can relate to Fig.~\ref{convex}. It shows the
local convexity region (upper curve) as a function of one single extensive
variable $X$ like energy; it also
indicates the spinodal region (lower curve) located in between $X_{min}$ and
$X_{max}$ and the mixed phaselike region in between $X_{1}$ and
$X_{2}$. This means that if a signature of spinodal instabilities is observed  
one must observe correlatively the bimodality of an order parameter
(canonical sampling), a negative microcanonical heat capacity and
a pressure constrained caloric curve with backbending, all related to the
resulting equilibrium entropy function with convexity (upper
curve). On the other hand, if spinodal instabilities are not responsible
for phaselike separation and if a phase transition occurs, the other 
signatures must be correlatively observed. 
Moreover, for hot nuclei, critical behaviours are expected to be
observed in the coexistence region due to the finite size of the system
which leads to the same effect as a diverging correlation length in
an infinite system.

From the body of phase transition signals discussed in this section,
a full coherence of phase transition signals is observed.
Starting from a weak signature of the phase transition dynamics (spinodal
instabilities), the three expected correlated signals: bimodality of an order
parameter (canonical sampling), negative microcanonical heat capacity
and pressure constrained caloric curve with backbending have been
observed. Moreover, critical exponents are also observed in the
coexistence region. From quasifusion hot nuclei produced in central Xe+Sn
collisions, for which information is the most complete, the thermal
excitation energy domain 3.5 - 9.5 MeV per nucleon corresponds to 
the spinodal - coexistence region.

From universal fluctuations ($\Delta$-scaling) the size of the
largest fragment of each partition was determined as an order parameter
in coherence with the bimodality observed. The determination of such
an order parameter which corresponds to an aggregation
scenario, in addition to the observed system mass and bombarding
energy dependence of the associated probability distributions which
was linked to the onset and importance of radial collective energy due
to expansion, also agrees with the process of spinodal fragmentation.

\section{Conclusions}

When the first study programs concerning a phase transition for hot
nuclei were proposed, many pessimistic comments were made related with
the difficulties to overcome, for example:\\
(i) - produced by collisions, hot nuclei in themselves are transient in
nature;\\
(ii) - the highly dynamical nature of collisions between objects with
a small number of constituents and some direct processes involved
between nucleons seem incompatible with the application  of well
known thermodynamic concepts used to describe phase transitions for
macroscopic systems;\\ 
(iii) - phase transitions cannot be defined for finite systems, even
talking about ``phases'' for small systems like nuclei makes no sense.

The difficulties have been largely overcome on
both theoretical and experimental sides. The key words for that have
been: thermodynamics of non-additive systems, stochastic mean field approach,
chaoticity and large covering of phase space, statistical models,
universal fluctuations, sophisticated 4$\pi$ detectors, data selection,
homogeneous event samples,
exclusive analysis, quasicomplete events, collective
variables on an event-by-event basis, intra-event correlations.
Close collaboration between theorists and experimentalists has also
been very valuable in defining strategies for analysing data.

As we have seen all along this review an enormous progress has been done
even if some points can be deeper investigated.

It is important to recall here one last time that signals of phase transition
for finite systems are only meaningful at the level of statistical
ensembles constructed from the outcome of many carefully selected
collisions, and this fact must always be borne in mind for correct
interpretation. The properties at the freeze-out instant of a very
large number of similar collisions reveal the fingerprints of the
phase transition.

Whatever the results of a limited number of experiments given the
cumbersome nature of the experiments and the analyses, the expected coherence
between signals indicating a first-order phase transition has been
observed. There is also now a good understanding of the observation of critical
exponents in the region of coexistence compatible with the class of
universality. The theory of universal fluctuations ($\Delta$-scaling)
was used to
determine that an order parameter of the phase transition for hot
nuclei is the size of the largest fragment
of each partition which is coherent with the bimodality observed for
this fragment size.
For the phase transition dynamics the presence of spinodal
fluctuations was definitively established in coherence with aggregation
process deduced from fluctuations of the largest fragment size.
However the weakness of the signature could also indicate that chaoticity
is dominating to produce fragments.

To progress further, besides a few proposed new signatures to be
confronted to data, the main effort has to be made from the experimental
side by identifying not only charge but also mass of fragments.  
Knowing $Z$ and $A$ one would be able to reduce or suppress hypotheses 
made to evaluate freeze-out properties which are, as we have seen, key
points to derive quantitative information. 
It would become possible to disentangle for fragments Coulomb repulsion from
collective radial expansion and consequently to experimentally access
this energy component obtained up to now from a parameter of statistical models
fitting well the data. 
By largely varying the proportions of neutrons and protons involved in
collisions new signatures related to the phase transition for nuclei are predicted: 
the distillation, which makes the `gas' phase
more asymmetric than the `liquid' phase (even more asymmetric for spinodal
decomposition as compared to phase equilibrium). 
By measuring the isotopic composition of all the fragments it would become 
possible to better identify the two phaselike forms and determine by a
new way the dynamics of the transition.
Finally, on the theoretical side a full quantal description of collisions would be
the supreme outcome; for example at present radial collective
energy due to thermal pressure cannot be calculated.



\end{document}